\documentclass[prd,aps,showpacs,nofootinbib,tightenlines]{revtex4}
\usepackage{mathrsfs}
\usepackage{amsmath}
\usepackage{amssymb}
\usepackage{epsfig}
\usepackage{graphicx}
\usepackage{booktabs}
\usepackage{multirow}
\usepackage{subfigure}
\usepackage[section]{placeins}
\usepackage{float}
\usepackage{slashed}
\begin{document}
\newcommand{\psl}{ p \hspace{-1.8truemm}/ }
\newcommand{\nsl}{ n \hspace{-2.2truemm}/ }
\newcommand{\vsl}{ v \hspace{-2.2truemm}/ }
\newcommand{\epsl}{\epsilon \hspace{-1.8truemm}/\,  }

\title{Charmonium decays of  beauty baryons in the perturbative QCD approach } 
\author{Zhou Rui$^1$}\email[Corresponding  author: ]{jindui1127@126.com}
\author{Zhi-Tian Zou$^2$}\email[Corresponding  author: ]{zouzt@ytu.edu.cn}
\affiliation{$^1$College of Sciences, North China University of Science and Technology,
Tangshan 063009,  China}
\affiliation{$^2$Department of Physics, Yantai University, Yantai 264005, China}
\date{\today}
\begin{abstract}

Motivated by recent advances in experimental measurements of heavy baryon decays,
the charmonium decays of single $b$ baryon are investigated systematically in the framework of perturbative QCD approach.
We calculate the decay branching ratios  and helicity amplitudes as well as many pertinent decay asymmetry parameters that characterize the angular decay distributions.
In particular, we estimate the fragmentation fractions of $b$ quark to $b$ baryons based on the current world averages of the branching ratio multiplied by fragmentation fraction.
Some of the featured results are compared with the predictions of other approaches and experimental data.
We also evaluate the branching ratios and angular asymmetries of the $\Sigma_b$ and $\Xi'_b$ decay modes,
which have neither been measured experimentally nor calculated theoretically.
Our results could be helpful for a future experimental search for them
and provide deeper insights in the understanding of the dynamics of heavy-flavor weak decay processes.

\end{abstract}



\maketitle

\section{Introduction}
 Single heavy flavor baryons  consisting  of  one  heavy quark  and two light quarks within the quark model
can be described in terms of SU(3) flavor multiplets~\cite{Roberts:2007ni}.  
In the bottom baryon sector, the lowest-lying $\frac{1}{2}$ states contain a flavor SU(3) antitriplet $\bar {\textbf{3}}$ and a sextet $ \textbf{6}$~\cite{Li:2021kfb}.
The antitriplet $\bar {\textbf{3}}$ means  the two light quarks are in an $S$-wave state with $S=0$ (e.g., $\Lambda_b$ and $\Xi_b$),
while the sextet $\textbf{6}$  with $S=1$ (e.g., $\Sigma_b, \Xi'_b$ and $\Omega_b$).
$\Lambda_b$ and $\Xi_b$ are the lightest baryons in their respective quark contents, so they can only decay via weak interactions.
$\Sigma_b(\Xi'_b)$ is heavier than $\Lambda_b(\Xi_b)$,
which should decay predominantly strongly through the $P$-wave one pion transition $\Sigma_b(\Xi'_b)\rightarrow\Lambda_b(\Xi_b)\pi$~\cite{CDF:2007oeq,LHCb:2014nae}.
Since a scalar $ss$ diquark is forbidden by the Pauli principle, the $b$ baryon with $bss$ quark components  must be sextet.
While the lowest-lying sextet $\Omega_b$ lies below the threshold for the decays into  $\Xi_bK$, it only decays weakly.
As a tremendous amount of $b$-flavored baryons is produced at the LHC~\cite{Dainese:2019rgk},
their weak decays provide a complementary laboratory to search for effects beyond the Standard Model (SM) and
offer rich angular structures compared to $B$ meson decays~\cite{Shen:2023eln,Wang:2022fih}.

Many hadronic $b$ baryon decays have been observed to date~\cite{pdg2022}.
Among the numerous beauty baryon decay channels,
the category involving the $J/\psi$ final state, proceeds through $b\rightarrow sc\bar c$  and $b\rightarrow dc\bar c$  transitions,
is more easily reconstructed due to the narrow peak of $J/\psi$ and the high purity of $J/\psi\rightarrow l^+l^-$, thus is particularly interesting.
So far various collaborations have reported the observation of the decays $\Lambda_b\rightarrow \Lambda J/\psi$~\cite{UA1:1991vse,CDF:1996rvy,D0:2011pqa},
 $\Lambda_b\rightarrow \Lambda \psi(2S)$~\cite{ATLAS:2015hik,LHCb:2019fim},
$\Xi_b\rightarrow \Xi J/\psi$~\cite{D0:2007gjs,CDF:2009sbo}, $\Xi_b\rightarrow \Lambda J/\psi$~\cite{LHCb:2019aci}, $\Omega_b\rightarrow \Omega J/\psi$~\cite{CDF:2009sbo,D0:2008sbw,LHCb:2023qxn,Nicolini:2023stq}.
However, the absolute branching fractions cannot be accessed at the moment  due to a lack of knowledge of the fragmentation fractions,
which quantify the probabilities for a bottom quark to hadronize into a certain weakly decaying $b$ hadron.
Instead, the product of branching ratio and the relevant fragmentation fraction is determined
relative to the corresponding values for the topologically similar normalization channel.
The current experimental knowledge of the fragmentation fractions multiplied by the branching fraction related to the charmonium decays of $b$ baryon
are listed in Table~\ref{tab:measure}.
These efforts provide a great platform for us to deeply understand the dynamics involved in the heavy-flavor baryon weak decays.

\begin{table}[!htbh]
	\caption{
A summary of measured fragmentation fractions multiplied by the branching fractions related to the charmonium decays of $b$ baryon.
The first three rows are taken from the 2023 update of the Particle Data Group (PDG)~\cite{pdg2022},
while the remaining are quoted from collaborations,  where the first uncertainty is statistical the second is systematic.   } 
\newcommand{\tabincell}[2]{\begin{tabular}{@{}#1@{}}#2\end{tabular}}
	\label{tab:measure}
	\begin{tabular}[t]{lcc}
		\hline\hline
 Parameter  & Measurement   & Source  \\ \hline		
$f_{\Lambda_b}\mathcal{B}(\Lambda_b\rightarrow\Lambda J/\psi)$       &$(5.8\pm0.8)\times 10^{-5}$                       & PDG~\cite{pdg2022} \\ \hline	
$f_{\Xi_b^-}\mathcal{B}(\Xi_b^-\rightarrow \Xi^- J/\psi)$            &$(1.02^{+0.26}_{-0.21})\times 10^{-5}$            & PDG~\cite{pdg2022}\\ \hline	
$f_{\Omega_b^-}\mathcal{B}(\Omega_b^-\rightarrow \Omega^- J/\psi)$   &$(2.9^{+1.1}_{-0.8})\times 10^{-6}$               & PDG~\cite{pdg2022} \\ \hline	

$\frac{f_{\Xi_b^-}}{f_{\Lambda_b}}\frac{\mathcal{B}(\Xi_b^-\rightarrow\Xi^- J/\psi)}
{\mathcal{B}(\Lambda_b\rightarrow\Lambda J/\psi)}$
&\tabincell{c}{$0.28\pm0.09^{+0.09}_{-0.08}$\\$0.167^{+0.037}_{-0.025}\pm0.012$\\$(10.8\pm0.9\pm0.8)\times10^{-2}[\sqrt{s}=7,8 TeV]$\\$(13.1\pm1.1\pm1.0)\times10^{-2}[\sqrt{s}=13 TeV]$}
& \tabincell{c}{D0~\cite{D0:2007gjs}\\  CDF~\cite{CDF:2009sbo}\\  LHCb~\cite{LHCb:2019sxa} \\  LHCb~\cite{LHCb:2019sxa}  } \\ \hline	
$\frac{f_{\Omega_b^-}}{f_{\Lambda_b}}\frac{\mathcal{B}(\Omega_b^-\rightarrow\Omega^- J/\psi)}
{\mathcal{B}(\Lambda_b\rightarrow\Lambda J/\psi)}$                   &$0.045^{+0.017}_{-0.012}\pm0.004$            &CDF~\cite{CDF:2009sbo}\\\hline	
$\frac{f_{\Omega_b^-}}{f_{\Xi_b^-}}\frac{\mathcal{B}(\Omega_b^-\rightarrow\Omega^- J/\psi)}
{\mathcal{B}(\Xi_b^-\rightarrow\Xi^- J/\psi)}$        &\tabincell{c}{$0.80\pm0.32^{+0.14}_{-0.22} $ \\$0.120\pm0.008\pm0.008$ \\$0.27\pm0.12\pm0.01$ }
                                                      &\tabincell{c}{D0~\cite{D0:2008sbw} \\LHCb~\cite{LHCb:2023qxn,Nicolini:2023stq}\\CDF~\cite{CDF:2009sbo} }  \\
		\hline\hline
	\end{tabular}
\end{table}

Stimulated by the promising prospect of experiments specially at LHC,
there have been a number of theoretical works on the heavy baryons decays in the literature~\cite{Mohanta:1998iu,prd531457,prd58014016,prd575632,mpla13181,prd80094016,ijmpa271250016,prd65074030,prd99054020,prd96013003,prd88114018,prd92114008,plb614165,Leitner:2004dt},
but they mainly focus on the $\Lambda_b$ modes.
More systematic studies of the charmonium decays involved other $b$ baryon multiplets
have been explored using the factorization ansatz (FA)~\cite{Fayyazuddin:2017sxq}, the nonrelativistic quark model (NQM)~\cite{Cheng:1996cs},
covariant confined quark model (CCQM)~\cite{Gutsche:2018utw}, generalized factorization approach (GFA)~\cite{Hsiao:2015cda},  and light-front quark model (LFQM)~\cite{Hsiao:2021mlp}.
The SU(3) flavor analysis for beauty baryon decays has been performed in~\cite{Dery:2020lbc}, in which some sum rules in the SU(3) limit are presented.
Furthermore, beauty baryon fragmentation fractions in the charmonium decays of $b$ baryon have been discussed in~\cite{Hsiao:2021mlp,plb751127,Jiang:2018iqa}.
These calculations  are important to check if the results in the SM are consistent with the experimental measurements.

The perturbative QCD approach (PQCD) is a powerful theoretical tool in exploring the weak decays of heavy-flavored hadrons and
predicts many features of $B$ meson decays~\cite{Keum:2000wi,Lu:2000em}.
The application of the PQCD formalism extension to baryon decays has also achieved a preliminary success~\cite{prd59094014,prd61114002,cjp39328,prd65074030,prd74034026,prd80034011,220204804,220209181,prd106053005,221015357,230213785,Rui:2023fpp}.
The hadronic LCDAs, which describe the momentum fraction distribution of valence quarks inside hadrons,
are primary nonperturbative quantities for calculating the heavy baryon decays based on the PQCD approach.
As of today, many efforts have been made to construct the heavy-baryon LCDAs~\cite{plb665197,jhep112013191,epjc732302,plb738334,jhep022016179,Ali:2012zza}.
Meanwhile,  the LCDAs of light baryons have been studied using QCD sum rules~\cite{zpc42569,Farrar:1988vz,Liu:2014uha,Liu:2008yg,Braun:2000kw,Braun:1999te},
lattice QCD~\cite{QCDSF:2008qtn,Gockeler:2008xv,jhep020702016,prd89094511,epja55116},  the chiral quark-soliton model~\cite{Kim:2021zbz},
and the Dyson-Schwinger equations~\cite{Mezrag:2017znp,Mezrag:2018hkk}.
Quite recently, the one-loop perturbative contributions to LCDA of a light baryon have been derived in large-momentum effective theory~\cite{Deng:2023csv},
which provides a first step toward deriving the LCDA from first principle lattice QCD calculations in the future.
Facing these exciting status, one has reason to believe that it is a suitable time to investigate the charmonium decays of $b$ baryon, 
which is the focus of this presentation.

In this article, we systematically analyze the weak nonleptonic decays of $\mathcal{B}_b\rightarrow \mathcal{B} J/\psi$  using PQCD approach,
where $\mathcal{B}_b$ represents the lowest-lying spin $\frac{1}{2}$ ground-state bottom baryons,
 while $\mathcal{B}$ stands for the light baryon octet and decuplet. 
More higher bottom baryons, such as the spin $\frac{3}{2}$ partners decay strongly or electromagnetically,
are beyond the scope of the present work.
Furthermore, the vertex corrections~\cite{Beneke:1999br,Beneke:2000ry,Beneke:2003zv} to the factorizable amplitude have also been taken into account,
whose effects are combined in the Wilson coefficients similar to the cases of hadronic charmonium $B$ meson decays.

The paper is arranged as follows.
We first classify the decays according to the SU(3) flavor representation of the initial and final states in Sec~\ref{sec:framework}.
Then with the introduction of the light-cone distribution amplitudes of baryons,
we will present the general formulas of the amplitudes, decay rates, and various asymmetry observables.
In Sec~\ref{sec:results}, we give the numerical results for the branching ratios
and use them to extract the baryon fragmentation fractions.
Subsequently, numerical results for various asymmetries are discussed and compared in detail.
The last section contains our summary.
As representative examples,  the factorization formula for $\Omega_b\rightarrow\Omega J/\psi$ decay, are presented in Appendix~\ref{sec:for}.

\section{Theoretical framework}\label{sec:framework}

\subsection{Classification of decays}\label{sec:class}
\begin{figure}[!htbh]
	\begin{center}
	    \vspace{0.01cm} \centerline{\epsfxsize=10cm \epsffile{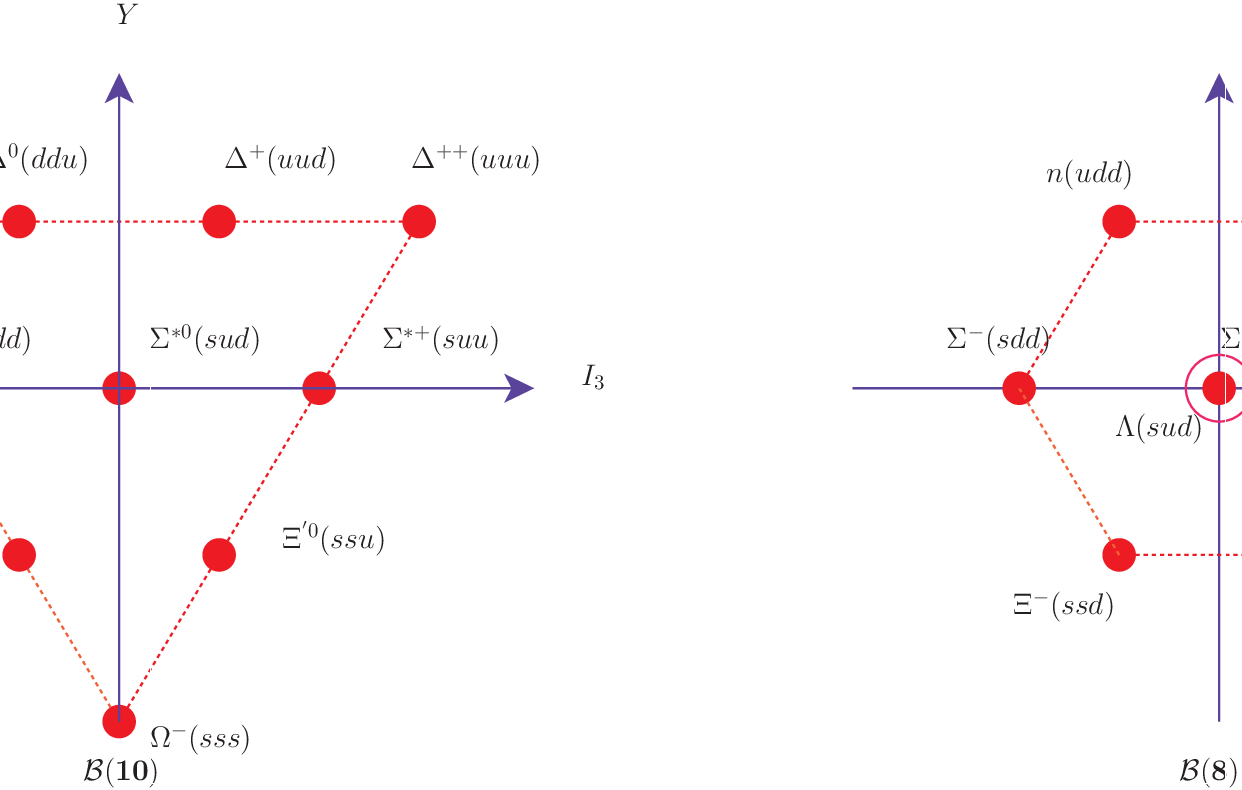}}
		\setlength{\abovecaptionskip}{1.0cm}
		\caption{The SU(3) flavor multiplets of decuplet (left panel) and octet (right panel) in the $Y-I_3$ plane,
where $Y$ and $I_3$ denote the hypercharge and the third component of isospin, respectively.}
		\label{fig:picF1}
	\end{center}
\end{figure}

\begin{figure}[!htbh]
	\begin{center}
		\vspace{0.01cm} \centerline{\epsfxsize=15cm \epsffile{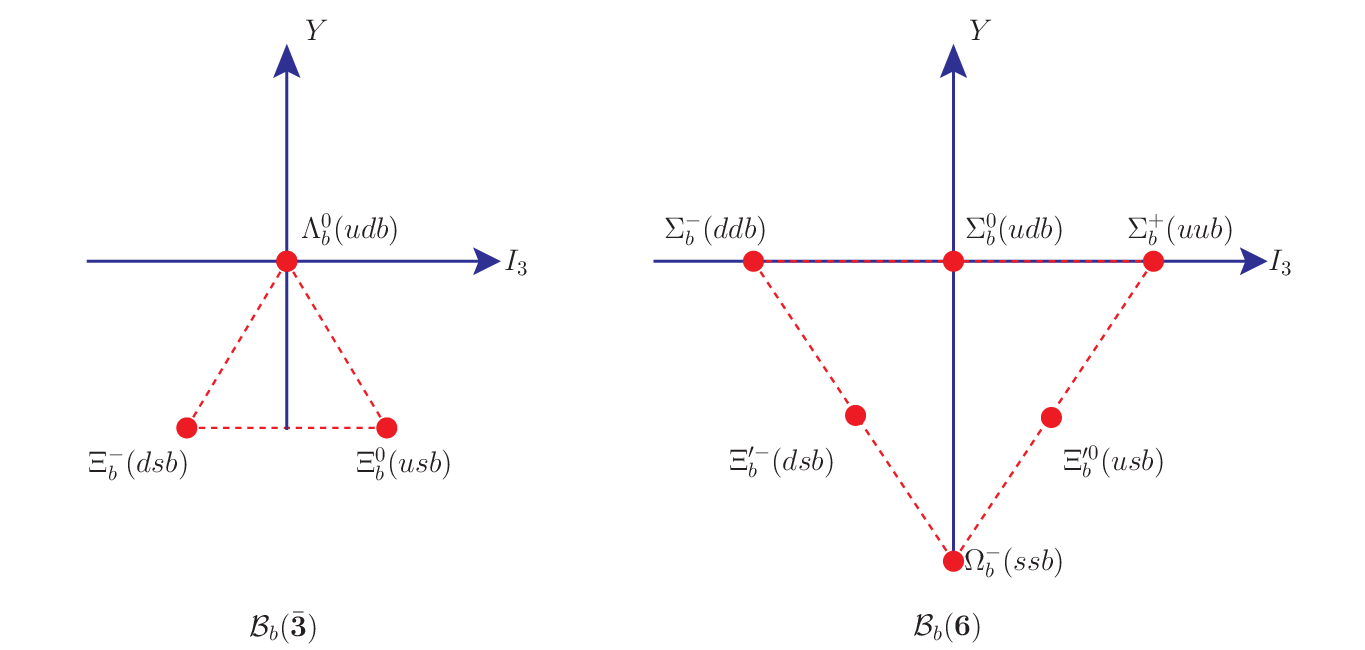}}
		\setlength{\abovecaptionskip}{1.0cm}
\caption{The antitriplet (left) and sextet (right) representations of the ground-state single $b$ baryons.}
		\label{fig:picF2}
	\end{center}
\end{figure}
Baryon and meson are the two basic bound states in the conventional quark model.
The former is composed of  three quarks, while the latter is made of a quark and an antiquark.
They are all in color singlet states.
In the language of group theory, three light quarks $u,d,s$ and their antiparticle partners belong to $\textbf{3}$ and $\bar{\textbf{3}}$ representations, respectively.
The combination of $\textbf{3}$ and $\bar{\textbf{3}}$ can be decomposed into a singlet and an octet,
\begin{eqnarray}
\textbf{3}\otimes\bar{\textbf{3}} =\textbf{1}\oplus \textbf{8}.
\end{eqnarray}
The charmonium is a bound state of heavy quarkonium  $c\bar c$, which is outside the three flavors $u,d,s$ of SU(3).
Likewise, the combination of three quarks in SU(3) flavor symmetry gives a totally antisymmetric singlet state,
  two mixed symmetry octets, and a symmetric decuplet,
\begin{eqnarray}
\textbf{3}\otimes\textbf{3}\otimes\textbf{3}=\textbf{1}\oplus \textbf{8}_A\oplus\textbf{8}_S \oplus\textbf{10},
\end{eqnarray}
where the subscripts $A$ and $S$ signify the mixed antisymmetry and symmetry, respectively.
In Fig.~\ref{fig:picF1}, we display the observed octet and decuplet of light baryon states in the $Y-I_3$ plane,
where $Y$ and $I_3$ denote the hypercharge and the third component of isospin, respectively.
They are related by $Y=2(Q-I_3)$ with $Q$ being the charge.
In the following,
the ground octet and decuplet baryons will be labeled by $\mathcal{B}(\textbf{8})$ and $\mathcal{B}(\textbf{10})$,
corresponding to $J^P=\frac{1}{2}^+$ and $\frac{3}{2}^+$, respectively.

The baryons containing a single bottom quark can also be described in terms of SU(3) flavor multiplets.
The combination of the two light quarks can be written as $\textbf{3}\otimes\textbf{3}=\bar{\textbf{3}}\oplus \textbf{6}$.
The antitriplet $\bar {\textbf{3}}$ means the spin of the two light quarks is in $S=0$ state;
thus, its spin degree of freedom is antisymmetric.
The sextet $\textbf{6}$ denotes the spin of the two light quarks is in $S=1$ state and thus is symmetric.
In the case of the lowest-lying ground state, which has zero orbital angular momentum $l=0$,
 the orbital degree of freedom is symmetric.
Remembering that  the color wave function of any baryon is antisymmetric invariably,
the Pauli principle requires that $\bar {\textbf{3}}$ and $\textbf{6}$ should have the antisymmetric and symmetric flavor structures, respectively.
The nine ground-state single $b$ baryons with spin-parity $J^P=\frac{1}{2}^+$
involve an isospin singlet $\Lambda_b$, a triplet $\Sigma_b$,  two strange doublets $\Xi_b$ and $\Xi'_b$, and a doubly strange state $\Omega_b$~\cite{pdg2022}.
They can form an  antitriplet $\Lambda_b,\Xi_b^0,\Xi_b^-$ and a  sextet $\Sigma_b^{0,\pm},\Xi_b^{'0},\Xi_b^{'-},\Omega_b$, as depicted in Fig~\ref{fig:picF2}.

According to the SU(3) representations of the initial and final baryons,
all the possible two-body charmnium decays of bottom baryon that comply with the charge conservation can be classified into the following four groups:
\begin{itemize}
  \item $\mathcal{B}_b(\bar {\textbf{3}})\rightarrow\mathcal{B}(\textbf{8})J/\psi$:\\
  \quad $(\Lambda_b,\Xi_b^0)\rightarrow (\Lambda,n,\Sigma^0,\Xi^0)J/\psi$, \quad $\Xi_b^-\rightarrow (\Xi^-,\Sigma^-)J/\psi$,
  \item $\mathcal{B}_b(\bar {\textbf{3}})\rightarrow\mathcal{B}(\textbf{10})J/\psi$:\\
  \quad $(\Lambda_b,\Xi_b^0)\rightarrow (\Delta^0,\Sigma^{*0},\Xi^{*0})J/\psi$, \quad $\Xi_b^-\rightarrow (\Sigma^{*-},\Xi^{*-},\Delta^-,\Omega^-)J/\psi$,
  \item $\mathcal{B}_b(\textbf{6})\rightarrow\mathcal{B}(\textbf{8})J/\psi$:\\
  \quad $\Sigma_b^{+}\rightarrow(\Sigma^+,p)J/\psi$, \quad $(\Sigma_b^{0},\Xi_b^{'0})\rightarrow(\Lambda,\Xi^0,\Sigma^0,n)J/\psi$,
  \quad $(\Sigma_b^{-},\Xi_b^{'-},\Omega_b^-)\rightarrow(\Sigma^-,\Xi^-)J/\psi$,
  \item $\mathcal{B}_b(\textbf{6})\rightarrow\mathcal{B}(\textbf{10})J/\psi$:\\
  \quad $\Sigma_b^{+}\rightarrow(\Sigma^{*+},\Delta^+)J/\psi$, \quad $(\Sigma_b^{0},\Xi_b^{'0})\rightarrow(\Delta^0,\Xi^{*0},\Sigma^{*0})J/\psi$,
  \quad $(\Sigma_b^{-},\Xi_b^{'-},\Omega_b^-)\rightarrow(\Sigma^{*-},\Xi^{*-},\Delta^-,\Omega^-)J/\psi$.
\end{itemize}
Note that some modes, such as $\Lambda_b\rightarrow \Xi^{(*)0}J/\psi, \Xi^0_b\rightarrow (n,\Delta^0) J/\psi$,
only receive contributions from the vertical $W$-loop diagrams, which are prohibited in the baryon decays~\cite{Chau:1995gk}.
They may be admitted by two insertions of weak effective operators~\cite{Dery:2020lbc}, which  are, however, highly suppressed and not considered here.
Since the two spectator light quarks in $\mathcal{B}_b(\bar {\textbf{3}})$ are antisymmetry and symmetry in $\mathcal{B}(\textbf{10})$,
it is clear that the $W$-emission diagrams cannot contribute to $\mathcal{B}_b(\bar {\textbf{3}})\rightarrow\mathcal{B}(\textbf{10})J/\psi$ channels.
However, they can proceed via the $W$-exchange topologies albeit are dynamically suppressed
just like the isospin-violating decays of $\Lambda_b\rightarrow \Sigma^0 J/\psi$ and $\Sigma^0_b\rightarrow \Lambda J/\psi$.
The study of these modes can  improve our knowledge of the $W$-exchange mechanism.
The LCDAs of most decuplet baryons are still unknown yet, and we limit out attention to the $\Delta$ and $\Omega$ final states in the $\mathcal{B}(\textbf{10})$ modes.
Except $\Omega_b$,  all other sextet baryons will decay predominantly strongly or electromagnetically, and thus, their weak branching ratios will be very small.
Nevertheless, from another perspective, these rare decays may be more sensitive to new physics; therefore, deeper investigation might be warranted~\cite{Ke:2012wa}.
Moreover, those Cabibbo-Kobayashi-Maskawa (CKM)  favored decays, which are of experimental interest, will be  discussed  in detail in the part of numerical results.

\subsection{Light-cone distribution amplitudes}\label{sec:LCDAs}

Light-cone distribution amplitudes (LCDAs),
which can be constructed via the nonlocal matrix elements of the non-local light-ray operators between the vacuum and the hadronic state,
are essential quantities that describe the nonperturbative physics in high energy QCD exclusive processes.
Once reasonable models of the LCDAs are determined, we can take them as basic input ingredients to make predictions 
in a systematic ways with the PQCD approach.
We now summarize the definitions of the LCDAs for the initial and final states.

\subsubsection{$b$ baryon light-cone distribution amplitudes}
LCDAs of heavy baryons can be simplified in the heavy-quark symmetry limit.
The heavy $b$ quark decouples from the light quark pair (diquark) with aligned helicities  in the leading order of the heavy quark mass expansion and  can be regarded as a nonrelativistic particle.
Then the ground-state baryons  with the spin parity $J^P$ are characterized by the spin parity $j^p$ of the diquark.
A diquark can be either in the spin singlet $0^+$ or spin triplet states $1^+$.
The $0^+$ diquark combines with the $b$ quark to form one of the $\bar {\textbf{3}}$ multiplets as mentioned before, which is antisymmetric under the interchange of the two light quarks.
The set of the LCDAs in the rest frame of the heavy baryon can be parametrized as~\cite{epjc732302}
\begin{eqnarray}\label{eq:LCDAs0}
\langle0|[q_1(t_1)C\gamma_5\slashed{n}q_2(t_2)]Q(0)|H_b^{j=0}\rangle&=&\psi^{n}(t_1,t_2)f^{(1)} u_b,\nonumber\\
\langle0|[q_1(t_1)C\gamma_5\slashed{\bar n}q_2(t_2)]Q(0)|H_b^{j=0}\rangle&=&\psi^{\bar n}(t_1,t_2)f^{(1)} u_b,\nonumber\\
\langle0|[q_1(t_1)C\gamma_5q_2(t_2)]Q(0)|H_b^{j=0}\rangle&=&\psi^{1}(t_1,t_2)f^{(2)} u_b,\nonumber\\
\frac{i}{2}\langle0|[q_1(t_1)C\gamma_5\sigma_{\bar nn}q_2(t_2)]Q(0)|H_b^{j=0}\rangle&=&\psi^{\bar nn}(t_1,t_2)f^{(2)}u_b,
\end{eqnarray}
where $\sigma_{\bar nn}=\frac{i}{2}(\slashed{\bar n}\slashed{n}-\slashed{ n}\slashed{\bar n})$ %
 with $n,\bar n$ being two light-cone vectors satisfy $n^2=\bar n^2=0$ and  $n \cdot \bar n=2$.
 $C$ denotes the charge conjugation matrix.
Here, $Q(0)$ is the static heavy-quark field situated at the origin of the position-space frame, and
$q_{i}(t_i)$  is light quark field separated by a lightlike distance $t_i$
between the $i$th light quark and the origin along the direction of $n$.
$u_b$ is the Dirac spinor of the heavy $b$ quark, and its momentum dependence, spins, and spinor index of $u_b$ are suppressed in the notation.
 The SU(3) symmetry requites that $\psi^{\bar nn}$ is antisymmetric under permutation of two light quarks, but $\psi^{1,n,\bar n}$ are symmetric under the same operation.

$f^{(1,2)}(\mu)$ are the decay constants at some representative scale $\mu$, defined by the interpolating current~\cite{epjc732302}
\begin{eqnarray}\label{eq:LCDAsp0}
\epsilon^{ijk}\langle0|[q^{iT}_1(0)C\gamma_5q^{j}_2(0)]Q^k(0)|H_b^{j=0}\rangle&=&f^{(1)} u_b,\nonumber\\
\epsilon^{ijk}\langle0|[q^{iT}_1(0)C\gamma_5\slashed{v}q^{j}_2(0)]Q^k(0)|H_b^{j=0}\rangle&=&f^{(2)} u_b,
\end{eqnarray}
where $i$, $j$, and $k$ are color indices, and $\epsilon^{ijk}$ is the totally antisymmetric tensor.
As the interpolating currents for the baryons are not uniquely determined~\cite{Shi:2019hbf},
 different interpolating currents and input parameters will lead to distinct decay constants~\cite{Groote:1997yr,Groote:1996em,Wang:2010fq,Wang:2009cr}.
For instance,  the values from~\cite{Wang:2010fq,Wang:2009cr} are typical larger than those from~\cite{Groote:1997yr,Groote:1996em}.
Since in this work, we utilize the same interpolating currents as in~\cite{Groote:1997yr,Groote:1996em},
we should adopt their numerical values for the decay constants for consistency.
As argued in~\cite{220204804}, the assumption of $f^{(1)}=f^{(2)}$ may be not guaranteed in nondiagonal and mixed sum rules~\cite{Groote:1997yr}
as well as including the next-to-leading-order corrections~\cite{Groote:1996em}.
We favor to choose the values $f^{(1)}_{\Lambda_b}=f^{(2)}_{\Lambda_b}=0.022 \text{GeV}^3$ and
$f^{(1)}_{\Sigma_b}=f^{(2)}_{\Sigma_b}=0.031 \text{GeV}^3$ from the diagonal sum rule calculation at the leading-order level~\cite{Groote:1997yr}.
Because the decay constants of other $b$ baryons are less known in~\cite{Groote:1997yr,Groote:1996em},
to make our predictions,  we boldly envisage  that the decay constants of the baryons in the same SU(3) flavor multiplet coincide approximately.

The matrix element can be expressed in terms of the four LCDAs in the following way~\cite{Wang:2009hra,jhep112013191}:
\begin{eqnarray}\label{eq:LCDAs20}
\epsilon^{ijk}\langle0|q^i_{1\alpha}(t_1)q^j_{2\beta}(t_2)Q^k_\gamma(0)|\mathcal{B}_b(\bar {\textbf{3}})\rangle &=&\frac{ f^{(1)}}{8}
[(\slashed{\bar n}\gamma_5C)_{\alpha\beta}\psi^{n}(t_1,t_2)+(\slashed{ n}\gamma_5C)_{\alpha\beta}\psi^{\bar n}(t_1,t_2)](u_{\bar {\textbf{3}}})_\gamma \nonumber\\
&&+\frac{f^{(2)}}{4}[(\gamma_5C)_{\alpha\beta}\psi^{1}(t_1,t_2)-\frac{i}{2}(\sigma_{\bar nn}\gamma_5C)_{\alpha\beta}\psi^{\bar nn}(t_1,t_2)](u_{\bar {\textbf{3}}})_\gamma
\nonumber\\&=&\frac{1}{4}
\{f^{(2)}[M_1(t_1,t_2)\gamma_5C^T]_{\beta\alpha}
+f^{(1)}[M_2(t_1,t_2)\gamma_5C^T]_{\beta\alpha}\}(u_{\bar {\textbf{3}}})_\gamma,
\end{eqnarray}
with  $\alpha,\beta,\gamma$ being the Dirac indices.
In the case of $\bar {\textbf{3}}$ state, the bottom quark carries all of the angular momentum of the baryon, so the heavy-baryon spinor
$u_{\bar {\textbf{3}}}$ is nothing else but the heavy-quark spinor $u_b$.
The chiral-even and -odd projectors read as
\begin{eqnarray}
M_1(t_1,t_2)&=&\frac{1}{8}[\slashed{ n}\slashed{\bar n}\Psi_3^{+-}(t_1,t_2)+\slashed{\bar n}\slashed{ n}\Psi_3^{-+}(t_1,t_2)],\nonumber\\
M_2(t_1,t_2)&=&\frac{1}{2}[\slashed{\bar n}\Psi_2 (t_1,t_2)+\slashed{ n}\Psi_4 (t_1,t_2)],
\end{eqnarray}
respectively,
where
\begin{eqnarray}
\Psi_2(t_1,t_2)=\psi^{n}(t_1,t_2), \quad \Psi_4(t_1,t_2)=\psi^{\bar n}(t_1,t_2),\quad  \Psi_3^{\pm\mp}(t_1,t_2)=2[\psi^{1}(t_1,t_2)\pm\psi^{ \bar nn}(t_1,t_2)].
\end{eqnarray}	
The corresponding LCDAs up to twist-4 accuracy in the momentum space can be written as~\cite{jhep112013191}
\begin{eqnarray}
(\Psi_{\bar {\textbf{3}}})_{\alpha\beta\gamma}(x_i,\mu)=\frac{1}{8N_c}
\{f^{(1)}(\mu)[M_1(x_2,x_3)\gamma_5C^T]_{\beta\alpha}
+f^{(2)}(\mu)[M_2(x_2,x_3)\gamma_5C^T]_{\beta\alpha}\}[u_{\bar {\textbf{3}}}(p)]_\gamma,
\end{eqnarray}
with
\begin{eqnarray}
M_1(x_2,x_3)&=&\frac{\slashed{n}^-\slashed{n}^+}{4}\Psi_3^{+-}(x_2,x_3)
+\frac{\slashed{n}^+\slashed{n}^-}{4}\Psi_3^{-+}(x_2,x_3), \nonumber\\
M_2(x_2,x_3)&=&\frac{\slashed{n}^+}{\sqrt{2}}\Psi_2(x_2,x_3)
+\frac{\slashed{n}^-}{\sqrt{2}}\Psi_4(x_2,x_3),
\end{eqnarray}
where $n^+=(1,0,\textbf{0}_T)$ and $n^-=(0,1,\textbf{0}_T)$ are two dimensionless vectors on the light cone, satisfying $n^+\cdot n^-=1$.
$x_{2,3}$ are the light quark longitudinal momentum fractions   inside the  $b$ baryon.
 $N_c$ is the number of colors.

For the case of the sextet $b$ baryon with spin-parity $J^P=\frac{1}{2}^+$
in which the light diquark state is the axial-vector state with $j^p=1^+$,
one needs to consider the sextet baryons with the longitudinal and transverse polarizations separately.
The parallel LCDAs  in the rest frame of the heavy baryon have been given in Ref~\cite{epjc732302}:
\begin{eqnarray}\label{eq:LCDAs}
\bar{v}^\mu\langle0|[q_1(t_1)C\slashed{n}q_2(t_2)]Q(0)|H_b^{j=1}\rangle&=&\frac{1}{\sqrt{3}}\psi_{\parallel}^{n}(t_1,t_2)f^{(1)}\epsilon_{\parallel}^\mu u_b,\nonumber\\
-\bar{v}^\mu\langle0|[q_1(t_1)C\slashed{\bar n}q_2(t_2)]Q(0)|H_b^{j=1}\rangle&=&\frac{1}{\sqrt{3}}\psi_{\parallel}^{\bar n}(t_1,t_2)f^{(1)}\epsilon_{\parallel}^\mu u_b,\nonumber\\
\bar{v}^\mu\langle0|[q_1(t_1)Cq_2(t_2)]Q(0)|H_b^{j=1}\rangle&=&\frac{1}{\sqrt{3}}\psi_{\parallel}^{1}(t_1,t_2)f^{(2)}\epsilon_{\parallel}^\mu u_b,\nonumber\\
i\frac{\bar{v}^\mu}{2}\langle0|[q_1(t_1)C\sigma_{\bar nn}q_2(t_2)]Q(0)|H_b^{j=1}\rangle&=&\frac{1}{\sqrt{3}}\psi_{\parallel}^{\bar nn}(t_1,t_2)f^{(2)}\epsilon_{\parallel}^\mu u_b,
\end{eqnarray}
where $\epsilon_{\parallel}^\mu $ is the diquark polarization vector,  and $\bar v=\frac{n-\bar n}{2}$.
$\psi_{\parallel}^{n,\bar n,1,\bar n n}$ are four parallel LCDAs with  definite twists.
In the SU(3) flavor symmetry limit, the LCDAs $\psi_{\parallel}^{1}(t_1,t_2)$ is antisymmetric
under the exchange $t_1\leftrightarrow t_2$ and normalized as $\psi_{\parallel}^{1}(0,0)=0$,
and the remaining three ones are symmetric and hence satisfy the condition $\psi_{\parallel}^{i}(0,0)=1$ with $i=n,\bar n,\bar n n$.
The two decay constants  $f^{(1,2)}$ are defined in the Heavy quark effective theory (HQET) as~\cite{epjc732302}
\begin{eqnarray}\label{eq:LCDAsp}
\epsilon^{ijk}\langle0|[q^{iT}_1(0)C(\gamma^{\mu}-\slashed{v}v^\mu)q^{j}_2(0)]Q^k(0)|H_b^{j=1}\rangle&=&\frac{1}{\sqrt{3}}f^{(1)}\epsilon^\mu u_b,\nonumber\\
\epsilon^{ijk}\langle0|[q^{iT}_1(0)C(\gamma^{\mu}-\slashed{v}v^\mu)\slashed{v}q^{j}_2(0)]Q^k(0)|H_b^{j=1}\rangle&=&\frac{1}{\sqrt{3}}f^{(2)}\epsilon^\mu u_b.
\end{eqnarray}
with $v=\frac{n+\bar n}{2}$ being the four-velocity of the heavy baryon.
Note that  the product of the spinor and the polarization vector on the rhs of Eqs.~(\ref{eq:LCDAs}) and (\ref{eq:LCDAsp})
can be expanded in irreducible representations corresponding to physical baryon states with $J^P=\frac{1}{2}^+$ and $J^P=\frac{3}{2}^+$ using suitable projection operators~\cite{epjc732302,Falk:1991nq}.
Following the similar procedure as in Refs.~\cite{Hu:2022xzu,Aliev:2022gxi},
one can extract the purely spin $\frac{1}{2}$ component as
\begin{eqnarray}\label{eq:LCDAs1}
\langle0|[q_1(t_1)C\slashed{n}q_2(t_2)]Q(0)|\mathcal{B}_b(\textbf{6},\frac{1}{2}^+)\rangle&=&\psi_{\parallel}^{n}(t_1,t_2)f^{(1)}(\gamma_5 \slashed{\bar v} u^{\frac{1}{2}^+}_{\textbf{6}}),\nonumber\\
\langle0|[q_1(t_1)C\slashed{\bar n}q_2(t_2)]Q(0)|\mathcal{B}_b(\textbf{6},\frac{1}{2}^+)\rangle&=&-\psi_{\parallel}^{\bar n}(t_1,t_2)f^{(1)}(\gamma_5 \slashed{\bar v} u^{\frac{1}{2}^+}_{\textbf{6}}),\nonumber\\
\langle0|[q_1(t_1)Cq_2(t_2)]Q(0)|\mathcal{B}_b(\textbf{6},\frac{1}{2}^+)\rangle&=&\psi_{\parallel}^{1}(t_1,t_2)f^{(2)}(\gamma_5 \slashed{\bar v} u^{\frac{1}{2}^+}_{\textbf{6}}),\nonumber\\
\frac{i}{2}\langle0|[q_1(t_1)C\sigma_{\bar nn}q_2(t_2)]Q(0)|\mathcal{B}_b(\textbf{6},\frac{1}{2}^+)\rangle&=&\psi_{\parallel}^{\bar nn}(t_1,t_2)f^{(2)}(\gamma_5 \slashed{\bar v} u^{\frac{1}{2}^+}_{\textbf{6}}).
\end{eqnarray}
Now $u^{\frac{1}{2}^+}_{\textbf{6}}$  represents the spinor of the sextet baryons with quantum numbers $J^P=\frac{1}{2}^+$.
It is straightforward to rewrite Eq.~(\ref{eq:LCDAs1}) in the following compact form:
\begin{eqnarray}\label{eq:LCDAs2}
\epsilon^{ijk}\langle0|q^i_{1\alpha}(t_1)q^j_{2\beta}(t_2)Q^k_\gamma(0)|\mathcal{B}_b(\textbf{6},\frac{1}{2}^+)\rangle &=&\frac{ f^{(1)}}{8}
[(\slashed{\bar n}C^T)_{\alpha\beta}\psi_{\parallel}^{n}(t_1,t_2)-(\slashed{ n}C^T)_{\alpha\beta}\psi_{\parallel}^{\bar n}(t_1,t_2)](\gamma_5 \slashed{\bar v} u^{\frac{1}{2}^+}_{\textbf{6}})_\gamma \nonumber\\
&&\frac{f^{(2)}}{4}[(C^T)_{\alpha\beta}\psi_{\parallel}^{1}(t_1,t_2)+\frac{i}{2}(\sigma_{\bar nn}C^T)_{\alpha\beta}\psi_{\parallel}^{\bar nn}(t_1,t_2)](\gamma_5 \slashed{\bar v} u^{\frac{1}{2}^+}_{\textbf{6}})_\gamma.
\end{eqnarray}
In the momentum space, the LCDAs of the sextet $b$ baryon with spin-parity $J^P=\frac{1}{2}^+$ reads
\begin{eqnarray}
(\Psi_{ \textbf{6}})_{\alpha\beta\gamma}(x_i,\mu)=&=&\frac{1 }{6N_c}\{\frac{f^{(1)}}{4\sqrt{2}}[(\psi_{\parallel}^{n}(x_2,x_3)\slashed{ n}^+
-\psi_{\parallel}^{\bar n}(x_2,x_3)\slashed{ n}^-)C^T]_{\alpha\beta}\nonumber\\&&+\frac{f^{(2)}}{4}[(\psi_{\parallel}^{1}(x_2,x_3)+i\psi_{\parallel}^{\bar nn}(x_2,x_3)
\sigma_{n^+n^-})C^T]_{\alpha\beta}\}[\gamma_5 \frac{\slashed{n}^--\slashed{n}^+}{\sqrt{2}} u^{\frac{1}{2}^+}_{\textbf{6}}]_\gamma.
\end{eqnarray}
where a factor $1/(6N_c)$  is introduced so that 
the relation in Eq.~(\ref{eq:LCDAsp}) holds for sextet baryons.

The general structure  of the model functions for the $b$ baryon LCDAs is governed by their scale evolution and can be composed of the
exponential component relating to the heavy-light interaction and the Gegenbauer polynomials pertaining to the light-light interaction~\cite{Parkhomenko:2017exu}.
Several asymptotic models for the various twist LCDAs of $\Lambda_b$ have been proposed in Refs.~\cite{plb665197,jhep112013191,Ali:2012zza}
but  mostly focused  on the $\Lambda_b$ sector.
In~\cite{Ali:2012zza},  the authors constructed simple model functions expanded by Gegenbauer polynomial for the LCDAs of $\Lambda_b$ and $\Xi_b$
in the heavy quark limit with the moments being derived in QCD sum rules.
Later, the similar analysis was extended to all the ground-state $b$ baryons with both the spin parities $J^P=\frac{1}{2}^+$ and $J^P=\frac{3}{2}^+$~\cite{epjc732302}.
As one can see, in the next section, this parametrization exhibits a considerably smaller parametric uncertainty.
On the other hand, from a systematic and consistent standpoint,
it is also preferable to
adopt the same framework of model functions for the $b$ baryon LCDAs to systematically calculate the decay branching ratios.
The Fourier transform of the model functions into the momentum space are  given by
\begin{eqnarray}\label{eq:twust}
\psi_{\parallel}^{n}(x_2,x_3)&=&M^4x_2x_3\sum_{l=0}^2\frac{a_l}{\epsilon_l^4}\frac{c_l^{3/2}(\frac{x_2-x_3}{x_2+x_3})}{|c_l^{3/2}|^2}e^{-\frac{\omega}{\epsilon_l}},\nonumber\\
\psi_{\parallel}^{1, \bar nn}(x_2,x_3)&=&\frac{1}{2}M^3(x_2+x_3)\sum_{l=0}^2\frac{a_l}{\epsilon_l^3}\frac{c_l^{1/2}(\frac{x_2-x_3}{x_2+x_3})}{|c_l^{1/2}|^2}e^{-\frac{\omega}{\epsilon_l}},\nonumber\\
\psi_{\parallel}^{\bar n}(x_2,x_3)&=&M^2\sum_{l=0}^2\frac{a_l}{\epsilon_l^2}\frac{c_l^{1/2}(\frac{x_2-x_3}{x_2+x_3})}{|c_l^{1/2}|^2}e^{-\frac{\omega}{\epsilon_l}},
\end{eqnarray}
where $\omega=(x_2+x_3)M$ with $M$ being the mass of $b$ baryon. 
$c_l$ are the Gegenbauer polynomials and read as~\cite{epjc732302}
\begin{eqnarray}
[c_0^{z}(x),c_1^{z}(x), c_2^{z}(x)]&=&[1, 2zx,2z(1+z)x^2-z],\nonumber\\
(|c_0^{1/2}|^2,|c_1^{1/2}|^2,|c_2^{1/2}|^2)&=&(1,\frac{1}{3},\frac{1}{5}),\nonumber\\
(|c_0^{3/2}|^2,|c_1^{3/2}|^2,|c_2^{3/2}|^2)&=&(1,3,6).
\end{eqnarray}	

\begin{table}[!htbh]
	\caption{ Shape parameters entering the transverse LCDAs of $b$ baryons at the scale  $\mu=1.0$ GeV. The longitudinal ones can be obtained by replacing $A$ to $1-A$.
The corresponding parameters for the $\Lambda_b$ and $\Xi_b$ have the same forms as the parallel $\Sigma_b$ and $\Xi_b'$ ones, respectively.}
	\label{tab:LCDAss}
	\begin{tabular}[t]{lcccccccc}
	\hline\hline
&  & Twist   & $a_0$    & $a_1$ & $a_2$ & $\epsilon_0$[GeV] &$\epsilon_1$[GeV]  &$\epsilon_2$[GeV] \\ \hline

&$\Sigma_b$  & $\psi_{\parallel}^{n}$ & $1$ & $-$    & $\dfrac{6.4A}{A+0.44}$ & $\dfrac{1.4A+0.6}{A+5.7}$     &$-$ & $\dfrac{0.32A}{A-0.17}$ \\		
&  & $\psi_{\parallel}^{\bar nn}$ & $1$ & $-$    & $\dfrac{0.12A-0.08}{A-1.4}$ & $\dfrac{0.56A-0.77}{A-2.6}$      &$-$ & $\dfrac{0.25A-0.16}{A+0.41}$ \\
&  & $\psi_{\parallel}^{1}$ & $-$ & $1$    & $-$ & $-$      &$\dfrac{0.35A-0.43}{A-1.2}$ & $-$ \\
&  & $\psi_{\parallel}^{\bar n}$ & $1$ & $-$    & $\dfrac{-0.07A-0.05}{A+0.34}$ & $\dfrac{0.65A+0.22}{A+1}$      &$-$ & $\dfrac{5.5A+3.8}{A+29}$ \\
	
&$\Xi'_b$  & $\psi_{\parallel}^{n}$ & $1$ & $\dfrac{0.25A+0.46}{A+0.68}$    & $\dfrac{6.6A+0.6}{A+0.68}$ & $\dfrac{1.4A+1}{A+6.7}$      &$\dfrac{0.57A+1.1}{A+4}$ & $\dfrac{0.36A+0.03}{A-0.02}$ \\
&  & $\psi_{\parallel}^{\bar nn}$ & $1$ & $\dfrac{0.04A-0.14}{A-1.6}$    & $\dfrac{0.12A-0.09}{A-1.6}$ & $\dfrac{0.56A-0.91}{A-2.9}$      &$\dfrac{-27A+92}{160}$ & $\dfrac{0.3A-0.24}{A+0.54}$ \\
&  & $\psi_{\parallel}^{1}$ & $\dfrac{-0.16A+0.16}{A-1.3}$ & $1$    & $\dfrac{0.17A-0.17}{A-1.3}$ & $\dfrac{0.11A-0.11}{A-1}$      &$\dfrac{0.39A-0.49}{A-1.3}$ & $\dfrac{0.33A-0.33}{A-1}$ \\
&  & $\psi_{\parallel}^{\bar n}$ & $1$ & $\dfrac{0.03A+0.11}{A+0.16}$    & $\dfrac{-0.1A-0.03}{A+0.61}$ & $\dfrac{0.63A+0.38}{A+1.3}$      &$\dfrac{-0.82A-3.1}{A-3.9}$ & $\dfrac{1.2A+0.34}{A+4.1}$ \\
	
&$\Omega_b$ & $\psi_{\parallel}^{n}$ & $1$ & $-$    & $\dfrac{8A+1}{A+1}$ & $\dfrac{1.3A+1.3}{A+6.9}$      &$-$ & $\dfrac{0.41A+0.06}{A+0.11}$ \\
&  & $\psi_{\parallel}^{\bar nn}$ & $1$ & $-$    & $\dfrac{0.17A-0.16}{A-2}$ & $\dfrac{0.56A-1.1}{A-3.22}$      &$-$ & $\dfrac{0.44A-0.43}{A+0.27}$ \\
&  & $\psi_{\parallel}^{1}$ & $-$ & $1$    & $-$ & $-$      &$\dfrac{0.45A-0.63}{A-1.4}$ & $-$ \\
&  & $\psi_{\parallel}^{\bar n}$ & $1$ & $-$    & $\dfrac{-0.10A-0.01}{A+1}$ & $\dfrac{0.62A+0.62}{A+1.62}$      &$-$ & $\dfrac{0.87A-0.07}{A+2.53}$ \\
		\hline\hline
	\end{tabular}
\end{table}	
The two shape parameters $a_l$ and $\epsilon_l$ in Eq.~(\ref{eq:twust})
dependence on the free parameter $A$ have been determined in~\cite{epjc732302},
and we collect them in Table~\ref{tab:LCDAss}  to make the paper self-contained.
Note that the parameters given in Table~\ref{tab:LCDAss} correspond to the transverse LCDAs,
and parallel counterparts can be obtained by the replacement $A\rightarrow 1-A$.
For the numerical calculations, we take $A=0.5\pm 0.2$~\cite{epjc732302}.
In principle, the parameters $\epsilon_i$ should be strictly positive to satisfy the asymptotic behavior.
However, as can be seen from Table~\ref{tab:LCDAss}, all the second-order term $\epsilon_2$ in $\psi_{\parallel}^{\bar nn}$ are negative when $A=0.5$,
which causes the corresponding LCDAs to diverge in the endpoint region.
To obtain finite results, the authors of~\cite{Hu:2022xzu} choose a smaller value of $A$,
but we favor to ignore the second-order terms  of $\psi_{\parallel}^{\bar nn}$ because their contributions are usually subleading in the Gegenbauer expansion.

\subsubsection{Light baryon light-cone distribution amplitudes}
The leading twist LCDAs of baryon octet can be defined through the matrix element of the three-quark operator on the light cone~\cite{zpc42569}
\begin{eqnarray}\label{eq:LCDAs8}
4\langle0|\epsilon^{ijk}q^i_{1\alpha}(z_1)q^j_{2\beta}(z_2)q^k_{3\gamma}(z_3)|\mathcal{B}(\textbf{8}(p'))\rangle &=&f_{\mathcal{B}(\textbf{8})}
(\slashed{p}'C)_{\alpha\beta}[\gamma_5u_{\textbf{8}}({p}')]_\gamma\mathcal{V}(z_lp')
+f_{\mathcal{B}(\textbf{8})}(\slashed{p}'\gamma_5C)_{\alpha\beta}[u_{\textbf{8}}({p}')]_\gamma\mathcal{A}(z_lp') \nonumber\\
&&+f_{\mathcal{B}(\textbf{8})}^T(i\sigma_{\mu\nu}{p'}^\nu C)_{\alpha\beta}[\gamma^\mu\gamma_5u_{\textbf{8}}({p}')]_\gamma\mathcal{T}(z_lp'),
\end{eqnarray}
where $q_l(z_l)$  are light quark field operators of the given flavor, chosen to match the valence quark content of the specific baryon.
$u_{\textbf{8}}(p')$ is the spinor characterizing the octet that satisfies the Dirac equation $\slashed{p}' u_{\textbf{8}}(p')= mu_{\textbf{8}}(p')$
and normalizes to $\bar{u}_{\textbf{8}}(p') u_{\textbf{8}}(p')=2m$
with $p'$ and $m$ being the momentum and mass of octet, respectively.
$f_{\textbf{8}}$ and $f_{\textbf{8}}^T$ are two normalization constants that determine the values of the matrix element at the origin.
The two couplings coincide for the nucleon due to isospin symmetry.
$\mathcal{V}$, $\mathcal{A}$, and $\mathcal{T}$ are the vector, axial-vector, and tensor structure  LCDAs, respectively.
The scale dependence will be suppressed from now on, unless it is explicitly needed.

In this work, we adopt the Chernyak-Ogloblin-Zhitnitsky (COZ) model for the light baryon LCDAs proposed in Refs.~\cite{zpc42569,Farrar:1988vz},
whose expressions are shown in Table~\ref{tab:VAT} for completeness.
The leading contribution $\phi_{as}=120x_1x_2x_3$  are usually referred to as the asymptotic shape.

\begin{table}[!htbh]
	\caption{The  vector, axial vector, and tensor LCDAs of octet baryons in COZ model,
where $\phi_{as}=120x_1x_2x_3$  denotes the asymptotic shape at infinitely large scales.
The last column is  the numerical values of the corresponding  decay constants given in units of $(10^{-3}\text{GeV}^2)$.  }
	\label{tab:VAT}
\newcommand{\tabincell}[2]{\begin{tabular}{@{}#1@{}}#2\end{tabular}}
	\begin{tabular}[t]{lcccc}
	\hline\hline
Octet  & $\frac{\mathcal{V}}{\phi_{as}}$   & $\frac{\mathcal{A}}{\phi_{as}}$ & $\frac{\mathcal{T}}{\phi_{as}}$  &$f_{\mathcal{B}(\textbf{8})}^{(T)}$\\ \hline
$N$
         & \tabincell{c}{$18.396(x_1^2+x_2^2)$\\$+6.174x_3^2+5.88x_3-7.098$}
         & \tabincell{c}{ $-5.418(x_1^2-x_2^2)$}
         &\tabincell{c}{ $10.836(x_1^2+x_2^2)+5.88x_3^2$\\$-8.316x_1x_2-11.256x_3(x_1+x_2)$}   & $f=f^T=5.0$\\ \\
$\Lambda$
         & $42[0.18(x_1^2-x_2^2)-0.1(x_1-x_2)]$
         &\tabincell{c}{$-42[0.26(x_1^2+x_2^2)+0.34x_3^2$\\$-0.56x_1x_2-0.24x_3(x_1+x_2)]$}
         &$42[1.2(x_2^2-x_1^2)+1.4(x_1-x_2)]$                                                  & \tabincell{c}{$f=6.3$\\$f^T=0.63$}\\\\

$\Sigma$ & \tabincell{c}{$42[0.3(x_1^2+x_2^2)+0.14x_3^2$\\$-0.54x_1x_2-0.16x_3(x_1+x_2)]$}
         & $-42[0.06(x_1^2-x_2^2)+0.05(x_1-x_2)]$
         &\tabincell{c}{$42[0.32(x_1^2+x_2^2)+0.16x_3^2$\\$-0.47x_1x_2-0.24x_3(x_1+x_2)]$}        &\tabincell{c}{$f=5.1$\\$f^T=4.9$} \\\\

$\Xi$    & \tabincell{c}{$42[0.29(x_1^2+x_2^2)+0.16x_3^2$\\$-0.26x_1x_2-0.3x_3(x_1+x_2)]$}
         & $42[0.09(x_1^2-x_2^2)+0.02(x_1-x_2)]$
         &\tabincell{c}{$42[0.28(x_1^2+x_2^2)+0.18x_3^2$\\$-0.16x_1x_2-0.35x_3(x_1+x_2)]$}        &\tabincell{c}{$f=5.3$\\$f^T=5.4$}\\

		\hline\hline
	\end{tabular}
\end{table}	

Following Refs.~\cite{Farrar:1988vz,Braun:1999te,Mannel:2011xg}, the LCDAs of baryon decuplet to the leading twist accuracy are defined as
\begin{eqnarray}\label{eq:LCDAs10}
4\langle0|\epsilon^{ijk}q^i_{1\alpha}(z_1)q^j_{2\beta}(z_2)q^k_{3\gamma}(z_3)|\mathcal{B}(\textbf{10}(p'))\rangle &=&\lambda^{1/2}_{\mathcal{B}(\textbf{10})}
[(\gamma_\mu C)_{\alpha\beta}R^\mu_\gamma \mathcal{V}(z_lp')+(\gamma_\mu \gamma_5C)_{\alpha\beta}(\gamma_5R^\mu)_\gamma \mathcal{A}(z_lp')
\nonumber\\&&-\frac{\mathcal{T}(z_lp')}{2}(i \sigma_{\mu\nu}C)_{\alpha\beta}(\gamma^\mu R^\nu)_\gamma]\nonumber\\&&-f^{3/2}_{\mathcal{B}(\textbf{10})}(i \sigma_{\mu\nu}C)_{\alpha\beta}
(p'^{\mu} R^\nu-\frac{1}{2}m\gamma^\mu R^\nu)_\gamma \mathcal{\phi}(z_lp'),
\end{eqnarray}
where $f^{3/2}_{\mathcal{B}(\textbf{10})}=\sqrt{\frac{2}{3}}\frac{\lambda^{1/2}_{\mathcal{B}(\textbf{10})}}{m}$,
 and $R^\mu$ is the Rarita-Schwinger vector spinor that satisfies the subsidiary conditions
\begin{eqnarray}
\slashed{p}'R^{\mu}(p')=mR^{\mu}(p'), \quad p_\mu'R^{\mu}(p')=\gamma_\mu R^{\mu}(p')=0,\quad \bar R^\mu R_\mu=-2m.
\end{eqnarray}
The dimensionless amplitudes $\mathcal{V,A,T} $ determine the distribution of quarks in the helicity-1/2 state,
while the $\phi$ determines the one in the helicity-3/2 state.
Model wave functions for $\Delta$ have been investigated using the QCD sum rules, which yield~\cite{Farrar:1988vz}
\begin{eqnarray}
\mathcal{V}&=&\phi_{as}[3.36(x_1^2+x_2^2)-3.15(x_1+x_2)x_3+4.2x_3^2+0.42x_1x_2],\nonumber\\
\mathcal{A}&=&\phi_{as}[-0.84(x_1^2-x_2^2)+3.57(x_1-x_2)x_3],\nonumber\\
\mathcal{T}&=&\phi_{as}[4.2(x_1^2+x_2^2)+0.42(x_1+x_2)x_3+2.52x_3^2-6.72x_1x_2],\nonumber\\
\phi&=&120x_1x_2x_3,
\end{eqnarray}
with the normalizations
\begin{eqnarray}
\int_0^1dx_1dx_2dx_3\delta(1-x_1-x_2-x_3)(\mathcal{V,T,\phi})=1.
\end{eqnarray}
Determination of the $\Omega$ one is similar, and the resulting structures are much like the asymptotic one~\cite{Farrar:1988vz}
\begin{eqnarray}
\mathcal{V, T,\phi}=120x_1x_2x_3, \quad \mathcal{A}=0.
\end{eqnarray}
The values of the decay constants are set as~\cite{Farrar:1988vz}
\begin{eqnarray}
|f_{\Delta}^{1/2}|&=&(1.1\pm0.2)\times 10^{-2} \text{GeV}^2,\quad |f_{\Delta}^{3/2}|=1.4\times 10^{-2} \text{GeV}^2,\nonumber\\
|f_{\Omega}^{1/2}|&=&(1.6\pm0.4)\times 10^{-2} \text{GeV}^2,\quad |f_{\Omega}^{3/2}|=(1.7\pm0.4)\times 10^{-2} \text{GeV}^2.
\end{eqnarray}
Model wave functions for other decuplet are still lacking at the current stage;
thus, these decay modes are excluded from the calculations in this paper.

\subsubsection{Charmonium light-cone distribution amplitudes}
For the LCDAs of the charmonium state, we adopt the harmonic oscillator models  derived in the previous work~\cite{Sun:2008ew,prd90114030,epjc75293},
which are successful in describing various charmonium decays of the $B$ meson
\cite{epjc76564,epjc77199,epjc77610,prd97033006,prd98113003,prd99093007,epjc79792,cpc44073102,prd101016015}.
Up to twist-3, the longitudinally and transversely polarized LCDAs for $J/\psi$ are decomposed into~\cite{prd71114008}
\begin{eqnarray}
\Psi_{L}&=&\frac{1}{\sqrt{2N_c}}(m_\psi\slashed{\epsilon}_L\psi^L(y,b)+\slashed{\epsilon}_L\slashed{q}\psi^t(y,b)),\nonumber\\
\Psi_{T}&=&\frac{1}{\sqrt{2N_c}}(m_\psi\slashed{\epsilon}_T\psi^V(y,b)+\slashed{\epsilon}_T\slashed{q}\psi^T(y,b)),
\end{eqnarray}
where $m_\psi$, $q$ and $\epsilon_{L,T}$ are the mass, momentum, and polarization vectors of $J/\psi$ meson, respectively.
Various twist functions of $\psi^{L,T,V,t}(y,b)$ can be parametrized as~\cite{Sun:2008ew,prd90114030}
\begin{eqnarray}
\psi^{L,T}(y,b)&=&\frac{f_{\psi}}{2\sqrt{2N_c}}N^{L,T}y\bar{y}
\exp\left\{-\frac{m_c}{\omega_c}y\bar{y}[(\frac{y-\bar{y}}{2y\bar{y}})^2+\omega_c^2b^2]\right\} ,\nonumber\\
\psi^t(y,b)&=&\frac{f_{\psi}}{2\sqrt{2N_c}}N^t(y-\bar{y})^2
\exp\left\{-\frac{m_c}{\omega_c}y\bar{y}[(\frac{y-\bar{y}}{2y\bar{y}})^2+\omega_c^2b^2]\right\},\nonumber\\
\psi^V(y,b)&=&\frac{f_{\psi}}{2\sqrt{2N_c}}N^V[1+(y-\bar{y})^2]
\exp\left\{-\frac{m_c}{\omega_c}y\bar{y}[(\frac{y-\bar{y}}{2y\bar{y}})^2+\omega_c^2b^2]\right\},
\end{eqnarray}
with $m_c$ being the charm quark mass.
$y(\bar y)$ is the momentum fraction associated with the (anti) charm quark and satisfies $y+\bar y=1$,
while $b$ is the corresponding transverse momentum in the $b$ space.
We take the shape parameter $\omega_c=0.6$ GeV~\cite{prd90114030} and decay constant $f_{\psi}=0.363$ GeV~\cite{prd106053005} for $J/\psi$ meson in the following analysis.
$N^{L,T,V,t}$  are the normalization constants which obey the normalization conditions
\begin{eqnarray}
\int_0^1\psi^{L,T,V,t}(x,0)dx=\frac{f_\psi}{2\sqrt{2N_c}}.
\end{eqnarray}

\subsection{ Kinematics and observables}\label{sec:kin}
\begin{figure}[!htbh]
	\begin{center}
		\vspace{0.01cm} \centerline{\epsfxsize=15cm \epsffile{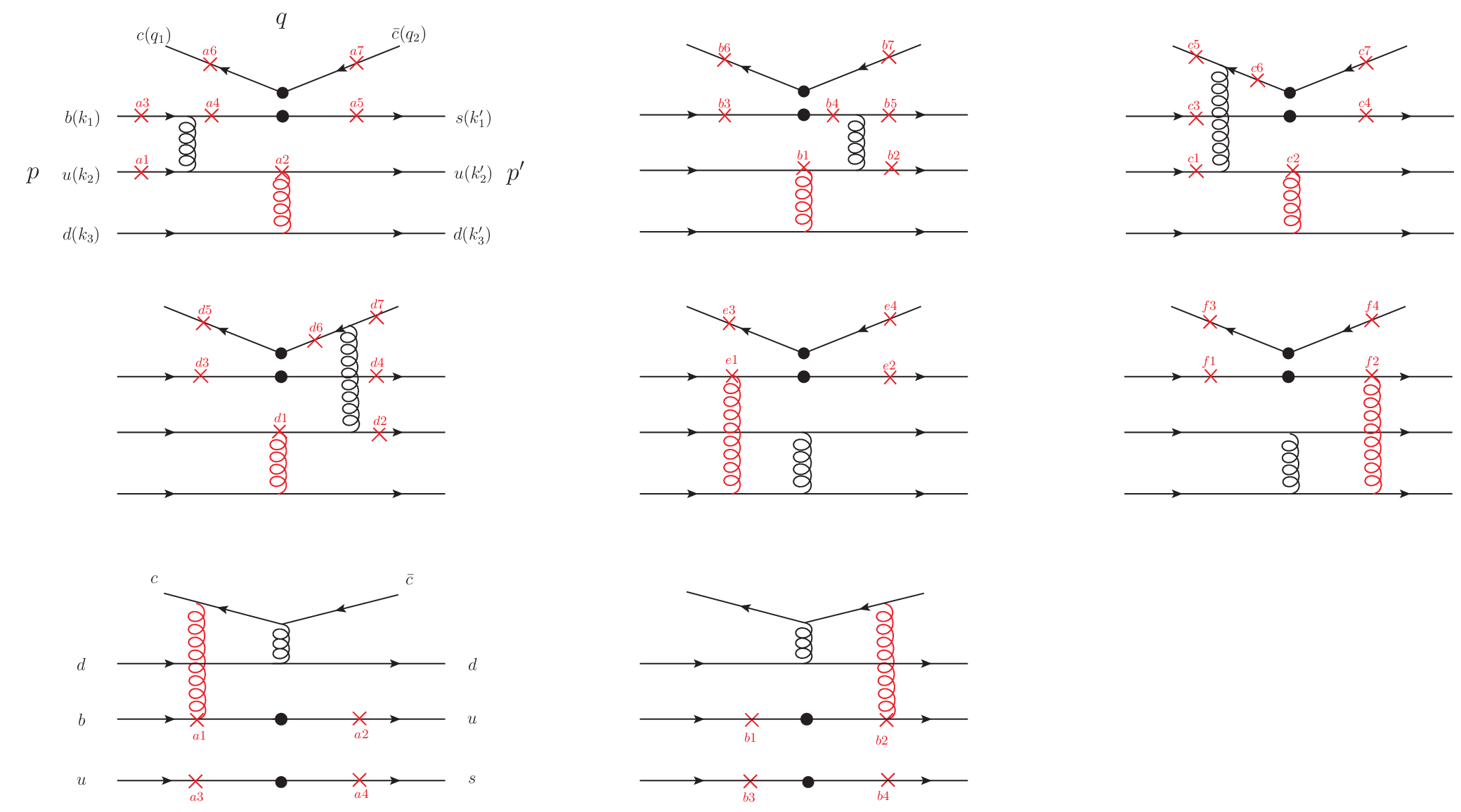}}
		\setlength{\abovecaptionskip}{1.0cm}
\caption{ Topological diagrams for the $\Lambda_b\rightarrow\Lambda J/\psi$ decay.
The first two rows are called $W$-emission diagrams characterized by weak $b\rightarrow s c\bar c$ quark transition,
while the last two are the $W$-exchange ones induced by the $bu\rightarrow su$ transition.
The triple-gluon vertex diagrams are not shown here 
since their color rearrangement factors are zero in the present case.}
		\label{fig:Feynman}
	\end{center}
\end{figure}
With the LCDAs obtained in the last subsection, we can perform the PQCD calculations on the decay amplitudes.
The corresponding Feynman diagrams (taking the $\Lambda_b\rightarrow \Lambda J/\psi$ as an example here) are displayed in Fig.~\ref{fig:Feynman}.
Note that the last two diagrams in Fig.~\ref{fig:Feynman} are not forbidden by the Yang theorem,
which holds exactly when the two gluons are on shell, i.e., collinear.
In PQCD, however, the two gluons must be hard to produce an energetic $c \bar c$ that forms a color singlet and further hadronizes to a $J/\psi$.
The decay amplitude of $\mathcal{B}_b\rightarrow \mathcal{B}J/\psi$ can be
described by sandwiching $\mathcal{H}_{eff}$  with the initial and final states,
\begin{eqnarray}\label{eq:amp}
\mathcal{M}=\langle \mathcal{B} J/\psi|\mathcal{H}_{eff}|\mathcal{B}_b\rangle,
\end{eqnarray}
with the weak effective Hamiltonian~\cite{Buchalla:1995vs}
\begin{eqnarray}
\mathcal{H}_{eff}&=&\frac{G_F}{\sqrt{2}} \{V_{cb}V^*_{cq}[C_1(\mu)O_1(\mu)+C_2(\mu)O_2(\mu)]-\sum_{k=3}^{10}V_{tb}V^*_{tq}C_k(\mu)O_k(\mu)\}+\mathrm{H.c.},
\end{eqnarray}
where $q=d,s$. $G_F$ is the Fermi constant, and $V_{ij}$ are the CKM matrix elements,
$C_l(\mu)$ denotes the Wilson coefficients evaluated at the renormalization scale $\mu$,
and  $O_l$ are the local four-quark operators, defined by
\begin{eqnarray}
O_1&=& \bar{c}_i \gamma_\mu(1-\gamma_5) b_j  \otimes \bar{q}_j  \gamma^\mu(1-\gamma_5) c_i, \nonumber\\
O_2&=& \bar{c}_i \gamma_\mu(1-\gamma_5) b_i \otimes \bar{q}_j  \gamma^\mu(1-\gamma_5) c_j,  \nonumber\\
O_3&=& \bar{q}_i  \gamma_\mu(1-\gamma_5) b_i  \otimes \sum_{q'} \bar{q}'_j \gamma^\mu(1-\gamma_5) q'_j, \nonumber\\
O_4&=& \bar{q}_i  \gamma_\mu(1-\gamma_5) b_j \otimes \sum_{q'} \bar{q}'_j \gamma^\mu(1-\gamma_5) q'_i,  \nonumber\\
O_5&=& \bar{q}_i  \gamma_\mu(1-\gamma_5) b_i  \otimes \sum_{q'} \bar{q}'_j \gamma^\mu(1+\gamma_5) q'_j, \nonumber\\
O_6&=& \bar{q}_i  \gamma_\mu(1-\gamma_5) b_j \otimes \sum_{q'} \bar{q}'_j \gamma^\mu(1+\gamma_5) q'_i,  \nonumber\\
O_7&=&   \frac{3}{2}\bar{q}_i \gamma_\mu(1-\gamma_5) b_i  \otimes \sum_{q'} e_{q'}\bar{q}'_j \gamma^\mu(1+\gamma_5) q'_j, \nonumber\\
O_8&=&   \frac{3}{2}\bar{q}_i \gamma_\mu(1-\gamma_5) b_j \otimes \sum_{q'} e_{q'}\bar{q}'_j \gamma^\mu(1+\gamma_5) q'_i,  \nonumber\\
O_9&=&   \frac{3}{2}\bar{q}_i \gamma_\mu(1-\gamma_5) b_i  \otimes \sum_{q'} e_{q'}\bar{q}'_j \gamma^\mu(1-\gamma_5) q'_j, \nonumber\\
O_{10}&=&\frac{3}{2}\bar{q}_i \gamma_\mu(1-\gamma_5) b_j \otimes \sum_{q'} e_{q'}\bar{q}'_j \gamma^\mu(1-\gamma_5) q'_i,
\end{eqnarray}
where the sum over $q'$ runs over the quark fields that are active at the scale $\mu=\mathcal{O}(m_b)$.

Following~\cite{prd106053005,221015357},
the general decay amplitude for a $\frac{1}{2}^+\rightarrow \frac{1}{2}^+ + 1^- $ process can be expanded with the Dirac spinors and polarization vector as
\begin{eqnarray}\label{eq:kq}
\mathcal{M}^L&=&\bar {u}_{\textbf{8}} (p')\epsilon^{\mu*}_{L}[A_1^L\gamma_{\mu}\gamma_5+A_2^L\frac{p'_{\mu}}{M}\gamma_5+B_1^L\gamma_\mu+B_2^L\frac{p'_{\mu}}{M}]u_{\mathcal{B}_b}(p),
\nonumber\\
\mathcal{M}^T&=&\bar {u}_{\textbf{8}} (p')\epsilon^{\mu*}_{T}[A_1^T\gamma_{\mu}\gamma_5+B_1^T\gamma_\mu]u_{\mathcal{B}_b}(p),
\end{eqnarray}
where $A^{L,T}_{1,2}$ and $B^{L,T}_{1,2}$ are the so-called invariant amplitudes
with $L$ and $T$ in the superscripts denoting the longitudinal and transverse components, respectively.
$u_{\mathcal{B}_b}$ is either $u_{\bar{\textbf{3}}}$ or  $u^{\frac{1}{2}^+}_{\textbf{6}}$.

Similarly, the decay amplitudes for decays to daughter baryons with $J^P=\frac{3}{2}^+$
can also be separated into two parts~\cite{Korner:1992wi,Chua:2019yqh}
\begin{eqnarray}\label{eq:kqp}
\mathcal{M}^L&=&\bar {R}^\nu (p')\epsilon^{\mu*}_{L}[g_{\mu\nu}(C_1^L+D_1^L\gamma_5)+q_{\nu}\gamma_{\mu}(C_2^L+D_2^L\gamma_5)+q_{\nu}p_{\mu}(C_3^L+D_3^L\gamma_5)]u_{\mathcal{B}_b}(p),
\nonumber\\
\mathcal{M}^T&=&\bar {R}^\nu (p')\epsilon^{\mu*}_{T}[g_{\mu\nu}(C_1^T+D_1^T\gamma_5)+q_{\nu}\gamma_{\mu}(C_2^T+D_2^T\gamma_5)]u_{\mathcal{B}_b}(p),
\end{eqnarray}
 corresponding to the longitudinally and transversely polarizations,  respectively.
 Taking a spin average of the initial baryon and summing over the final state polarizations,
we get the square of the amplitude
 \begin{eqnarray}\label{eq:ampli}
|\mathcal{M}|^2=\frac{1}{2}\sum_{\sigma=L,T}|\mathcal{M}^\sigma|^2.
\end{eqnarray}

To derive the decay amplitudes, one has to specify the kinematics of initial and final states.
We carry out the calculations in the parent baryon  rest frame
and write the momenta of parent and  daughter baryons in the light-cone coordinates as
\begin{eqnarray}\label{eq:pq}
p=\frac{M}{\sqrt{2}}(1,1,\textbf{0}_{T}), \quad p'=\frac{M}{\sqrt{2}}(f^+,f^-,\textbf{0}_{T}),
\end{eqnarray}
with the factors
\begin{eqnarray}
f^\pm=\frac{1}{2}\left(1-r^2+r'^2 \pm \sqrt{(1-r^2+r'^2)^2-4r'^2}\right),
\end{eqnarray}
and the mass ratios $r=m_{\psi}/M$ and $r'=m/M$.
The $J/\psi$  meson momentum is then given by $q=p-p'$ and
the corresponding polarization vectors  read as
\begin{eqnarray}
\epsilon_L=\frac{1}{\sqrt{2(1-f^+)(1-f^-)}}\left(f^+-1,1-f^-,\textbf{0}_{T}\right),
\quad \epsilon_T=\left(0,0,\textbf{1}_{T}\right).
\end{eqnarray}
The momenta of eight valence quarks as illustrated in Fig.~\ref{fig:Feynman} are parametrized as
\begin{eqnarray}
k_1&=&\left(\frac{M}{\sqrt{2}},\frac{M}{\sqrt{2}}x_1,\textbf{k}_{1T}\right),\quad
k_2=\left(0,\frac{M}{\sqrt{2}}x_2,\textbf{k}_{2T}\right),\quad
k_3=\left(0,\frac{M}{\sqrt{2}}x_3,\textbf{k}_{3T}\right),\nonumber\\
k_1'&=&\left(\frac{M}{\sqrt{2}}f^+x_1',0,\textbf{k}'_{1T}\right),\quad
k_2'=\left(\frac{M}{\sqrt{2}}f^+x_2',0,\textbf{k}'_{2T}\right),\quad
k_3'=\left(\frac{M}{\sqrt{2}}f^+x_3',0,\textbf{k}'_{3T}\right),\nonumber\\
q_1&=&yq+\textbf{q}_{T},\quad q_2=\bar y q-\textbf{q}_{T},
\end{eqnarray}
where $x^{(')}_l$ with $l=1,2,3$ and $y$ are the parton longitudinal momentum fractions,
while $\textbf{k}^{(')}_{lT}$ and $\textbf{q}_T$ are the corresponding transverse momenta.
They satisfy  the momentum conservation conditions: 
\begin{eqnarray}
\sum_{l=1}^3x^{(')}_l=1,\quad \sum_{l=1}^3\textbf{k}^{(')}_{lT}=0.
\end{eqnarray}

Usually,  these invariant amplitudes are converted into the helicity amplitudes $H_{\lambda_\mathcal{B}\lambda_\psi}$,
which are convenient for expressing various observable quantities in the angular distributions.
Here, $\lambda_\mathcal{B}$  and $\lambda_\psi$ are the  helicities of the baryon and meson in the final state, respectively.
Angular momentum conservation for a $\frac{1}{2}^+$ baryon decay imposes $|\lambda_\mathcal{B}-\lambda_\psi|\leq\frac{1}{2}$
such that the possible helicity configurations are
$H_{\pm\frac{1}{2}\pm1},H_{\pm\frac{1}{2}0}$ for $\frac{1}{2}^+\rightarrow \frac{1}{2}^+ + 1^- $
and $H_{\pm\frac{3}{2}\pm1}, H_{\pm\frac{1}{2}\pm1}, H_{\pm\frac{1}{2}0}$ for $\frac{1}{2}^+\rightarrow \frac{3}{2}^+ + 1^- $ modes.
The explicit relations between the helicity amplitudes  and the invariant amplitudes  are~\cite{Cheng:1996cs,Chua:2019yqh,Ebert:2006rp,Korner:1992wi}
\begin{eqnarray}\label{eq:helicity1}
H_{\pm\frac{1}{2}\pm1}&=&\mp\sqrt{Q_+}A_1^T-\sqrt{Q_-}B_1^T,\nonumber\\
H_{\pm\frac{1}{2}0}&=& \frac{1}{\sqrt{2}m_\psi}[\pm\sqrt{Q_+}(M-m)A_1^L\mp\sqrt{Q_-}P_cA_2^L+\sqrt{Q_-}(M+m)B_1^L+\sqrt{Q_+}P_cB_2^L],
\end{eqnarray}
for $\frac{1}{2}^+\rightarrow \frac{1}{2}^+ + 1^- $, and
\begin{eqnarray}\label{eq:helicity2}
H_{\pm\frac{3}{2}\pm1}&=&\frac{1}{\sqrt{2}}\left(-\sqrt{Q_+}C_1^T\pm\sqrt{Q_-}D_1^T\right),\nonumber\\
H_{\pm\frac{1}{2}\pm1}&=&\frac{1}{\sqrt{6}}\left(-\sqrt{Q_+}C_1^T+\frac{\sqrt{Q_+}Q_-}{m}C_2^T\mp \sqrt{Q_-}D_1^T\pm\frac{\sqrt{Q_-}Q_+}{m}D_2^T\right),
\nonumber\\
H_{\pm\frac{1}{2}0}&=&\frac{1}{\sqrt{12}mm_\psi}\left[-\sqrt{Q_+}(M^2-m^2-m^2_\psi)C_1^L-\sqrt{Q_+}Q_-(M+m)C_2^L-\frac{\sqrt{Q_+}Q_+Q_-}{2}C_3^L \right.\nonumber\\&&\left.
\pm\sqrt{Q_-}(M^2-m^2-m^2_\psi)D_1^L\mp\sqrt{Q_-}Q_+(M-m)D_2^L\pm\frac{\sqrt{Q_-}Q_+Q_-}{2}D_3^L\right],
\end{eqnarray}
for $\frac{1}{2}^+\rightarrow \frac{3}{2}^+ + 1^- $.
In these expressions, we use the abbreviations $Q_{\pm}=(M\pm m)^2-m_\psi^2$ and
$P_c=\frac{\sqrt{Q_+Q_-}}{2M}$ as the magnitude of the three-momentum of the daughter baryon in the rest frame of the parental baryon.
Summing over all the allowed squared helicity amplitudes, one also get the modulus squares of amplitude
\begin{eqnarray}\label{eq:hn}
|\mathcal{M}|^2= \sum_{\lambda_\mathcal{B},\lambda_\psi} |H_{\lambda_\mathcal{B}\lambda_\psi}|^2,
\end{eqnarray}
which is equivalent to Eq.~(\ref{eq:ampli}).

The decay rate and up-down asymmetries read~\cite{Cheng:1996cs,Chua:2019yqh}
\begin{eqnarray}\label{eq:two}
\Gamma&=&\frac{P_c}{8\pi M^2}|\mathcal{M}|^2,\nonumber\\
\alpha_b (\frac{1}{2}^+\rightarrow \frac{1}{2}^+ + 1^- ) &=&|\hat{H}_{\frac{1}{2}0}|^2-|\hat{H}_{-\frac{1}{2}0}|^2+|\hat{H}_{-\frac{1}{2}-1}|^2-|\hat{H}_{\frac{1}{2}1}|^2,\nonumber\\
\alpha_b (\frac{1}{2}^+\rightarrow \frac{3}{2}^+ + 1^- ) &=&|\hat{H}_{\frac{1}{2}0}|^2-|\hat{H}_{-\frac{1}{2}0}|^2+|\hat{H}_{-\frac{1}{2}-1}|^2-|\hat{H}_{\frac{1}{2}1}|^2
+|\hat{H}_{\frac{3}{2}1}|^2-|\hat{H}_{-\frac{3}{2}-1}|^2,
\end{eqnarray}
where the hatted helicity amplitudes $\hat{H}_{\lambda_\mathcal{B}\lambda_\psi}=H_{\lambda_\mathcal{B}\lambda_\psi}/|\mathcal{M}|$  are normalized to 1.
Following~\cite{Gutsche:2018utw,plb72427,prd88114018,221015357},
we also express some asymmetry observables for the  $\frac{1}{2}^+\rightarrow \frac{1}{2}^+ + 1^-$ decay in terms of the hatted helicity amplitudes
\begin{eqnarray}\label{eq:alp}
r_0 &=&|\hat{H}_{\frac{1}{2}0}|^2+|\hat{H}_{-\frac{1}{2}0}|^2, \nonumber\\
r_1 &=&|\hat{H}_{\frac{1}{2}0}|^2-|\hat{H}_{-\frac{1}{2}0}|^2,\nonumber\\
\alpha_{\lambda_\mathcal{B}}&=&|\hat{H}_{\frac{1}{2}0}|^2+|\hat{H}_{\frac{1}{2}1}|^2-|\hat{H}_{-\frac{1}{2}-1}|^2-|\hat{H}_{-\frac{1}{2}0}|^2, \nonumber\\
\alpha_{\lambda_\psi}&=&|\hat{H}_{\frac{1}{2}0}|^2+|\hat{H}_{-\frac{1}{2}0}|^2-|\hat{H}_{\frac{1}{2}1}|^2-|\hat{H}_{-\frac{1}{2}-1}|^2,
\end{eqnarray}
where $r_0 $ and $r_1$ are the longitudinal unpolarized and polarized parameters, respectively.
$\alpha_{\lambda_\mathcal{B}}$ denotes the longitudinal polarization of the daughter baryon,
and $\alpha_{\lambda_\psi}$ represents the asymmetry between the longitudinal and transverse polarizations of the charmonium state.
In the same vein, for the  $\frac{1}{2}^+\rightarrow \frac{3}{2}^+ + 1^-$ case,
some useful polarization parameters that describe the angular decay distribution are defined as
\begin{eqnarray}\label{eq:alpp}
U^L &=&|\hat{H}_{\frac{1}{2}0}|^2+|\hat{H}_{-\frac{1}{2}0}|^2, \nonumber\\
P^L &=&|\hat{H}_{\frac{1}{2}0}|^2-|\hat{H}_{-\frac{1}{2}0}|^2, \nonumber\\
U^T_{\frac{3}{2}} &=&|\hat{H}_{\frac{3}{2}1}|^2+|\hat{H}_{-\frac{3}{2}-1}|^2, \nonumber\\
P^T_{\frac{3}{2}} &=&|\hat{H}_{\frac{3}{2}1}|^2-|\hat{H}_{-\frac{3}{2}-1}|^2, \nonumber\\
U^T_{\frac{1}{2}} &=&|\hat{H}_{\frac{1}{2}1}|^2+|\hat{H}_{-\frac{1}{2}-1}|^2, \nonumber\\
P^T_{\frac{1}{2}} &=&|\hat{H}_{\frac{1}{2}1}|^2-|\hat{H}_{-\frac{1}{2}-1}|^2, \nonumber\\
\alpha_{\lambda_\mathcal{B}}&=&|\hat{H}_{\frac{1}{2}0}|^2+|\hat{H}_{\frac{1}{2}1}|^2+|\hat{H}_{\frac{3}{2}1}|^2
-|\hat{H}_{-\frac{1}{2}-1}|^2-|\hat{H}_{-\frac{1}{2}0}|^2-|\hat{H}_{-\frac{3}{2}-1}|^2, \nonumber\\
\alpha_{\lambda_\psi}&=&|\hat{H}_{\frac{1}{2}0}|^2+|\hat{H}_{-\frac{1}{2}0}|^2-|\hat{H}_{\frac{1}{2}1}|^2-|\hat{H}_{-\frac{1}{2}-1}|^2
-|\hat{H}_{\frac{3}{2}1}|^2-|\hat{H}_{-\frac{3}{2}-1}|^2,
\end{eqnarray}
where $U$ and $P$ describe the unpolarized parameters and polarization asymmetries, respectively.

\section{Numerical results}\label{sec:results}
The primary aim of this section is to carry out the numerical calculations and discussions
to the branching ratios, baryon fragmentation fractions, helicity amplitudes, and various asymmetries of the decays under consideration.
Below we begin with collecting the necessary input parameters in further numerical analysis.

\subsection{Input parameters}
We choose the $b$ baryon masses (GeV) and lifetimes (ps) from the new edition of the PDG~\cite{pdg2022}
\begin{eqnarray}
M_{\Lambda_b}&=&5.620, \quad M_{\Xi^-_b}=5.797, \quad M_{\Sigma_b}=5.811, \quad M_{\Xi_b'}=5.935, \quad M_{\Omega_b}=6.045,\nonumber\\
\tau_{\Lambda_b}&=&1.464, \quad \tau_{\Xi^-_b}=1.572, \quad \tau_{\Xi^0_b}=1.480,  \quad \tau_{\Omega_b}=1.64.
\end{eqnarray}
The masses (GeV) of the light baryons are taken to be~\cite{pdg2022}
\begin{eqnarray}
m_{N}=0.938, \quad m_{\Lambda}=1.116, \quad m_{\Xi}=1.315, \quad m_{\Sigma}=1.193, \quad m_{\Omega}=1.672,\quad m_{\Delta}=1.210.
\end{eqnarray}
Because the isospin splitting in the baryons are rather small, 
we neglect the mass differences between the isomultiplets.
The CKM matrix elements are chosen as~\cite{pdg2022}
\begin{eqnarray}
\left(\begin{array}{ccc}
V_{ud}&V_{us}&V_{ub}\\
V_{cd}&V_{cs}&V_{cb}\\
V_{td}&V_{ts}&V_{tb}\\
\end{array}\right)=
\left(\begin{array}{ccc}
1-\lambda^2/2&\lambda &A\lambda^3(\rho-i\eta)\\
-\lambda&1-\lambda^2/2&A\lambda^2\\
A\lambda^3(1-\rho-i\eta)&-A\lambda^2&1\\
\end{array}\right)
\end{eqnarray}
with the Wolfenstein parameters
\begin{eqnarray}
\lambda =0.22650, \quad  A=0.790,  \quad \bar{\rho}=0.141, \quad \bar{\eta}=0.357.
\end{eqnarray}
The heavy quark masses are taken from the previous work~\cite{prd106053005}: $m_b=4.8$ GeV and $m_c=1.275$ GeV.

\subsection{Branching ratios}
\begin{table}[!htbh]
	\caption{The predicted amplitudes, decay widths, and branching ratios of charmonium two-body decays of $b$ baryons,
where the theoretical uncertainties are due to the hadronic parameter  and the hard scale  (see text).
Numerical results of the branching ratios from~\cite{Gutsche:2018utw,Hsiao:2015cda,Hsiao:2021mlp} are listed in the last two columns for comparisons.}
	\label{tab:branching1}
	\begin{tabular}[t]{lcccccc}
		\hline\hline
		Mode &Transition & $|\mathcal{M}|(\text{GeV})$ &$\Gamma$(\text{GeV}) & $\mathcal{B}$   &CCQM~\cite{Gutsche:2018utw}  &GFA~\cite{Hsiao:2015cda}  \\ \hline	
		$\mathcal{B}_b(\bar {\textbf{3}})\rightarrow\mathcal{B}(\textbf{8})\psi$ &&&&&\\
		$ \Lambda_b\rightarrow \Lambda J/\psi$  &$b\rightarrow s c\bar c$
		&$3.4^{+0.4+0.3}_{-0.3-0.0}\times10^{-7}$
		&$2.6^{+0.6+0.5}_{-0.4-0.0}\times10^{-16}$  &$5.8^{+1.2+1.0}_{-0.9-0.1}\times10^{-4}$   &$8.3\times10^{-4}$&$(3.3\pm2.0)\times10^{-4}$\\
		$ \Lambda_b\rightarrow n J/\psi$        &$b\rightarrow d c\bar c$ & $9.9^{+1.0+0.7}_{-0.6-0.2}\times10^{-8}$ &    $2.2^{+0.5+0.3}_{-0.3-0.1}\times10^{-17}$     &    $4.9^{+1.0+0.7}_{-0.6-0.2}\times10^{-5}$   &     $4.0\times10^{-5}$ &     $\cdots$\\
		$ \Lambda_b\rightarrow \Sigma^0 J/\psi$ &$bu\rightarrow su$    & $2.5^{+0.2+0.4}_{-0.2-0.7}\times10^{-9}$     &  $1.3^{+0.2+0.4}_{-0.2-0.7}\times10^{-20}$  &    $2.9^{+0.5+0.9}_{-0.5-1.6}\times10^{-8}$   &    $\cdots$ &     $\cdots$\\
		$ \Xi_b^0\rightarrow \Xi^0 J/\psi$      &$b\rightarrow s c\bar c$  & $4.0^{+0.3+0.1}_{-0.2-0.0}\times10^{-7}$ &$3.4^{+0.5+0.1}_{-0.4-0.1}\times10^{-16}$&    $7.5^{+1.2+0.3}_{-0.1-0.2}\times10^{-4}$  &    $4.4\times10^{-4}$ &    $(4.9\pm3.0)\times10^{-4}$\\
		$ \Xi_b^0\rightarrow \Lambda J/\psi$    &$b\rightarrow d c\bar c$ & $3.0^{+0.3+0.3}_{-0.4-0.2}\times10^{-8}$ &$2.0^{+0.3+0.4}_{-0.5-0.3}\times10^{-18}$&    $4.4^{+0.8+0.8}_{-1.0-0.7}\times10^{-6}$  &    $3.1\times10^{-6}$ &    $(4.7\pm2.9)\times10^{-6}$\\
		$ \Xi_b^0\rightarrow \Sigma^0 J/\psi$   &$b\rightarrow d c\bar c$ & $6.7^{+0.6+0.2}_{-0.5-0.1}\times10^{-8}$ &$9.8^{+1.7+0.1}_{-1.4-0.1}\times10^{-18}$&    $2.2^{+0.4+0.2}_{-0.3-0.0}\times10^{-5}$   &    $7.0\times10^{-6}$ &    $(1.4\pm0.8)\times10^{-5}$\\
		$\Xi_b^-\rightarrow \Sigma^- J/\psi$   &$b\rightarrow d c\bar c$ & $9.5^{+0.8+0.3}_{-0.7-0.1}\times10^{-8}$ &$1.9^{+3.4+0.2}_{-2.8-0.2}\times10^{-17}$&    $4.7^{+0.8+0.4}_{-0.6-0.1}\times10^{-5}$   &    $2\times10^{-5}$   &    $(2.9\pm1.8)\times10^{-5}$\\
		$ \Xi_b^-\rightarrow \Xi^- J/\psi$      &$b\rightarrow s c\bar c$ & $4.0^{+0.3+0.1}_{-0.2-0.0}\times10^{-7}$ &$3.4^{+0.5+0.1}_{-0.4-0.1}\times10^{-16}$&    $8.0^{+1.3+0.3}_{-0.1-0.2}\times10^{-4}$   &    $4.6\times10^{-4}$ &    $(5.1\pm3.2)\times10^{-4}$\\
		$\mathcal{B}_b(\bar {\textbf{3}})\rightarrow\mathcal{B}(\textbf{10})\psi$ & &&&&\\
		$ \Lambda_b\rightarrow \Delta^0 J/\psi$ &$bu\rightarrow du $      & $2.9^{+0.4+0.3}_{-0.6-0.7}\times10^{-9}$  &$1.8^{+0.5+0.4}_{-0.7-0.7}\times10^{-20}$   & $4.5^{+1.2+1.1}_{-1.7-1.8}\times10^{-8}$   &    $\cdots$ & $\cdots$  \\
		$\mathcal{B}_b( \textbf{6})\rightarrow\mathcal{B}(\textbf{8})\psi$ &&&&&\\
		$ \Sigma_b^+\rightarrow pJ/\psi$         &$b\rightarrow d c\bar c$ & $5.0^{+0.0+0.2}_{-0.1-0.0}\times10^{-8}$ &$5.8^{+0.0+0.4}_{-0.3-0.2}\times10^{-18}$   & $\cdots$    &    $\cdots$ & $\cdots$  \\
		$ \Sigma_b^+\rightarrow \Sigma^+ J/\psi$ &$b\rightarrow s c\bar c$ & $2.5^{+0.1+0.2}_{-0.0-0.0}\times10^{-7}$ &$1.4^{+0.1+0.2}_{-0.0-0.0}\times10^{-16}$   & $\cdots$    &    $\cdots$ & $\cdots$  \\
		$ \Sigma_b^0\rightarrow n J/\psi$        &$b\rightarrow d c\bar c$ & $1.7^{+0.0+0.0}_{-0.0-0.0}\times10^{-8}$ &$6.9^{+0.2+0.3}_{-0.0-0.0}\times10^{-19}$   & $\cdots$    &    $\cdots$ & $\cdots$  \\
		$ \Sigma_b^0\rightarrow \Lambda J/\psi$  &$bu\rightarrow su$ & $4.7^{+0.7+0.7}_{-0.3-0.8}\times10^{-10}$ &$4.8^{+1.5+1.6}_{-0.5-1.5}\times10^{-22}$   & $\cdots$    &    $\cdots$ & $\cdots$  \\
		$ \Sigma_b^0\rightarrow \Sigma^0 J/\psi$ &$b\rightarrow s c\bar c$ & $1.3^{+0.1+0.1}_{-0.0-0.0}\times10^{-7}$ &$3.5^{+0.3+0.5}_{-0.0-0.0}\times10^{-17}$  & $\cdots$    &    $\cdots$ & $\cdots$  \\
		$ \Xi_b^{'0}\rightarrow \Lambda J/\psi$  &$b\rightarrow d c\bar c$ & $8.0^{+0.4+0.5}_{-0.4-0.3}\times10^{-9}$ &$1.4^{+0.1+0.1}_{-0.1-0.1}\times10^{-19}$   & $\cdots$    &    $\cdots$ & $\cdots$  \\
		$ \Xi_b^{'0}\rightarrow \Sigma^0 J/\psi$ &$b\rightarrow d c\bar c$ & $7.3^{+0.5+0.1}_{-0.6-0.4}\times10^{-9}$ &$1.2^{+0.2+0.0}_{-0.2-0.1}\times10^{-19}$   & $\cdots$    &    $\cdots$ & $\cdots$  \\
		$ \Xi_b^{'0}\rightarrow \Xi^0 J/\psi$    &$b\rightarrow s c\bar c$ & $8.5^{+0.6+0.3}_{-0.5-0.2}\times10^{-8}$ &$1.6^{+0.2+0.1}_{-0.2-0.1}\times10^{-17}$   & $\cdots$    &    $\cdots$ & $\cdots$  \\
		$ \Xi_b^{'-}\rightarrow \Xi^- J/\psi$    &$b\rightarrow s c\bar c$ &  $8.5^{+0.6+0.3}_{-0.5-0.2}\times10^{-8}$ &$1.6^{+0.2+0.1}_{-0.2-0.1}\times10^{-17}$   & $\cdots$   &    $\cdots$ & $\cdots$  \\
		$ \Xi_b^{'-}\rightarrow \Sigma^- J/\psi$ &$b\rightarrow d c\bar c$ & $1.6^{+0.1+0.0}_{-0.3-0.1}\times10^{-8}$ &$5.0^{+0.4+0.1}_{-1.4-1.2}\times10^{-19}$   & $\cdots$    &    $\cdots$ & $\cdots$  \\
		$ \Sigma_b^-\rightarrow \Sigma^- J/\psi$ &$b\rightarrow s c\bar c$  & $2.5^{+0.1+0.2}_{-0.0-0.0}\times10^{-7}$ &$1.4^{+0.1+0.2}_{-0.0-0.0}\times10^{-16}$    & $\cdots$   &    $\cdots$ & $\cdots$  \\
		$ \Omega_b^-\rightarrow \Xi^- J/\psi$    &$b\rightarrow d c\bar c$ & $5.1^{+0.1+0.4}_{-0.0-0.0}\times10^{-8}$ &$5.6^{+0.3+0.8}_{-0.0-0.0}\times10^{-18}$& $1.4^{+0.1+0.2}_{-0.0-0.0}\times10^{-5}$     &    $1.8\times10^{-6}$ &$\cdots$  \\
		$\mathcal{B}_b( \textbf{6})\rightarrow\mathcal{B}(\textbf{10})\psi$ &&&&&\\
		$ \Sigma_b^0\rightarrow \Delta^0 J/\psi$  &$b\rightarrow d c\bar c$ & $6.5^{+0.0+0.5}_{-0.3-0.0}\times10^{-8}$  &$9.1^{+0.9+1.3}_{-0.0-0.0}\times10^{-18}$   & $\cdots$    &    $\cdots$ & $\cdots$  \\
		$ \Sigma_b^+\rightarrow \Delta^+ J/\psi$  &$b\rightarrow d c\bar c$ & $9.3^{+0.5+0.4}_{-0.2-0.3}\times10^{-8}$  &$1.9^{+0.2+0.2}_{-0.1-0.1}\times10^{-17}$   & $\cdots$    &    $\cdots$ & $\cdots$  \\
		$ \Omega_b^-\rightarrow \Omega^- J/\psi$ &$b\rightarrow s c\bar c$  & $3.8^{+0.1+0.3}_{-0.0-0.1}\times10^{-7}$  &$2.8^{+0.2+0.4}_{-0.0-0.1}\times10^{-16}$& $6.9^{+0.5+1.0}_{-0.0-0.3}\times10^{-4}$   &    $8.1\times10^{-4}$ &    $\cdots$\\
		\hline\hline
	\end{tabular}
\end{table}

The numerical results on the amplitudes, decay widths, and branching ratios are collected in Table~\ref{tab:branching1},
where the first and second uncertainties arise from the shape parameter $A=0.5\pm 0.2$ in the $b$ baryon LCDAs
and hard scale $t$ varying from 0.8$t$ to 1.2$t$, respectively.
The second column in Table~\ref{tab:branching1} lists the corresponding quark level transitions.
It is seen that the CKM favored $b\rightarrow s c\bar c$ induced channels like $\Lambda_b\rightarrow \Lambda J/\psi, \Xi_b\rightarrow \Xi J/\psi, \Omega_b\rightarrow \Omega J/\psi$
have large branching ratios, typically at the order of $10^{-4}$.
These modes have been observed by experiment and the relative branching fractions multiplied by fragmentation fractions have been measured as shown in Table~\ref{tab:measure}.
The decay widths and branching ratios of the  CKM suppressed $b\rightarrow d c \bar c$ processes are generally an order of magnitude smaller
due to the suppression of $|\frac{V_{cd}}{V_{cs}}|^2\sim\lambda^2$.
The pure  exchange type decays mediated  by the $bu\rightarrow du$ and $bu\rightarrow su$ transitions have even lower rates at the order of $10^{-8}$.

It is worth to mentioning that the two modes of $\Lambda_b\rightarrow \Lambda J/\psi$ and  $\Lambda_b\rightarrow \Sigma^0 J/\psi$
have been studied in our previous work~\cite{prd106053005,Rui:2023fpp},
where a simple exponential model with one shape parameter $\omega$ was applied to simulate the $\Lambda_b$ LCDAs.
This treatment makes the results extremely dependent on the variation of $\omega$.
For example, from the Table V of~\cite{prd106053005}, one see that a $20\%$ change in the $\omega$ results in a $50\%$ variation in $\mathcal{B}(\Lambda_b\rightarrow \Lambda J/\psi)$.
In contrast, the variation of leading twist-2 LCDAs with respect to the parameter $A$ has been studied in~\cite{epjc732302},
which shows good stability of the LCDAs on $A$.
As can be seen from Table~\ref{tab:branching1}, when $A$ varies in [0.3, 0.7], the branching ratios change by about $20\%$.
Due to the updated LCDAs of $b$ baryon, 
the central value of $\mathcal{B}(\Lambda_b\rightarrow \Lambda J/\psi)$ is reduced from $7.75\times 10^{-4}$ in~\cite{prd106053005} to the current value of $5.8\times 10^{-4}$,
whereas the mode of $\Lambda_b\rightarrow \Sigma^0 J/\psi$ exhibits an opposite trend,
with the branching ratio increasing by an order of magnitude, yield $10^{-8}$.
The evident enhancement implies that pure exchange type decay is more sensitive to the nonperturbation LCDAs.
They thus  provide an appropriate platform for discriminating various model functions of LCDAs
once experimental information on these decays becomes available in the future.

The decay widths of some CKM favored $\Sigma_b$  channels can reach $10^{-16}$ GeV, which  are comparable to that of the favored $\Omega_b$ decay.
However, the lifetimes of $\Sigma_b$ and $\Xi_b'$ are dominant by the strong or electromagnetical decay modes,
and thus their weak decays are rare and the corresponding branching ratios are not shown in Table~\ref{tab:branching1}.
Despite the fact that some channels have lager decay widths, they have received little attention in the literature.
These modes may provide an opportunity to search for new physics beyond the SM
and deserve further investigation from both the experimental and theoretical sides.

In Table~\ref{tab:branching1}, we also compare our branching ratios with those in the literature~\cite{Gutsche:2018utw,Hsiao:2015cda,Hsiao:2021mlp}.
A number of nonleptonic heavy baryon decays were studied systematically in~\cite{Gutsche:2018utw} based on CCQM.
Their results tend to  be smaller except for the two modes of $\Lambda_b\rightarrow \Lambda J/\psi$ and $\Omega_b\rightarrow \Omega J/\psi$.
In particular, our prediction on the branching ratio of the $\Omega_b^-\rightarrow \Xi^- J/\psi$  is larger by a factor of 7.8.
In~\cite{Hsiao:2015cda}, the GFA was used to analyze two-body antitriplet $b$ baryon decays,
in which the decay amplitude is factorized as the baryonic transition form factor multiplied by the meson decay constant.
The baryonic transition form factors for different decay modes are related by the SU(3) flavor and SU(2) spin symmetries
and can be expressed in terms of the reduced parameter $C_{\parallel}$,
which was extracted from the data of $\mathcal{B}(\Lambda_b\rightarrow p \pi, pK)$~\cite{Hsiao:2014mua}.
It can be seen that our results are consistent with their predictions by considering the theoretical uncertainties.
Recently, a study about the $\Omega_b \rightarrow \Omega$ form factor based on the light-front quark model was presented in~\cite{Hsiao:2021mlp}.
The branching ratio of $\Omega_b \rightarrow \Omega J/\psi$ was estimated to be $5.3^{+3.3+3.8}_{-2.1-2.7}\times 10^{-4}$,
which is compatible with our calculations.
The value of the branching ratio for $\Omega_b \rightarrow \Omega J/\psi$ given in~\cite{Fayyazuddin:2017sxq},
spans a wide range, $(0.8-4.5)\times 10^{-5}$, which is far too  small.
However, the decay width of $\Gamma(\Omega_b \rightarrow \Omega J/\psi)$ evaluated in~\cite{Cheng:1996cs}
is $3.15a_2^2\times10^{10}\text{s}^{-1}$ with $a_2$ being the effective Wilson coefficient.
If we take $a_2=0.28$ to account for the nonfactorizable effects as that in $B \rightarrow J/\psi K^{(*)}$ decays~\cite{Cheng:1996xy},
the yielded branching ratio is  $\mathcal{B}(\Omega_b \rightarrow \Omega J/\psi)=4.1\times10^{-3}$,
which is the largest value among these currently available predictions.
In addition, the result in~\cite{prd99054020} gives the branching ratio $\mathcal{B}(\Lambda_b\rightarrow n J/\psi)=2.06^{+0.30+0.39+0.21}_{-0.13-0.42-0.18}\times 10^{-5}$,
  which is smaller than ours by a factor of 2.4.
Experimental investigations on these decays  would help to discriminate between the proposed models.

Isospin symmetry is a useful tool for the phenomenological analysis of heavy quark decays involving particles that carry isospin.
In Table~\ref{tab:branching1}, one can identify the following isospin relations in PQCD calculations:
\begin{eqnarray}\label{eq:iso}
\mathcal{M}(\Xi_b^{(')-}\rightarrow \Xi^- \psi)&=&\mathcal{M}(\Xi_b^{(')0}\rightarrow \Xi^0 \psi),\nonumber\\
\mathcal{M}(\Xi_b^{(')-}\rightarrow \Sigma^- \psi)&=&\sqrt{2}\mathcal{M}(\Xi_b^{(')0}\rightarrow \Sigma^0 \psi),\nonumber\\
\mathcal{M}(\Sigma_b^{\pm}\rightarrow \Sigma^{\pm} \psi)&=&2\mathcal{M}(\Sigma_b^0\rightarrow \Sigma^0 \psi).
\end{eqnarray}
If the isospin symmetry is extended to the SU(3) symmetry, more amplitude relations hold~\cite{Dery:2020lbc}
\begin{eqnarray}\label{eq:suu}
\mathcal{R}_1&=&\left|\frac{\mathcal{M}(\Xi_b^{0}\rightarrow \Xi^0\psi)}{\mathcal{M}(\Lambda_b\rightarrow \Lambda\psi)}\right|=\sqrt{\frac{3}{2}},\nonumber\\
\mathcal{R}_2&=&\left|\frac{\mathcal{M}(\Xi^0_b\rightarrow \Lambda\psi)}{\mathcal{M}(\Xi_b^{0}\rightarrow \Xi^0\psi)}\right|=\sqrt{\frac{1}{6}}\left|\frac{V_{cd}}{V_{cs}}\right|,\nonumber\\
\mathcal{R}_3&=&\left|\frac{\mathcal{M}(\Xi^0_b\rightarrow \Sigma^0\psi)}{\mathcal{M}(\Xi_b^{0}\rightarrow \Lambda\psi)}\right|=\sqrt{3},\nonumber\\
\mathcal{R}_4&=&\left|\frac{\mathcal{M}(\Lambda_b\rightarrow n\psi)}{\mathcal{M}(\Xi^0_b\rightarrow \Sigma^0\psi)}\right|=\sqrt{2}.
\end{eqnarray}
Note that we neglect the CKM-suppressed amplitudes here.
For more details of the complete SU(3) sum rules including the subleading amplitudes, we refer the reader to Ref.~\cite{Dery:2020lbc}.
Utilizing the results of the amplitudes from Table~\ref{tab:branching1},
one obtains the amplitude ratios defined in Eq.~(\ref{eq:suu}), whose numbers are displayed in Table~\ref{tab:rrr}.
The two errors are the same as for branching ratios in Table~\ref{tab:branching1}.
For comparison, we also list the results from the CCQM~\cite{Gutsche:2018utw}, GFA~\cite{Hsiao:2015cda} as well as the values in the SU(3) limit~\cite{Dery:2020lbc}  in the table.
Note that the decay amplitudes in~\cite{Gutsche:2018utw,Hsiao:2015cda} are not explicitly given,
but one can infer them from the corresponding branching ratios.
Obviously, the values from GFA match well the requirement from the SU(3) limit as shown in Table~\ref{tab:rrr}.
Our results deviate by $3\%-29\%$ from the expectations in the SU(3) limit,
which are the typical sizes of SU(3) breaking effects.
The CCQM predictions for the $\mathcal{R}_1$ and $\mathcal{R}_4$ apparently deviate from the SU(3) limit values amount up to $39\%$ and $67\%$, respectively.
 The large SU(3) symmetry breaking effects imply that the branching ratio of $\Lambda_b \rightarrow \Lambda J/\psi$ may be overestimated,
while that of $\Xi^0_b \rightarrow \Sigma^0 J/\psi$ is underestimated  in the CCQM as can be seen in Table~\ref{tab:branching1}.

Experimentally, the LHCb collaboration has published measurements of the
isospin amplitudes in $\Lambda_b\rightarrow J/\psi \Lambda (\Sigma^0)$ and $\Xi_b^0 \rightarrow J/\psi \Xi^0 (\Lambda)$ decays~\cite{LHCb:2019aci}.
 The isospin amplitude ratio $|A(\Xi_b^0 \rightarrow J/\psi \Lambda)/A(\Xi_b^0 \rightarrow J/\psi \Xi^0)|$ was determined to be $0.37\pm 0.06 \pm 0.02$,
where  the uncertainties are statistical and systematic, respectively.
It should be noted that the measured ratio does not contain the relative Cabibbo suppression factor of $\frac{V_{cd}}{V_{cs}}$.
After correcting for this factor, the PQCD prediction on this ratio is 0.33, which is in good agreement with the LHCb measurement.
An upper limit on another isospin amplitude ratio $|A(\Lambda_b\rightarrow \Sigma^0 J/\psi)/A(\Lambda_b\rightarrow \Lambda J/\psi)|$
was measured to be 1/21.8 at $95\%$ confidence level~\cite{LHCb:2019aci}.
From Table~\ref{tab:branching1}, we obtain the isospin amplitude ratio
$|A(\Lambda_b\rightarrow \Sigma^0 J/\psi)/A(\Lambda_b\rightarrow \Lambda J/\psi)|=7.4\times 10^{-3}$,
about one order below the present bound.
This leaves plenty of room for $\Lambda-\Sigma$ mixing effects and new physics contributions in $\Lambda_b\rightarrow \Sigma^0 J/\psi$ decay,
which may be an intriguing topic for future investigation.

\begin{table}[!htbh]
	\caption{The comparison of different theoretical predictions for the ratios between decay amplitudes defined in Eq.~(\ref{eq:suu}).
The theoretical uncertainties are the same as for branching ratios in Table~\ref{tab:branching1}.
		The expectations in the SU(3) limit are given in the last row.}
	\label{tab:rrr}
	\begin{tabular}[t]{lcccc}
		\hline\hline
		&$\mathcal{R}_1$&$\mathcal{R}_2$&$\mathcal{R}_3$&$\mathcal{R}_4$\\ \hline
		PQCD     & $1.18^{+0.05+0.00}_{-0.05-0.07}$ &$0.075^{+0.002+0.005}_{-0.007-0.005}$ &$2.23^{+0.15+0.13}_{-0.02-0.14}$ &$1.48^{+0.02+0.06}_{-0.00-0.01}$\\
		CCQM~\cite{Gutsche:2018utw}         & 0.74 &0.082 &1.51 &2.35    \\
		GFA~\cite{Hsiao:2015cda}             & 1.23 &0.096 &1.74 &$\cdots$\\
		SU(3) limit~\cite{Dery:2020lbc}     & 1.22 &0.092 &1.73 &1.41    \\
		\hline\hline
	\end{tabular}
\end{table}		

\subsection{$b$ baryon fragmentation fractions}

Combining the obtained branching ratios in Table~\ref{tab:branching1} and the PDG average values in Table~\ref{tab:measure},
it is straightforward to calculate  the fragmentation fractions
\begin{eqnarray}\label{eq:fragmentation}
f_{\Lambda_b}=0.100^{+0.018+0.002+0.014}_{-0.017-0.015-0.014}, \quad f_{\Xi_b}=0.013^{+0.000+0.000+0.003}_{-0.002-0.001-0.003}, \quad f_{\Omega_b}=4.2^{+0.0+0.2+1.6}_{-0.3-0.5-1.2}\times10^{-3},
\end{eqnarray}
where the first two errors are the same as for branching ratios in Table~\ref{tab:branching1},
while the last one rises from data in Table~\ref{tab:measure}.
It is apparent that our prediction $f_{\Lambda_b}=0.100^{+0.018+0.002+0.014}_{-0.017-0.015-0.014}$ overlaps $f_{\Lambda_b}=0.07$ from the LEP experiment~\cite{Galanti:2015pqa} within errors,
and $f_{\Xi_b}=0.013^{+0.000+0.000+0.003}_{-0.002-0.001-0.003}$ is consistent with $f_{\Xi_b}=0.011\pm 0.005$ from the DELPHI Collaboration~\cite{DELPHI:2003pao}.
The estimations in~\cite{plb751127} with the fragmentation fractions $f_{\Lambda_b}=0.175\pm0.106$ and $f_{\Xi_b}=0.019\pm 0.013$
present larger central values.
Recalling that their predictions suffer large uncertainties from the nonperturbative parameters,
so within one standard deviation tolerance, one still can count them as being consistent.
The fragmentation of double strange $\Omega_b$ baryon  has not been determined experimentally.
Until now, only the calculation based on the light-front quark model~\cite{Hsiao:2021mlp} was available.
The resulting value $f_{\Omega_b}=0.54^{+0.34+0.39+0.21}_{-0.22-0.28-0.15}\%$ is compatible with the PQCD prediction.

Summing of all the weakly decaying $b$ baryons, one can get the total fragmentation fraction of $b$ baryon,
$f_{\text{baryons}}=f_{\Lambda_b}+2 f_{\Xi_b}+f_{\Omega_b}=0.13^{+0.02+0.00+0.02}_{-0.02-0.02-0.02}$,
which lies between the experimental data $0.084\pm 0.011$ from the LEP measurements on $Z$ decays
and $0.198\pm0.046$ from the Tevatron Collaboration~\cite{pdg2022}.
Note that here isospin invariance is assumed in the production of $\Xi_b^0$ and $\Xi_b^-$.
Our number is somewhat lower than the value $0.213 \pm 0.108$ in~\cite{plb751127}.

From the values of the fragmentation fractions in Eq.~(\ref{eq:fragmentation}), we obtain three fragmentation ratios
\begin{eqnarray}\label{eq:ratio}
\frac{f_{\Xi_b}}{f_{\Lambda_b}}=0.130^{+0.003+0.011+0.010}_{-0.020-0.003-0.014}, \quad \frac{f_{\Omega_b}}{f_{\Lambda_b}}=0.042^{+0.005+0.002+0.009}_{-0.006-0.000-0.007}, \quad \frac{f_{\Omega_b}}{f_{\Xi_b}}=0.323^{+0.032+0.015+0.040}_{-0.000-0.015-0.023}.
\end{eqnarray}
 Among these fragmentation ratios, $\frac{f_{\Xi_b}}{f_{\Lambda_b}}$ is particularly interesting,
 and its status has been analyzed in detail in~\cite{Jiang:2018iqa} by means of the relevant measurements.
 The first way to determine $\frac{f_{\Xi_b}}{f_{\Lambda_b}}$ is utilizing the experimental data of
  $\Xi_b^-\rightarrow \Xi^- J/\psi$ and  $\Lambda_b \rightarrow \Lambda J/\psi$ as displayed in Table~\ref{tab:measure}.
 Since $\Xi_b$ and $\Lambda_b$ belong to the same SU(3) antitriplet and the final baryons are an octet as mentioned above,
 the two decay rates are related through SU(3) flavor symmetry.
  By invoking the SU(3) symmetry relation of $\mathcal{B}(\Lambda_b\rightarrow \Lambda J/\psi)/\mathcal{B}(\Xi^-_b\rightarrow \Xi^- J/\psi)=2/3$
 and the experimental data in Table~\ref{tab:measure}, the authors of~\cite{Voloshin:2015xxa} obtain the ratio of $\frac{f_{\Xi_b}}{f_{\Lambda_b}}=0.11\pm 0.03$,
 which indicates good agreement with our prediction and the value $0.108\pm0.034$ given in~\cite{plb751127}.
 Likewise, assuming SU(3) symmetry, the ratio of fragmentation fraction  $\frac{f_{\Xi_b}}{f_{\Lambda_b}}$ is
determined to be $(6.7\pm 0.5_{stat}\pm0.5_{syst}\pm2.0_{SU(3)})\times 10^{-2}$ at 7, 8 TeV and
$(8.2\pm 0.7_{stat}\pm0.6_{syst}\pm2.5_{SU(3)})\times 10^{-2}$ at 13 TeV data samples by LHCb~\cite{LHCb:2019sxa}.
The last uncertainty of $30\%$ is the typical size of SU(3) breaking effects.
Although the measured central values are generally below the theoretical predictions,
they are still consistent within one standard deviation considering the SU(3) breaking effects.

Another independent determination of $\frac{f_{\Xi_b}}{f_{\Lambda_b}}$ can be performed through the heavy-flavor-conserving decay.
Very recently,  LHCb published  a more precise measurement of the hadronic weak decay $\Xi_b^-\rightarrow \Lambda_b^0\pi^-$
with the relative production ratio being~\cite{LHCb:2023tma}
\begin{eqnarray}
\frac{f_{\Xi_b}}{f_{\Lambda_b}} \mathcal{B}(\Xi_b^-\rightarrow \Lambda_b^0\pi^-)=(7.3\pm0.8\pm0.6)\times10^{-4},
\end{eqnarray}
which improves three times better statistical precision than the previous measurement in 2015~\cite{LHCb:2015une}.
In this case, theoretical estimate of $\mathcal{B}(\Xi_b^-\rightarrow \Lambda_b^0\pi^-)$ is necessary.
However, the predicted branching ratio of $\Xi_b^-\rightarrow \Lambda_b^0\pi^-$ spans a wide interval from
$0.14\%$ to $8\%$~\cite{Voloshin:2000et,Gronau:2015jgh,Niu:2021qcc,Cheng:2015ckx,Cheng:2022kea,Cheng:2022jbr,Faller:2015oma,Li:2014ada},
leading to $\frac{f_{\Xi_b}}{f_{\Lambda_b}}$ lies in a lager range from $9.1\times10^{-3}$ to $0.52$,
which even covers  the range between 0.1 and 0.3 estimated in~\cite{LHCb:2015une},
based on the measured production rates of other strange particles relative to their nonstrange counterparts.
More precise theoretical predictions on the branching ratio of the heavy-flavor-conserving decay would be highly desirable.

The third method to constrain $\frac{f_{\Xi_b}}{f_{\Lambda_b}}$ is using the LHCb data~\cite{LHCb:2014chk}
\begin{eqnarray}
\frac{f_{\Xi_b}}{f_{\Lambda_b}} \frac{\mathcal{B}(\Xi_b^0\rightarrow \Xi_c^+\pi^-)}{\mathcal{B}(\Lambda_b^0\rightarrow \Lambda_c^+\pi^-)}
\frac{\mathcal{B}(\Xi_c^+ \rightarrow pK^-\pi^+)}{\mathcal{B}( \Lambda_c^+\rightarrow pK^-\pi^+)}=(1.88\pm0.04\pm0.03)\times10^{-2},
\end{eqnarray}
which is much higher precision than the above two measurements.
In~\cite{LHCb:2014chk}, the two ratios  of $\frac{\mathcal{B}(\Xi_b^0\rightarrow \Xi_c^+\pi^-)}{\mathcal{B}(\Lambda_b^0\rightarrow \Lambda_c^+\pi^-)}$ and
$\frac{\mathcal{B}(\Xi_c^+ \rightarrow pK^-\pi^+)}{\mathcal{B}( \Lambda_c^+\rightarrow pK^-\pi^+)}$
are approximately equal to 1 and 0.1, respectively, and in the assumption of naive Cabibbo factors,
it is then expected that $\frac{f_{\Xi_b}}{f_{\Lambda_b}}\sim 0.2$~\cite{LHCb:2014chk}.
Nonetheless, with the branching fraction of $\mathcal{B}( \Lambda_c^+\rightarrow pK^-\pi^+)$  obtained under the $U$-spin symmetry,
the fragmentation ratio $\frac{f_{\Xi_b}}{f_{\Lambda_b}}$  was determined as $0.054\pm 0.020$ in~\cite{Jiang:2018iqa} and $0.065\pm0.020$ in~\cite{Wang:2019dls},
which are three times lower than the naive expectation and poses an interesting challenge for theoretical interpretations.

 More potential extractions of $\frac{f_{\Xi_b}}{f_{\Lambda_b}}$  are the measurements of  multibody charmless decays of the $\Lambda_b$ and $\Xi_b$ baryons,
 such as the three-body $\Lambda_b/\Xi_b\rightarrow\Lambda hh'$~\cite{LHCb:2016rja} and four-body $\Lambda_b/\Xi_b\rightarrow pK hh'$~\cite{LHCb:2017gjj} decays,
 where $h^{(')}$ denotes a kaon or pion.
 Although they have been observed by LHCb, the measurements suffer from large uncertainties.
 Moreover, accurate calculations of the multibody decays are quite challenging for theory,
 since a factorization formalism that describes multibody decays in full phase space is not yet available at the moment~\cite{Rui:2021kbn}.
 In any case, this fragmentation ratio deserves further theoretical and experimental investigations in the future.

Although CDF~\cite{CDF:2009sbo}, D0~\cite{D0:2008sbw}, and LHCb~\cite{LHCb:2023qxn,Nicolini:2023stq}
have published, the measurements of the relative production rate of the $\Omega_b$ and $\Lambda_b(\Xi_b)$ baryons as presented in Table~\ref{tab:measure},
it is difficult to determine $\frac{f_{\Omega_b}}{f_{\Lambda_b}}$ and $\frac{f_{\Omega_b}}{f_{\Xi_b}}$ experimentally
because the $\Omega$ baryon is not a member of the same baryon multiplet as the $\Xi$ and $\Lambda$ baryons,
and there is no SU(3) symmetry relations among these modes.
Further theoretical inputs on the branching ratios are needed to directly access the two fragmentation ratios.
By using the PQCD predictions on the branching ratios listed in Table~\ref{tab:branching1},
we can  demonstrate the possible range for the two fragmentation ratios in accordance with the measurements
by CDF~\cite{CDF:2009sbo}, D0~\cite{D0:2008sbw}, and LHCb~\cite{LHCb:2023qxn,Nicolini:2023stq}, yielding
\begin{eqnarray}\label{eq:s4}
\frac{f_{\Omega_b}}{f_{\Lambda_b}}&=&0.038^{+0.004+0.001+0.014+0.003}_{-0.000-0.006-0.010-0.004
} \quad \quad\quad\quad \quad \text{CDF}, \nonumber\\ 
\frac{f_{\Omega_b}}{f_{\Xi_b}}&=&\left\{
\begin{aligned}
&0.313^{+0.026+0.006+0.139+0.011}_{-0.004-0.029-0.139-0.012} \quad\quad\quad  &\text{CDF}\\
&0.928^{+0.077+0.018+0.371+0.162}_{-0.012-0.088-0.371-0.258} \quad            &\text{D0} \\
&0.139^{+0.012+0.003+0.009+0.009}_{-0.002-0.013-0.009-0.009} \quad\quad\quad  &\text{LHCb}\\
\end{aligned}\right.,
\end{eqnarray}
where the first two errors are the same as for branching ratios in Table~\ref{tab:branching1},
while the last two rise from data in Table~\ref{tab:measure}.
The values derived from the CDF~\cite{CDF:2009sbo} are in accordance with those in Eq.~(\ref{eq:ratio}).
The extraction from the LHCb data gives $\frac{f_{\Omega_b}}{f_{\Xi_b}}=0.139^{+0.012+0.003+0.009+0.009}_{-0.002-0.013-0.009-0.009}$
and is consistent with the estimation of $15\%$ that was obtained in Ref~\cite{LHCb:2016qpe}.
Nevertheless, the two measurements on $\frac{f_{\Omega_b}}{f_{\Xi_b}}\frac{\mathcal{B}(\Omega_b^-\rightarrow\Omega^- J/\psi)}
{\mathcal{B}(\Xi_b^-\rightarrow\Xi^- J/\psi)}$   from D0~\cite{D0:2008sbw} and LHCb~\cite{LHCb:2023qxn,Nicolini:2023stq} are not in good agreement as seen in Table~\ref{tab:branching1}.
This leads to a large spread in $\frac{f_{\Omega_b}}{f_{\Xi_b}}$ as shown in Eq.~(\ref{eq:s4}).
The big value extracted from D0 is disfavored
since the production of $\Omega_b$ should be further suppressed by an additional strange quark with respect to the $\Xi_b$.
 Alternatively, one can determine $\frac{f_{\Omega_b}}{f_{\Xi_b}}$ via the decays of the $\Xi_b^-$ and $\Omega_b^-$ baryons to the charmless three-body final states $pK^-K^-$.
However, no significant signal for $\Omega_b^-\rightarrow pK^-K^-$ decay is found at present,
only an upper limit on $\frac{f_{\Omega_b}}{f_{\Xi_b}}\frac{\mathcal{B}(\Omega_b^-\rightarrow pK^-K^-)}{\mathcal{B}(\Xi_b^-\rightarrow pK^-K^-)}$ was set by LHCb~\cite{LHCb:2021enr}.

\subsection{Helicity amplitudes  and  asymmetry parameters}

\begin{table}[!htbh]
	\caption{Helicity amplitudes (GeV) and the normalized helicity amplitudes squared for the octet channels.
		For those modes satisfying the isospin relations in Eq.~(\ref{eq:iso}), we only list one of them in this table. Only central values are presented here.}
	\label{tab:helicity}
	\begin{tabular}[t]{lccccc}
		\hline\hline
		Mode	&$\lambda_{\mathcal{B}}\lambda_{\psi}$  &$\frac{1}{2}1$ & $-\frac{1}{2}-1$ & $\frac{1}{2}0$ & $-\frac{1}{2}0$ \\ \hline		
		$\Lambda_b\rightarrow \Lambda J/\psi$ 	&$H_{\lambda_{\mathcal{B}}\lambda_{\psi}}(10^{-8})$
		& $-1.7+i5.8$  & $6.3-i19.8$   & $5.1-i13.0$   & $5.6-i22.1$   \\	
		&	$|\hat{H}_{\lambda_{\mathcal{B}}\lambda_{\psi}}|^2$  &$0.031$  &$0.364$  &$0.165$  &$0.440$ \\
		
		$\Lambda_b \to nJ/\psi$ 	&$H_{\lambda_{\mathcal{B}}\lambda_{\psi}}(10^{-9})$
		& $-0.2+i5.1$  & $-22.3+i59.3$   & $-4.9+i7.2$   & $-11.6+i74.2$   \\	
		&	$|\hat{H}_{\lambda_{\mathcal{B}}\lambda_{\psi}}|^2$  &$0.003$  &$0.412$  &$0.008$  &$0.578$ \\
		
		$\Lambda_b \to \Sigma^0J/\psi$    &$H_{\lambda_{\mathcal{B}}\lambda_{\psi}}(10^{-10})$
		& $-12.2-i13.2$  & $7.6+i8.3$   & $7.7+i8.7$   & $-4.2-i1.6$   \\	
		&	$|\hat{H}_{\lambda_{\mathcal{B}}\lambda_{\psi}}|^2$  &$0.534$  &$0.210$  &$0.222$  &$0.034$ \\
		
		$\Omega^-_b \to \Xi^-J/\psi$     &$H_{\lambda_{\mathcal{B}}\lambda_{\psi}}(10^{-8})$
		& $0.5-i1.4$  & $-0.8+i2.2$   & $0.9-i2.6$   & $1.1-i3.1$   \\	
		&	$|\hat{H}_{\lambda_{\mathcal{B}}\lambda_{\psi}}|^2$  &$0.083$  &$0.205$  &$0.284$  &$0.428$ \\
		
		$\Sigma^0_b \to \Lambda J/\psi$   &$H_{\lambda_{\mathcal{B}}\lambda_{\psi}}(10^{-10})$
		& $3.1+i2.7$  & $-0.8-i0.6$   & $-1.0-i0.8$   & $1.3+i0.5$   \\	
		&	$|\hat{H}_{\lambda_{\mathcal{B}}\lambda_{\psi}}|^2$  &$0.792$  &$0.046$  &$0.076$  &$0.085$ \\
		
		$\Sigma^0_b \to nJ/\psi$         &$H_{\lambda_{\mathcal{B}}\lambda_{\psi}}(10^{-9})$
		& $1.5-i3.8$  & $-2.4+i6.6$   & $3.5-i10.2$   & $3.8-i10.2$   \\	
		&	$|\hat{H}_{\lambda_{\mathcal{B}}\lambda_{\psi}}|^2$  &$0.057$  &$0.162$  &$0.386$  &$0.392$ \\
		
		$\Sigma^+_b \to pJ/\psi$         &$H_{\lambda_{\mathcal{B}}\lambda_{\psi}}(10^{-8})$
		& $0.4-i1.1$  & $-0.7+i1.9$   & $1.1-i3.0$   & $1.0-i3.0$   \\	
		&	$|\hat{H}_{\lambda_{\mathcal{B}}\lambda_{\psi}}|^2$  &$0.053$  &$0.156$  &$0.398$  &$0.329$ \\
		
		$\Sigma^+_b \to \Sigma^+J/\psi$  &$H_{\lambda_{\mathcal{B}}\lambda_{\psi}}(10^{-8})$
		& $-2.3+i6.5$  & $3.6-i10.3$   & $-4.6+i13.1$   & $-5.7+i16.0$   \\	
		&	$|\hat{H}_{\lambda_{\mathcal{B}}\lambda_{\psi}}|^2$  &$0.073$  &$0.185$  &$0.298$  &$0.443$ \\
		
		$\Xi^0_b \to \Lambda J/\psi$      &$H_{\lambda_{\mathcal{B}}\lambda_{\psi}}(10^{-9})$
		& $1.2+i0.4$  & $-7.9+i16.3$   & $-1.9+i3.6$   & $-2.1+i23.1$   \\	
		&	$|\hat{H}_{\lambda_{\mathcal{B}}\lambda_{\psi}}|^2$  &$0.002$  &$0.370$  &$0.019$  &$0.609$ \\
		
		$\Xi^0_b \to \Sigma^0J/\psi$    &$H_{\lambda_{\mathcal{B}}\lambda_{\psi}}(10^{-9})$
		& $1.0+i4.4$  & $-15.6+i38.6$   & $-3.3+i5.7$   & $-6.8+i51.4$   \\	
		&	$|\hat{H}_{\lambda_{\mathcal{B}}\lambda_{\psi}}|^2$  &$0.005$  &$0.387$  &$0.010$  &$0.599$ \\
		
		$\Xi^0_b \to \Xi^0J/\psi$          &$H_{\lambda_{\mathcal{B}}\lambda_{\psi}}(10^{-8})$
		& $0.6-i3.5$  & $8.4-i22.9$   & $2.7-i3.4$   & $7.1-i30.1$   \\	
		&	$|\hat{H}_{\lambda_{\mathcal{B}}\lambda_{\psi}}|^2$  &$0.008$  &$0.376$  &$0.012$  &$0.605$ \\
		
		$\Xi'^0_b \to \Lambda J/\psi$      &$H_{\lambda_{\mathcal{B}}\lambda_{\psi}}(10^{-9})$
		& $-0.4+i1.3$  & $1.6-i4.7$   & $1.6-i3.0$   & $-1.7+i4.8$   \\	
		&	$|\hat{H}_{\lambda_{\mathcal{B}}\lambda_{\psi}}|^2$  &$0.028$  &$0.389$  &$0.180$  &$0.404$ \\
		
		$\Xi'^0_b \to \Sigma^0J/\psi$   &$H_{\lambda_{\mathcal{B}}\lambda_{\psi}}(10^{-9})$
		& $0.4-i1.3$  & $-1.6+i4.4$   & $0.03+i0.08$   & $1.7-i5.2$   \\	
		&	$|\hat{H}_{\lambda_{\mathcal{B}}\lambda_{\psi}}|^2$  &$0.035$  &$0.412$  &$0.001$  &$0.553$ \\
		
		$\Xi'^0_b \to \Xi^0J/\psi$        &$H_{\lambda_{\mathcal{B}}\lambda_{\psi}}(10^{-8})$
		& $-0.5+i1.8$  & $1.5-i5.1$   & $-0.3-i1.0$   & $-1.9+i6.0$   \\	
		&	$|\hat{H}_{\lambda_{\mathcal{B}}\lambda_{\psi}}|^2$  &$0.046$  &$0.393$  &$0.016$  &$0.545$ \\
		
		\hline\hline
	\end{tabular}
\end{table}	

\begin{table}[!htbh]
	\caption{The same as Table~\ref{tab:helicity} but for the decuplet modes.}
	\label{tab:helicity1}
	\begin{tabular}[t]{lccccccc}
		\hline\hline
		Mode	&$\lambda_{\mathcal{B}}\lambda_{\psi}$
		& $\frac{1}{2}0$ & $-\frac{1}{2}0$ &$\frac{3}{2}1$ & $-\frac{3}{2}-1$  &$\frac{1}{2}1$ & $-\frac{1}{2}-1$ \\ \hline	
	$\Lambda_b\rightarrow \Delta^0 J/\psi$ &$H_{\lambda_{\mathcal{B}}\lambda_{\psi}}(10^{-10})$
		&22.7+i8.9  & 5.8+i4.5 & 10.2+i2.7 &5.2+i0.2 &2.7+i0.0& 5.5+i3.1 \\	
		&$|\hat{H}_{\lambda_{\mathcal{B}}\lambda_{\psi}}|^2$  &$0.712$  &$0.065$  &$0.133$  &$0.033$ &$0.008$&$0.048$\\
		$\Omega^-_b\rightarrow \Omega^- J/\psi$ &$H_{\lambda_{\mathcal{B}}\lambda_{\psi}}(10^{-8})$
		& 3.8-i9.8  &7.1-i17.1 & 0.8-i0.8 & -5.3+i20.1 & 1.3-i1.3& -7.1+i22.0 \\	
		&$|\hat{H}_{\lambda_{\mathcal{B}}\lambda_{\psi}}|^2$  &$0.077$  &$0.241$  &$0.001$  &$0.303$ &$0.002$&$0.375$\\
		$\Sigma^0_b\rightarrow \Delta^0 J/\psi$ &$H_{\lambda_{\mathcal{B}}\lambda_{\psi}}(10^{-8})$
&-1.8+i5.8  & -3.3+i10.2 & -0.8+i2.1 & -0.1-i0.8 & -1.4+i3.2 & 2.2-i6.5 \\	
&$|\hat{H}_{\lambda_{\mathcal{B}}\lambda_{\psi}}|^2$  &$0.072$  &$0.150$  &$0.001$  &$0.441$ &$0.015$&$0.321$\\
		$\Sigma_b^+\rightarrow \Delta^+ J/\psi$&$H_{\lambda_{\mathcal{B}}\lambda_{\psi}}(10^{-8})$
		& -0.8+i2.9  & -1.9+i5.8 & -0.7+i2.1 & -1.2+i3.2 & -1.4+i3.8 & 0.9-i2.7 \\	
		&$|\hat{H}_{\lambda_{\mathcal{B}}\lambda_{\psi}}|^2$    &$0.106$  &$0.420$  &$0.055$ &$0.137$ &$0.191$ &$0.091$  \\

		\hline\hline
	\end{tabular}
\end{table}	

The calculated helicity amplitudes for the  octet and decuplet modes are summarized in Tables~\ref{tab:helicity} and~\ref{tab:helicity1}, respectively.
From Table~\ref{tab:helicity}, one see that the helicity amplitudes of the two isospin-violating modes of $\Lambda_b\rightarrow \Sigma^0 J/\psi$ and $\Sigma^0_b\rightarrow \Lambda^0 J/\psi$
 are dominated by $H_{\frac{1}{2}1}$. The pattern is similar to the previous observation in~\cite{Rui:2023fpp}, and the explanation has been given there as well.
 This reflects that the normalized helicity amplitudes are less sensitive to the choice of the $\Lambda_b$ LCDAs with respect to the branching ratios.
 According to Eqs.~(\ref{eq:two}) and~(\ref{eq:alp}), the various asymmetries are defined by the combinations of normalized helicity amplitudes;
 they thus  are also insensitive to the LCDAs as we will see later.

 The $\Sigma_b$ and $\Omega_b$ decays, except for the isospin-violating one, 
 receive substantial contributions from $H_{\frac{1}{2}0}$ and $H_{-\frac{1}{2}0}$ as shown in Table~\ref{tab:helicity},
 which means the longitudinal polarization state for the $J/\psi$ meson is preferred in these decays.
 However, the $\Lambda_b$ and $\Xi_b^{(')}$ channels associated with $b \rightarrow d(s) c\bar c$ transitions
 are clearly governed by negative-helicity states for the daughter baryon,
 and the longitudinal and transverse components are  of similar sizes.
 Contributions from the $\lambda_{\mathcal{B}}=\frac{1}{2}$ helicity states are small, around $20\%$ or less.
 The pattern of the suppressed positive-helicity states of the daughter baryon in $\Lambda_b\rightarrow\Lambda J/\psi$ decay has been observed
 by LHCb~\cite{plb72427}, ATLAS~\cite{prd89092009}, and CMS~\cite{prd97072010} experiments.
More angular analysis of other channels, especially $\Xi_b\rightarrow \Xi J/\psi$, are expected to examine our predictions.

From Table~\ref{tab:helicity1}, it is observed that  different modes exhibit distinct behaviors for the relative importance  of helicity amplitudes.
The pure exchange mode of $\Lambda_b\rightarrow \Delta^0 J/\psi$
shows a significant contribution from $H_{\frac{1}{2}0}$, which occupies about $71.2\%$ of the full contribution.
The dominances of the negative helicity configurations were observed in $\Omega_b\rightarrow \Omega J/\psi$  and $\Sigma^0_b\rightarrow \Delta^0 J/\psi$ decays,
whereas the positive ones  are strongly suppressed, leading to less than $10\%$.
The  $\Sigma_b^+\rightarrow \Delta^+ J/\psi$ process is dominated by the helicity amplitude $H_{-\frac{1}{2}0}$,
while the contribution from $H_{\frac{3}{2}1}$ is predicted to be rather small.
Since these helicity amplitudes in the  decuplet modes have still received less attention
in both theory and experiment, we wait for future comparison.

\begin{table}[!htbh]
	\caption{Asymmetry parameters for the $\mathcal{B}_b \rightarrow \mathcal{B}(\textbf{8}) J/\psi$ decays.
The theoretical uncertainties are the same as for branching ratios in Table~\ref{tab:branching1}.}
	\label{tab:asy}
	\begin{tabular}[t]{lccccc}
		\hline\hline
		Mode    & $\alpha_b$ & $\alpha_{\lambda_\mathcal{B}}$ & $\alpha_{\lambda_\psi}$  &$r_0$  &$r_1$ \\ \hline		
		$\Lambda_b\rightarrow \Lambda J/\psi$ & $0.057^{+0.010+0.000}_{-0.024-0.026}$    & $-0.608^{+0.007+0.008}_{-0.028-0.002}$     & $0.209^{+0.028+0.036}_{-0.047-0.000}$      &$0.605^{+0.014+0.018}_{-0.023-0.000}$ & $-0.275^{+0.000+0.000}_{-0.009-0.009}$ \\
		$\Lambda_b \to nJ/\psi$          & $-0.161^{+0.000+0.003}_{-0.040-0.015}$    & $-0.980^{+0.000+0.006}_{-0.008-0.004}$   & $0.171^{+0.034+0.012}_{-0.000-0.000}$      &$0.586^{+0.017+0.006}_{-0.000-0.000}$ & $-0.570^{+0.000+0.004}_{-0.024-0.009}$ \\
		$\Lambda_b \to \Sigma^0J/\psi$   & $-0.135^{+0.063+0.091}_{-0.075-0.000}$    & $0.512^{+0.105+0.090}_{-0.000-0.000}$   & $-0.488^{+0.269+0.197}_{-0.021-0.000}$      &$0.256^{+0.135+0.098}_{-0.010-0.000}$ & $0.189^{+0.084+0.090}_{-0.005-0.000}$ \\
		$\Xi^0_b \to \Lambda J/\psi$      & $-0.222^{+0.092+0.116}_{-0.000-0.000}$    & $-0.958^{+0.022+0.054}_{-0.000-0.000}$   & $0.256^{+0.000+0.000}_{-0.090-0.089}$      &$0.628^{+0.000+0.000}_{-0.045-0.045}$ & $-0.590^{+0.049+0.071}_{-0.000-0.000}$ \\
		$\Xi^0_b \to \Sigma^0J/\psi$     & $-0.206^{+0.052+0.119}_{-0.007-0.001}$    & $-0.971^{+0.000+0.023}_{-0.006-0.000}$   & $0.217^{+0.002+0.001}_{-0.063-0.100}$      &$0.608^{+0.001+0.000}_{-0.031-0.050}$ & $-0.589^{+0.023+0.071}_{-0.006-0.000}$ \\
		$\Xi^0_b \to \Xi J/\psi$          & $-0.225^{+0.019+0.060}_{-0.000-0.051}$    & $-0.960^{+0.011+0.009}_{-0.000-0.000}$   & $0.233^{+0.005+0.059}_{-0.010-0.051}$      &$0.617^{+0.003+0.030}_{-0.005-0.025}$ & $-0.593^{+0.015+0.035}_{-0.000-0.024}$ \\
		$\Omega^-_b \to \Xi^-J/\psi$     & $-0.022^{+0.022+0.044}_{-0.028-0.000}$    & $-0.265^{+0.054+0.063}_{-0.054-0.000}$   & $0.424^{+0.063+0.071}_{-0.022-0.000}$      &$0.712^{+0.031+0.036}_{-0.011-0.000}$ & $-0.144^{+0.038+0.052}_{-0.041-0.000}$ \\
		$\Sigma^0_b \to \Lambda J/\psi$   & $-0.756^{+0.138+0.018}_{-0.000-0.000}$    & $0.738^{+0.034+0.000}_{-0.029-0.049}$   & $-0.677^{+0.209+0.056}_{-0.000-0.000}$      &$0.162^{+0.104+0.028}_{-0.000-0.000}$ & $-0.009^{+0.086+0.007}_{-0.000-0.024}$ \\
		$\Sigma^0_b \to nJ/\psi$         & $0.095^{+0.100+0.027}_{-0.082-0.009}$    & $-0.115^{+0.117+0.014}_{-0.123-0.000}$   & $0.562^{+0.038+0.084}_{-0.056-0.035}$      &$0.781^{+0.019+0.042}_{-0.028-0.018}$ & $-0.010^{+0.109+0.018}_{-0.103-0.000}$ \\
		$\Sigma^+_b \to pJ/\psi$         & $0.109^{+0.082+0.000}_{-0.095-0.043}$    & $-0.097^{+0.100+0.000}_{-0.119-0.044}$   & $0.581^{+0.035+0.021}_{-0.023-0.000}$      &$0.790^{+0.017+0.010}_{-0.011-0.000}$ & $0.006^{+0.091+0.000}_{-0.107-0.044}$ \\
		$\Sigma^+_b \to \Sigma^+J/\psi$  & $-0.033^{+0.058+0.000}_{-0.080-0.027}$    & $-0.257^{+0.093+0.000}_{-0.097-0.021}$   & $0.484^{+0.078+0.023}_{-0.000-0.000}$      &$0.742^{+0.039+0.012}_{-0.000-0.000}$ & $-0.145^{+0.075+0.000}_{-0.088-0.024}$ \\
		$\Xi'^0_b \to \Lambda J/\psi$     & $0.137^{+0.023+0.078}_{-0.118-0.062}$   & $-0.585^{+0.000+0.086}_{-0.141-0.078}$  & $0.167^{+0.000+0.000}_{-0.070-0.031}$  &$0.583^{+0.000+0.000}_{-0.035-0.015}$ & $-0.224^{+0.001+0.082}_{-0.129-0.070}$ \\
		$\Xi'^0_b \to \Sigma^0J/\psi$    & $-0.176^{+0.047+0.036}_{-0.056-0.000}$  & $-0.929^{+0.013+0.012}_{-0.000-0.000}$  & $0.106^{+0.058+0.000}_{-0.058-0.032}$  &$0.553^{+0.029+0.000}_{-0.029-0.016}$ & $-0.553^{+0.030+0.019}_{-0.021-0.000}$ \\
		$\Xi'^0_b \to \Xi J/\psi$         & $-0.183^{+0.000+0.000}_{-0.070-0.034}$    & $-0.876^{+0.000+0.000}_{-0.025-0.018}$   & $0.121^{+0.069+0.040}_{-0.000-0.000}$      &$0.561^{+0.034+0.020}_{-0.000-0.000}$ & $-0.529^{+0.000+0.000}_{-0.047-0.026}$ \\
		\hline\hline
	\end{tabular}
\end{table}		

\begin{table}[!htbh]
	\caption{Asymmetry parameters for the $\mathcal{B}_b \rightarrow \mathcal{B}(\textbf{10}) J/\psi$ decays.
The theoretical uncertainties are the same as for branching ratios in Table~\ref{tab:branching1}. }
	\label{tab:asyp}
	\begin{tabular}[t]{lcccc}
		\hline\hline
		Mode &$\Sigma^0_b\rightarrow \Delta^0 J/\psi$        &$\Sigma_b^+\rightarrow \Delta^+ J/\psi$           &$\Lambda_b\rightarrow \Delta^0 J/\psi$ &$\Omega^-_b\rightarrow \Omega^- J/\psi$ \\ \hline
		$\alpha_b$&$-0.213^{+0.051+0.007}_{-0.072-0.000}$&$-0.495^{+0.048+0.005}_{-0.039-0.008}$&$0.787^{+0.054+0.034}_{-0.000-0.029}$&$-0.094^{+0.000+0.000}_{-0.035-0.022}$\\
		
		$\alpha_{\lambda_\mathcal{B}}$	&$-0.824^{+0.057+0.062}_{-0.022-0.000}$&$-0.296^{+0.045+0.008}_{-0.057-0.016}$&$0.707^{+0.015+0.016}_{-0.004-0.093}$&$-0.840^{+0.017+0.000}_{-0.026-0.033}$\\
		
		$\alpha_{\lambda_\psi}$&$-0.557^{+0.201+0.145}_{-0.092-0.000}$&$0.053^{+0.050+0.011}_{-0.000-0.004}$&$0.555^{+0.000+0.001}_{-0.081-0.100}$ &$-0.364^{+0.116+0.007}_{-0.034-0.000}$\\
		
		$U^T_{\frac{3}{2}}$&$0.442^{+0.080+0.000}_{-0.114-0.040}$&$0.192^{+0.000+0.004}_{-0.007-0.007}$&$0.166^{+0.017+0.026}_{-0.015-0.011}$&$0.304^{+0.047+0.000}_{-0.067-0.007}$\\
		
		$P^T_{\frac{3}{2}}$&$-0.440^{+0.113+0.042}_{-0.079-0.000}$&$-0.082^{+0.009+0.005}_{-0.001-0.002}$&$0.100^{+0.041+0.037}_{-0.005-0.000}$ &$-0.302^{+0.067+0.006}_{-0.049-0.000}$\\
		
		$U^T_{\frac{1}{2}}$&$0.336^{+0.014+0.000}_{-0.033-0.032}$&$0.282^{+0.000+0.001}_{-0.015-0.009}$&$0.057^{+0.023+0.025}_{-0.000-0.000}$&$0.378^{+0.009+0.006}_{-0.030-0.002}$\\
		
		$P^T_{\frac{1}{2}}$&$-0.305^{+0.064+0.031}_{-0.037-0.000}$&$0.100^{+0.000+0.002}_{-0.009-0.004}$&$-0.040^{+0.000+0.000}_{-0.029-0.032}$&$-0.373^{+0.026+0.002}_{-0.001-0.006}$\\
		
		$U^L$&$0.222^{+0.101+0.073}_{-0.046-0.000}$&$0.527^{+0.025+0.006}_{-0.000-0.002}$&$0.777^{+0.000+0.000}_{-0.041-0.050}$&$0.318^{+0.058+0.003}_{-0.017-0.000}$\\
		
		$P^L$&$-0.078^{+0.072+0.000}_{-0.099-0.014}$&$-0.314^{+0.037+0.008}_{-0.046-0.017}$&$0.647^{+0.040+0.000}_{-0.017-0.086}$&$-0.164^{+0.039+0.000}_{-0.092-0.033}$\\
		
		\hline\hline
	\end{tabular}
\end{table}	

With the preparation of the obtained helicity amplitudes,
one can compute the various asymmetry parameters defined in Eqs.~(\ref{eq:alp}) and~(\ref{eq:alpp})  straightforwardly,
whose values are presented in Table~\ref{tab:asy} for the octet modes and Table~\ref{tab:asyp} for the decuplet ones.
The predicted asymmetries for the $\Lambda_b\rightarrow \Lambda J/\psi$ and $\Lambda_b\rightarrow \Sigma^0 J/\psi$ processes
are comparable with the previous PQCD calculations~\cite{prd106053005,Rui:2023fpp},
and the former is also consistent with the measurements by the LHCb Collaboration~\cite{plb72427} within errors.
A detailed comparison of the  $\Lambda_b\rightarrow \Lambda J/\psi$ decay among various theoretical and experimental results
is referred to our preceding paper~\cite{prd106053005}, which will not be duplicated here.

The up-down asymmetry is an intriguing quantity that reflects parity violation.
The predicted up-down asymmetries for most processes are negative,
while those of the $\Lambda_b\rightarrow \Lambda J/\psi$, $\Sigma_b\rightarrow (n,p) J/\psi$, and $\Xi'^0_b \to \Lambda J/\psi$ decays
differ in sign but  with small values in magnitude.
For the exchange type channel $\Lambda_b\rightarrow \Delta^0 J/\psi$, the decay asymmetry is found to be positive, of order 0.787.
By comparison, the result in~\cite{prd99054020} is $\alpha_b(\Lambda_b\rightarrow n J/\psi)=-0.21\pm 0.00$,
which is consistent with our value of $-0.161^{+0.000+0.003}_{-0.040-0.015}$.
The asymmetry parameter $\alpha_b(\Omega_b\rightarrow \Omega J/\psi)$ supplied in~\cite{Fayyazuddin:2017sxq} is zero
owing to the absence of other power-suppressed form factors and
the vanishing parity-violating form factor $F_1^V$ from the naive quark model~\cite{Fayyazuddin:1996iy}.
 Despite many approximations, this value is in good agreement with our prediction of $-0.094^{+0.000+0.000}_{-0.035-0.022}$,
but smaller than $\alpha(\Omega_b\rightarrow \Omega J/\psi)=-0.18$ in magnitude made in the nonrelativistic quark model~\cite{Cheng:1996cs}.
As the current available predictions on these up-down asymmetries are very limited,
our results are expected to be checked in the future.

It is interesting to compare our predictions on the unpolarized parameters of the $\Omega_b\rightarrow \Omega J/\psi$ mode
with the corresponding results presented in~\cite{Gutsche:2018utw,Hsiao:2021mlp}.
Although the main concern of~\cite{Gutsche:2018utw} was the branching ratios,
the authors also present results on the asymmetry parameters in their Table XIV.
As one reads off from the table  $(U^T_{\frac{3}{2}},U^T_{\frac{1}{2}},U^L)=(0.312,0.236,0.452)$ agrees with our values of $(0.304,0.378,0.318)$.
In~\cite{Hsiao:2021mlp}, the helicity amplitudes were split into the vector and axial pieces.
It is found that the axial part overwhelms the vector one, and the three unpolarized parameters for the axial component
were predicted to be $(U^T_{\frac{3}{2}},U^T_{\frac{1}{2}},U^L)=(0.540,0.224,0.236)$,
which shows the important contribution from $\lambda_{\mathcal{B}}=\frac{3}{2}$ helicity state.
However, due to the lack of experimental measurements, all the theoretical predictions given above are still waiting to
be tested in future experiments.

\section{conclusion}\label{sec:sum}
Decays of beauty baryons offer alternative robust ways to test the SM and look for new physics, complementing searches with $B$ meson decays.
Inspired by recent experimental observations of the bottom baryons and theoretical advances about the nonperturbative LCDAs,
we conduct a comprehensive analysis of charmonium decays of the single beauty baryons mediated by $b\rightarrow s(d) c \bar c$ and $bu \rightarrow s(d) u$ transitions with the PQCD approach.
Particularly, the $\Sigma_b$ and $\Xi'_b$ decays that have not been measured or studied before are investigated for the first time.

We first calculate the decay amplitudes, widths, and  branching ratios for the concerned decays.
It demonstrates that the branching ratios of the CKM favored processes that stem from $b\rightarrow sc \bar c$ can reach to $10^{-4}$.
The CKM suppressed  $b\rightarrow dc \bar c$ modes  are lower by an order of magnitude because of the smaller CKM matrix elements.
Those channels, proceeding only through the power-suppressed exchange topology, have even lower rate at the order of $10^{-8}$.
In addition, a set of SU(3) amplitude relations among these decays are discussed.
We find the PQCD predictions on the amplitude ratios agree with the expectations in the SU(3) limit within errors.
The SU(3) breaking effects are estimated to be  less than 30$\%$.
The obtained branching ratios are compatible with those in the literature when they are available.

The combination of experimentally measured ratios and our predicted branching ratios enables us to extract the $b$ baryon fragmentation fractions  directly.
The obtained fragmentation fractions and their ratios are compared with other theoretical and experimental results in detail.
It is found that the two fragmentation ratios $\frac{f_{\Xi_b}}{f_{\Lambda_b}}$ and $\frac{f_{\Omega_b}}{f_{\Xi_b}}$ span wide ranges,
which call for more in-depth explorations.

We have also estimated the numerical results of helicity amplitudes and  various asymmetry parameters.
To our knowledge many of them have neither been measured experimentally nor calculated theoretically, in particular for those of the pure exchange modes,
which exhibit discriminating features with respect to the emission type processes.
For example, the dominations of the helicity configuration $\lambda_{\mathcal{B}}=\frac{1}{2}$ are observed in all the three exchange channels.
Moreover, the up-down asymmetry $\alpha(\Lambda_b\rightarrow \Delta^0 J/\psi)$ is found to be positive, of order 0.787.
Future experimental measurements can examine these predictions.

It is worth remarking that the PQCD formalism for the $\frac{1}{2}^+ \rightarrow \frac{3}{2}^+ + 1^-$ processes
have been constructed in this work, which have a rather broad field of applications,
particularly to the calculations of charmless and  charmful decays of beauty baryons.
In any event, they deserve more investigation.

\begin{acknowledgments}
We would like to acknowledge Professor Hsiang-nan Li and  Professor Aliev for helpful discussions.
This work is supported by National Natural Science Foundation of China under Grants No. 12075086 and No. 11605060.
Z.R. is also supported in part by the Natural Science Foundation of Hebei Province under Grant No. A2021209002 and No. A2019209449.
Z.T.Z. is  supported by  the Natural Science Foundation of Shandong province under the Grant No. ZR2022MA035.
\end{acknowledgments}

\begin{appendix}
\section{FACTORIZATION FORMULAS}\label{sec:for}

There are a lot of factorization formulas in this work.
According to the classification of decays made in Sec.\ref{sec:class},
the concerned decays can be divided into two subclasses,
namely $\frac{1}{2}^+\rightarrow \frac{1}{2}^+ +1^-$ and $\frac{1}{2}^+\rightarrow \frac{3}{2}^+ +1^-$ transitions.
The former ones have been given in~\cite{prd106053005} and need not be repeated here.
In the following, we only present the factorization formulas for $\frac{1}{2}^+\rightarrow \frac{3}{2}^+ + 1^-$ by taking $\Omega_b\rightarrow \Omega J/\psi$ as an example.

The invariant amplitudes in Eq.~(\ref{eq:kqp}) can symbolically be written as
\begin{eqnarray}\label{eq:ab}
C(D)=\frac{4f_{\Omega_b}\pi^2 G_F}{27\sqrt{6}}\sum_{R_{ij}}
\int[\mathcal{D}x][\mathcal{D}b]_{R_{ij}}
\alpha_s^2(t_{R_{ij}})e^{-S_{R_{ij}}}\Omega_{R_{ij}}(b,b',b_q)\sum_{\sigma=LL,SP}a_{R_{ij}}^{\sigma}H^{\sigma}_{R_{ij}}(x,x',y),
\end{eqnarray}
with the integration measure
\begin{eqnarray}
[\mathcal{D}x]=[dx_1dx_2dx_3\delta(1-x_1-x_2-x_3)][dx'_1dx'_2dx'_3\delta(1-x'_1-x'_2-x'_3)]dy,
\end{eqnarray}
where the $\delta$ functions enforce momentum conservation.
The expressions of $[\mathcal{D}b]_{R_{ij}}$, $S_{R_{ij}}$, $\Omega_{R_{ij}}$, $t_{R_{ij}}$, and $a_{R_{ij}}^{\sigma}$  can be found in Ref~\cite{prd106053005}.
The hard functions $H^{\sigma}_{R_{ij}}$ for vector amplitudes $C_l^L$ with $l=1,2,3$ are displayed in Table~\ref{tab:amp},
where the daughter baryon mass have been neglected for simplicity.
For the axial-vector amplitudes $D_l^L$, they are related with the vector ones
\begin{eqnarray}
D_{1,3}^L&=&C_{1,3}^L|_{\psi_{\parallel}^{n,\bar n}\rightarrow -\psi_{\parallel}^{n,\bar n},\phi\rightarrow-\phi},\nonumber\\
D_{2}^L&=&-C_{2}^L|_{\psi_{\parallel}^{n,\bar n}\rightarrow -\psi_{\parallel}^{n,\bar n},\phi\rightarrow-\phi}.
\end{eqnarray}
Because the longitudinal and transverse LCDAs of $J/\psi$ have similar Lorentz  structures,
we can obtain $C(D)^T$ from $C(D)^L$ by the operations $\psi^L\rightarrow \psi^V$ and $\psi^t\rightarrow \psi^T $.
 It is worth noting that the interchange of two spectator strange quarks in the emission diagrams leads to equivalent amplitudes.
 As a result, only 18 independent emission type of amplitudes are shown in Table~\ref{tab:amp}.
 Furthermore, the amplitudes associated with Feynman diagrams in the last row of Fig.~\ref{fig:Feynman} are not shown since they  are vanishingly small.

\begin{table}[H]
	\scriptsize
	\centering
	\caption{The expressions of $H^{\sigma}_{R_{ij}}$ for $C_l^L$. }
	\newcommand{\tabincell}[2]{\begin{tabular}{@{}#1@{}}#2\end{tabular}}
	\label{tab:amp}
	\begin{tabular}[t]{lccc}
		\hline\hline
		$$&$\frac{C_1^L}{16M^3}$&$\frac{C_2^L}{16M^2}$&$\frac{C_3^L}{16M}$\\ \hline
		
		$H_{T_{a1}}^{LL(SP)}$
		&\tabincell{c}{$r (r^2-1) (1-x_1) \psi_{\parallel}^{1} \psi ^L (1-x_1') (\mathcal{A}+\mathcal{V})$}
		&\tabincell{c}{$\frac{1}{2} r (r^2-1) \psi_{\parallel}^{\bar n} \psi ^L x_3' (-\mathcal{A}+\mathcal{T}+\mathcal{V})$}
		&\tabincell{c}{$2 r \psi_{\parallel}^{\bar n} \psi ^L x_3' (-\mathcal{A}+\mathcal{T}+\mathcal{V})$}                       \\\\
		
		$H_{T_{a2}}^{LL(SP)}$
		&$0$
		&\tabincell{c}{$-\frac{1}{2} r x_3 \psi_{\parallel}^{\bar n} \psi ^L (-\mathcal{A}+\mathcal{T}+\mathcal{V})$}
		&\tabincell{c}{$\frac{1}{r^2-1}r x_3 \psi_{\parallel}^{\bar n} \psi ^L ((r^2-1)$\\$(1-x_1')+2) (\mathcal{A}-\mathcal{T}-\mathcal{V})$}      \\\\
		
		$H_{T_{a3}}^{LL(SP)}$
		&\tabincell{c}{$-\frac{1}{2}r\psi^L(x_2((r^2-1)(1-x_1')(\psi_{\parallel}^{\bar nn}+\psi_{\parallel}^{1})(\mathcal{A}+\mathcal{V})$\\$+\mathcal{A}(\psi_{\parallel}^{n}+2\psi_{\parallel}^{1})-2M(r^2-1)\psi_{\parallel}^{1}\phi+\psi_{\parallel}^{n}\mathcal{T}$\\$-\mathcal{V}(\psi_{\parallel}^{n}-2\psi_{\parallel}^{1}))-(r^2-1)(x_3'(\mathcal{A}(\psi_{\parallel}^{\bar nn}$\\$+\psi_{\parallel}^{1}-\psi_{\parallel}^{\bar n})-\psi_{\parallel}^{\bar n}\mathcal{T}+\mathcal{V}(\psi_{\parallel}^{\bar nn}+\psi_{\parallel}^{1}+\psi_{\parallel}^{\bar n}))$\\$+x_2'(\psi_{\parallel}^{\bar nn}+\psi_{\parallel}^{1})(\mathcal{A}+\mathcal{V})))$}
		&\tabincell{c}{$\frac{1}{2}r\psi^L((x_2-1)\psi_{\parallel}^{n}\mathcal{A}+\mathcal{T}((1-r^2)$\\$\psi_{\parallel}^{\bar n}x_3'+x_2(\psi_{\parallel}^{\bar n}-\psi_{\parallel}^{n})+\psi_{\parallel}^{n})$\\$-(x_2-1)\psi_{\parallel}^{n}\mathcal{V})$}
		&\tabincell{c}{$\frac{1}{r^2-1}r\psi^L(x_2((1-r^2)\psi_{\parallel}^{\bar n}x_2'(\mathcal{A}-\mathcal{V})$\\$+\mathcal{A}(\psi_{\parallel}^{n}-\psi_{\parallel}^{\bar n})+2M(r^2-1)\psi_{\parallel}^{1}\phi$\\$+\mathcal{T}(\psi_{\parallel}^{\bar n}-3\psi_{\parallel}^{n})+\mathcal{V}(\psi_{\parallel}^{\bar n}-\psi_{\parallel}^{n}))$\\$-(r^2-1)\psi_{\parallel}^{\bar n}x_3'(x_2(\mathcal{A}-\mathcal{V})$\\$+2\mathcal{T})+2\psi_{\parallel}^{n}(-\mathcal{A}+\mathcal{T}+\mathcal{V}))$}      \\\\
		
		$H_{T_{a5}}^{LL(SP)}$
		&\tabincell{c}{$-\frac{1}{2}r(r^2-1)x_3\psi^L(x_2'((\psi_{\parallel}^{\bar nn}+\psi_{\parallel}^{1})$\\$(\mathcal{A}+\mathcal{V})-\psi_{\parallel}^{\bar n}\mathcal{T})+\psi_{\parallel}^{\bar n}\mathcal{T})$}
		&\tabincell{c}{$-\frac{1}{2}rx_3\psi^L(\psi_{\parallel}^{\bar nn}-\psi_{\parallel}^{1})(\mathcal{A}+\mathcal{V})$}
		&\tabincell{c}{$\frac{1}{r^2-1}rx_3\psi^L((1-r^2)x_2'((\psi_{\parallel}^{\bar nn}+\psi_{\parallel}^{1})$\\$(\mathcal{A}+\mathcal{V})-\psi_{\parallel}^{\bar n}\mathcal{T})+(r^2x_3-r^2-1)$\\$(\psi_{\parallel}^{\bar nn}-\psi_{\parallel}^{1})(\mathcal{A}+\mathcal{V})$\\$+\psi_{\parallel}^{\bar n}(r^2x_3-r^2+1)\mathcal{T})$}                       \\\\
		
		$H_{T_{a6}}^{LL}$
		&\tabincell{c}{$\frac{1}{2}(r_c\psi^t((r^2-1)x_2'(\psi_{\parallel}^{\bar nn}+\psi_{\parallel}^{1})(\mathcal{A}+\mathcal{V})$\\$+r^2x_3\psi_{\parallel}^{\bar n}\mathcal{T})+rx_3\psi^L(y-x_3)$\\$(\psi_{\parallel}^{n}\mathcal{A}-M(r^2-1)\phi(\psi_{\parallel}^{\bar nn}+\psi_{\parallel}^{1})-\psi_{\parallel}^{n}\mathcal{V}))$}		
		&\tabincell{c}{$\frac{1}{2}(r_c\psi^t(\psi_{\parallel}^{\bar nn}-\psi_{\parallel}^{1})(\mathcal{A}+\mathcal{V})$\\$+r\psi^L(y-x_3)(x_3(\mathcal{A}-\mathcal{T}-\mathcal{V})$\\$\psi_{\parallel}^{n}+\mathcal{A}(\psi_{\parallel}^{1}-\psi_{\parallel}^{n}-\psi_{\parallel}^{\bar nn})$\\$+\psi_{\parallel}^{n}\mathcal{T}+\mathcal{V}(\psi_{\parallel}^{n}-\psi_{\parallel}^{\bar nn}+\psi_{\parallel}^{1})))$}		
		&\tabincell{c}{$-\frac{1}{r^2-1}r_c\psi^t(\mathcal{A}+\mathcal{V})((r^2x_3-r^2-1)$\\$(\psi_{\parallel}^{\bar nn}-\psi_{\parallel}^{1})-(r^2-1)x_2'(\psi_{\parallel}^{\bar nn}+\psi_{\parallel}^{1}))$\\$-r\psi^L(y-x_3)(x_3(\mathcal{A}(r^2(\psi_{\parallel}^{\bar nn}-\psi_{\parallel}^{1})+2\psi_{\parallel}^{n})$\\$-M(r^2-1)\phi(\psi_{\parallel}^{\bar nn}+\psi_{\parallel}^{1})+\mathcal{T}(r^2\psi_{\parallel}^{\bar n}-\psi_{\parallel}^{n})$\\$+\mathcal{V}(r^2(\psi_{\parallel}^{\bar nn}-\psi_{\parallel}^{1})-2\psi_{\parallel}^{n}))$\\$-(r^2-1)x_2'(\psi_{\parallel}^{\bar nn}+\psi_{\parallel}^{1})(\mathcal{A}+\mathcal{V})$\\$+\mathcal{A}(-(r^2+1)(\psi_{\parallel}^{\bar nn}-\psi_{\parallel}^{1})-2\psi_{\parallel}^{n})$\\$+\mathcal{V}(2\psi_{\parallel}^{n}-(r^2+1)(\psi_{\parallel}^{\bar nn}-\psi_{\parallel}^{1}))+2\psi_{\parallel}^{n}\mathcal{T})$}\\\\
		
		$H_{T_{a6}}^{SP}$
		&\tabincell{c}{$\frac{1}{2}(r_c\psi^t(x_3(\mathcal{A}(r^2(\psi_{\parallel}^{1}-\psi_{\parallel}^{\bar nn})+\psi_{\parallel}^{n})-M(r^2-1)\phi$\\$(\psi_{\parallel}^{\bar nn}+\psi_{\parallel}^{1})-\mathcal{V}(r^2(\psi_{\parallel}^{\bar nn}-\psi_{\parallel}^{1})+\psi_{\parallel}^{n}))+(r^2-1)$\\$\psi_{\parallel}^{\bar n}x_2'(\mathcal{A}-\mathcal{V})+\mathcal{A}(r^2(\psi_{\parallel}^{\bar nn}-\psi_{\parallel}^{1})+\psi_{\parallel}^{\bar n})+M(r^2-1)$\\$\phi(\psi_{\parallel}^{\bar nn}-\psi_{\parallel}^{1})+\mathcal{V}(r^2(\psi_{\parallel}^{\bar nn}-\psi_{\parallel}^{1})-\psi_{\parallel}^{\bar n})+\psi_{\parallel}^{n}\mathcal{T})$\\$+r\psi^L(x_3(\mathcal{A}(r^2y(\psi_{\parallel}^{\bar nn}-\psi_{\parallel}^{1})+\psi_{\parallel}^{\bar n})+M(r^2-1)\phi$\\$(\psi_{\parallel}^{\bar nn}-\psi_{\parallel}^{1})+\mathcal{T}(r^2y\psi_{\parallel}^{\bar n}+\psi_{\parallel}^{n})+\mathcal{V}(r^2y(\psi_{\parallel}^{\bar nn}-\psi_{\parallel}^{1})-\psi_{\parallel}^{\bar n}))$\\$+y(\mathcal{A}(-(r^2(\psi_{\parallel}^{\bar nn}-\psi_{\parallel}^{1})+\psi_{\parallel}^{\bar n}))-M(r^2-1)\phi(\psi_{\parallel}^{\bar nn}-\psi_{\parallel}^{1})$\\$+\mathcal{V}(r^2(\psi_{\parallel}^{1}-\psi_{\parallel}^{\bar nn})+\psi_{\parallel}^{\bar n})-\psi_{\parallel}^{n}\mathcal{T})-(r^2-1)x_3'(x_3\psi_{\parallel}^{\bar n}\mathcal{T}$\\$+(x_3-1)(\psi_{\parallel}^{\bar nn}-\psi_{\parallel}^{1})(\mathcal{A}+\mathcal{V}))+(r^2-1)x_2'(y-x_3)$\\$(\mathcal{A}(\psi_{\parallel}^{\bar nn}+\psi_{\parallel}^{1}-\psi_{\parallel}^{\bar n})+\mathcal{V}(\psi_{\parallel}^{\bar nn}+\psi_{\parallel}^{1}+\psi_{\parallel}^{\bar n}))))$}
		&\tabincell{c}{$\frac{1}{2}((r^2-1)\psi_{\parallel}^{\bar n}x_2'(\mathcal{A}-\mathcal{V})$\\$(r_c\psi^t-r\psi^L(y-x_3))-\psi_{\parallel}^{n}$\\$(x_3-1)r_c\psi^t(-\mathcal{A}+\mathcal{T}+\mathcal{V}))$}
		&\tabincell{c}{$\frac{1}{r^2-1}r_c\psi^t(x_3(2\psi_{\parallel}^{n}\mathcal{A}-M(r^2-1)\phi$\\$(\psi_{\parallel}^{\bar nn}+\psi_{\parallel}^{1})+\mathcal{T}(r^2\psi_{\parallel}^{\bar n}-\psi_{\parallel}^{n})-2\psi_{\parallel}^{n}\mathcal{V})$\\$+2(r^2-1)\psi_{\parallel}^{\bar n}x_2'(\mathcal{A}-\mathcal{V})+\mathcal{A}(\psi_{\parallel}^{\bar n}-\psi_{\parallel}^{n})$\\$+M(r^2-1)\phi(\psi_{\parallel}^{\bar nn}-\psi_{\parallel}^{1})+2\psi_{\parallel}^{n}\mathcal{T}$\\$+\mathcal{V}(\psi_{\parallel}^{n}-\psi_{\parallel}^{\bar n}))+r\psi^L(y-x_3)$\\$(2(1-r^2)\psi_{\parallel}^{\bar n}x_2'(\mathcal{A}-\mathcal{V})-(\psi_{\parallel}^{n}+\psi_{\parallel}^{\bar n})$\\$(\mathcal{A}-\mathcal{V})-M(r^2-1)\phi(\psi_{\parallel}^{\bar nn}-\psi_{\parallel}^{1}))$}\\

	    \hline\hline
	\end{tabular}
\end{table}

\begin{table}[H]
	\scriptsize
	\centering
	TABLE~\ref{tab:amp} (continued)
	\newcommand{\tabincell}[2]{\begin{tabular}{@{}#1@{}}#2\end{tabular}}
	\begin{tabular}[t]{lccc}
		\hline\hline
		$$&$\frac{C_1^L}{16M^3}$&$\frac{C_2^L}{16M^2}$&$\frac{C_3^L}{16M}$\\ \hline
			
		$H_{T_{a7}}^{LL}$
		&\tabincell{c}{$\frac{1}{2}(r\psi^L(x_3(\mathcal{A}(\psi_{\parallel}^{\bar n}-r^2(y-1)(\psi_{\parallel}^{\bar nn}-\psi_{\parallel}^{1}))$\\$+M(r^2-1)\phi(\psi_{\parallel}^{\bar nn}-\psi_{\parallel}^{1})+\mathcal{T}(\psi_{\parallel}^{n}-r^2$\\$(y-1)\psi_{\parallel}^{\bar n})-\mathcal{V}(r^2(y-1)(\psi_{\parallel}^{\bar nn}-\psi_{\parallel}^{1})$\\$+\psi_{\parallel}^{\bar n}))+(y-1)(\mathcal{A}(r^2(\psi_{\parallel}^{\bar nn}-\psi_{\parallel}^{1})+\psi_{\parallel}^{\bar n})$\\$+M(r^2-1)\phi(\psi_{\parallel}^{\bar nn}-\psi_{\parallel}^{1})+\mathcal{V}(r^2(\psi_{\parallel}^{\bar nn}-\psi_{\parallel}^{1})$\\$-\psi_{\parallel}^{\bar n})+\psi_{\parallel}^{n}\mathcal{T})+(r^2-1)x_3'(x_3(-\psi_{\parallel}^{\bar n})\mathcal{T}$\\$-(x_3-1)(\psi_{\parallel}^{\bar nn}-\psi_{\parallel}^{1})(\mathcal{A}+\mathcal{V}))-(r^2-1)$\\$x_2'(x_3+y-1)(\mathcal{A}(\psi_{\parallel}^{\bar nn}+\psi_{\parallel}^{1}-\psi_{\parallel}^{\bar n})$\\$+\mathcal{V}(\psi_{\parallel}^{\bar nn}+\psi_{\parallel}^{1}+\psi_{\parallel}^{\bar n})))-r_c\psi^t(x_3(\mathcal{A}(r^2$\\$(\psi_{\parallel}^{1}-\psi_{\parallel}^{\bar nn})+\psi_{\parallel}^{n})-M(r^2-1)\phi(\psi_{\parallel}^{\bar nn}+\psi_{\parallel}^{1})$\\$-\mathcal{V}(r^2(\psi_{\parallel}^{\bar nn}-\psi_{\parallel}^{1})+\psi_{\parallel}^{n}))+(r^2-1)$\\$\psi_{\parallel}^{\bar n}x_2'(\mathcal{A}-\mathcal{V})+\mathcal{A}(r^2(\psi_{\parallel}^{\bar nn}-\psi_{\parallel}^{1})+\psi_{\parallel}^{\bar n})$\\$+M(r^2-1)\phi(\psi_{\parallel}^{\bar nn}-\psi_{\parallel}^{1})$\\$+\mathcal{V}(r^2(\psi_{\parallel}^{\bar nn}-\psi_{\parallel}^{1})-\psi_{\parallel}^{\bar n})+\psi_{\parallel}^{n}\mathcal{T}))$}
		
		&\tabincell{c}{$\frac{1}{2}((r^2-1)\psi_{\parallel}^{\bar n}x_2'(\mathcal{A}-\mathcal{V})$\\$(r\psi^L(x_3+y-1)-r_c\psi^t)+\psi_{\parallel}^{n}$\\$(x_3-1)r_c\psi^t(-\mathcal{A}+\mathcal{T}+\mathcal{V}))$}
		
		&\tabincell{c}{$\frac{1}{r^2-1}r_c\psi^t(x_3(-2\psi_{\parallel}^{n}\mathcal{A}+M(r^2-1)\phi$\\$(\psi_{\parallel}^{\bar nn}+\psi_{\parallel}^{1})+\mathcal{T}(\psi_{\parallel}^{n}-r^2\psi_{\parallel}^{\bar n})+2\psi_{\parallel}^{n}\mathcal{V})$\\$-2(r^2-1)\psi_{\parallel}^{\bar n}x_2'(\mathcal{A}-\mathcal{V})+\mathcal{A}(\psi_{\parallel}^{n}-\psi_{\parallel}^{\bar n})$\\$-M(r^2-1)\phi(\psi_{\parallel}^{\bar nn}-\psi_{\parallel}^{1})-2\psi_{\parallel}^{n}\mathcal{T}$\\$+\mathcal{V}(\psi_{\parallel}^{\bar n}-\psi_{\parallel}^{n}))+r\psi^L(x_3+y-1)$\\$(2(r^2-1)\psi_{\parallel}^{\bar n}x_2'(\mathcal{A}-\mathcal{V})+(\psi_{\parallel}^{n}+\psi_{\parallel}^{\bar n})$\\$(\mathcal{A}-\mathcal{V})+M(r^2-1)\phi(\psi_{\parallel}^{\bar nn}-\psi_{\parallel}^{1}))$}\\\\

		$H_{T_{a7}}^{SP}$
		&\tabincell{c}{$\frac{1}{2}(r_c\psi^t(-(r^2-1)x_2'(\psi_{\parallel}^{\bar nn}+\psi_{\parallel}^{1})(\mathcal{A}+\mathcal{V})$\\$-r^2x_3\psi_{\parallel}^{\bar n}\mathcal{T})+rx_3\psi^L(x_3+y-1)$\\$(-\psi_{\parallel}^{n}\mathcal{A}+M(r^2-1)\phi(\psi_{\parallel}^{\bar nn}+\psi_{\parallel}^{1})+\psi_{\parallel}^{n}\mathcal{V}))$}
		&\tabincell{c}{$\frac{1}{2}(-r_c\psi^t(\psi_{\parallel}^{\bar nn}-\psi_{\parallel}^{1})(\mathcal{A}+\mathcal{V})$\\$-r\psi^L(x_3+y-1)((\mathcal{A}-\mathcal{T}-\mathcal{V})$\\$x_3\psi_{\parallel}^{n}+\mathcal{A}(\psi_{\parallel}^{1}-\psi_{\parallel}^{n}-\psi_{\parallel}^{\bar nn})$\\$+\psi_{\parallel}^{n}\mathcal{T}+\mathcal{V}(\psi_{\parallel}^{n}-\psi_{\parallel}^{\bar nn}+\psi_{\parallel}^{1})))$}
		&\tabincell{c}{$-\frac{1}{r^2-1}r_c\psi^t(\mathcal{A}+\mathcal{V})((r^2-1)x_2'(\psi_{\parallel}^{\bar nn}+\psi_{\parallel}^{1})$\\$-(r^2x_3-r^2-1)(\psi_{\parallel}^{\bar nn}-\psi_{\parallel}^{1}))+r\psi^L$\\$(x_3+y-1)(x_3(\mathcal{A}(r^2(\psi_{\parallel}^{\bar nn}-\psi_{\parallel}^{1})+2\psi_{\parallel}^{n})$\\$-M(r^2-1)\phi(\psi_{\parallel}^{\bar nn}+\psi_{\parallel}^{1})+\mathcal{T}(r^2\psi_{\parallel}^{\bar n}-\psi_{\parallel}^{n})$\\$+\mathcal{V}(r^2(\psi_{\parallel}^{\bar nn}-\psi_{\parallel}^{1})-2\psi_{\parallel}^{n}))$\\$-(r^2-1)x_2'(\psi_{\parallel}^{\bar nn}+\psi_{\parallel}^{1})(\mathcal{A}+\mathcal{V})$\\$+\mathcal{A}((r^2+1)(\psi_{\parallel}^{1}-\psi_{\parallel}^{\bar nn})-2\psi_{\parallel}^{n})$\\$+\mathcal{V}(2\psi_{\parallel}^{n}-(r^2+1)(\psi_{\parallel}^{\bar nn}-\psi_{\parallel}^{1}))+2\psi_{\parallel}^{n}\mathcal{T})$}\\\\
		
		$H_{T_{b1}}^{LL(SP)}$
		&$0$
		&$-r(1-x_1)^2\psi_{\parallel}^{1}\psi^L(\mathcal{A}+\mathcal{V})$
		&$-\frac{1}{r^2-1}2(r^3+r)(1-x_1)^2\psi_{\parallel}^{1}\psi^L(\mathcal{A}+\mathcal{V})$  \\\\
		
		$H_{T_{b2}}^{LL(SP)}$
		&\tabincell{c}{$0$}
		&\tabincell{c}{$-\frac{1}{2}r(r^2-1)x_3\psi_{\parallel}^{\bar n}\psi^L(\mathcal{A}-\mathcal{T}-\mathcal{V})$}
		&\tabincell{c}{$2rx_3\psi_{\parallel}^{\bar n}\psi^L(-\mathcal{A}+\mathcal{T}+\mathcal{V})$}                       \\\\
		
		$H_{T_{b4}}^{LL(SP)}$
		&\tabincell{c}{$0$}
		&\tabincell{c}{$\frac{1}{2}rx_2\psi^L((r^2-1)\psi_{\parallel}^{\bar n}(\mathcal{A}-\mathcal{V})$\\$+(1-x_1)(\psi_{\parallel}^{\bar nn}+\psi_{\parallel}^{1})(\mathcal{A}+\mathcal{V}))$}
		&\tabincell{c}{$\frac{1}{r^2-1}rx_2\psi^L((r^2+1)(1-x_1)(\psi_{\parallel}^{\bar nn}+\psi_{\parallel}^{1})$\\$(\mathcal{A}+\mathcal{V})+2(r^2-1)\psi_{\parallel}^{\bar n}(\mathcal{A}-\mathcal{V}))$}                       \\\\
		
		$H_{T_{b6}}^{LL}$
		&\tabincell{c}{$0$}
		&\tabincell{c}{$-\frac{1}{2}x_2(r_c\psi^t(\mathcal{A}(r^2\psi_{\parallel}^{\bar n}+\psi_{\parallel}^{\bar nn}+\psi_{\parallel}^{1})$\\$+\mathcal{V}(r^2(-\psi_{\parallel}^{\bar n})+\psi_{\parallel}^{\bar nn}+\psi_{\parallel}^{1}))$\\$+r\psi_{\parallel}^{\bar n}\psi^L(\mathcal{A}-\mathcal{V})((r^2-1)x_3'-r^2y))$}
		&\tabincell{c}{$-\frac{1}{r^2-1}x_2(r_c\psi^t(\mathcal{A}((r^2+1)(\psi_{\parallel}^{\bar nn}+\psi_{\parallel}^{1})$\\$+2r^2\psi_{\parallel}^{\bar n})+\mathcal{V}((r^2+1)(\psi_{\parallel}^{\bar nn}+\psi_{\parallel}^{1})-2r^2\psi_{\parallel}^{\bar n}))$\\$+2r\psi_{\parallel}^{\bar n}\psi^L(\mathcal{A}-\mathcal{V})((r^2-1)x_3'-r^2y))$}\\\\
		
		$H_{T_{b6}}^{SP}$
		&\tabincell{c}{$\frac{1}{2}(r_c\psi^t((r^2-1)\psi_{\parallel}^{\bar n}(x_3'-1)(\mathcal{A}-\mathcal{V})$\\$-r^2x_2(\psi_{\parallel}^{\bar nn}-\psi_{\parallel}^{1})(\mathcal{A}+\mathcal{V}))+\psi^L$\\$(rx_2(\psi_{\parallel}^{1}-\psi_{\parallel}^{\bar nn})(\mathcal{A}+\mathcal{V})((r^2-1)x_3'-r^2y)$\\$-r(r^2-1)\psi_{\parallel}^{\bar n}(x_3'-1)(y-x_3)(\mathcal{A}-\mathcal{V})))$}
		&\tabincell{c}{$\frac{1}{2}rx_2\psi^L(x_3-y)$\\$(\psi_{\parallel}^{\bar nn}+\psi_{\parallel}^{1})(\mathcal{A}+\mathcal{V})$}
		&\tabincell{c}{$\frac{1}{r^2-1}r_c\psi^t((r^2-1)\psi_{\parallel}^{\bar n}(x_3'-1)(\mathcal{A}-\mathcal{V})$\\$-r^2x_2(\mathcal{A}(2\psi_{\parallel}^{\bar nn}-\psi_{\parallel}^{\bar n})+\mathcal{V}(2\psi_{\parallel}^{\bar nn}+\psi_{\parallel}^{\bar n})))$\\$+r\psi^L(x_2((1-r^2)x_3'(\psi_{\parallel}^{\bar nn}-\psi_{\parallel}^{1})(\mathcal{A}+\mathcal{V})$\\$+x_3(\mathcal{A}(r^2\psi_{\parallel}^{\bar n}+\psi_{\parallel}^{\bar nn}+\psi_{\parallel}^{1})$\\$+\mathcal{V}(\psi_{\parallel}^{\bar nn}-r^2\psi_{\parallel}^{\bar n}+\psi_{\parallel}^{1}))+y\mathcal{A}((r^2-1)\psi_{\parallel}^{\bar nn}$\\$-r^2(\psi_{\parallel}^{1}+\psi_{\parallel}^{\bar n})-\psi_{\parallel}^{1})+y\mathcal{V}((r^2-1)\psi_{\parallel}^{\bar nn}$\\$-(r^2+1)\psi_{\parallel}^{1}+r^2\psi_{\parallel}^{\bar n}))-(r^2-1)$\\$\psi_{\parallel}^{\bar n}(x_3'-1)(y-x_3)(\mathcal{A}-\mathcal{V}))$}\\\\
		
		$H_{T_{b7}}^{LL}$
		&\tabincell{c}{$\frac{1}{2}(r_c\psi^t(r^2x_2(\psi_{\parallel}^{\bar nn}-\psi_{\parallel}^{1})(\mathcal{A}+\mathcal{V})$\\$-(r^2-1)\psi_{\parallel}^{\bar n}(x_3'-1)(\mathcal{A}-\mathcal{V}))$\\$+r\psi^L((r^2-1)\psi_{\parallel}^{\bar n}(x_3'-1)(x_3+y-1)$\\$(\mathcal{A}-\mathcal{V})-x_2(\psi_{\parallel}^{\bar nn}-\psi_{\parallel}^{1})$\\$(\mathcal{A}+\mathcal{V})((r^2-1)x_3'+r^2(y-1))))$}
		&\tabincell{c}{$\frac{1}{2}rx_2\psi^L(x_3+y-1)$\\$(\psi_{\parallel}^{\bar nn}+\psi_{\parallel}^{1})(\mathcal{A}+\mathcal{V})$}
		&\tabincell{c}{$-\frac{1}{r^2-1}r_c\psi^t((r^2-1)\psi_{\parallel}^{\bar n}(x_3'-1)(\mathcal{A}-\mathcal{V})$\\$-r^2x_2(\mathcal{A}(2\psi_{\parallel}^{\bar nn}-\psi_{\parallel}^{\bar n})+\mathcal{V}(2\psi_{\parallel}^{\bar nn}+\psi_{\parallel}^{\bar n})))$\\$-r\psi^L(x_2(-(r^2-1)x_3'(\psi_{\parallel}^{\bar nn}-\psi_{\parallel}^{1})(\mathcal{A}+\mathcal{V})$\\$+x_3(\mathcal{A}(r^2\psi_{\parallel}^{\bar n}+\psi_{\parallel}^{\bar nn}+\psi_{\parallel}^{1})$\\$+\mathcal{V}(r^2(-\psi_{\parallel}^{\bar n})+\psi_{\parallel}^{\bar nn}+\psi_{\parallel}^{1}))+(y-1)(\mathcal{A}(r^2(-\psi_{\parallel}^{\bar nn})$\\$+r^2(\psi_{\parallel}^{1}+\psi_{\parallel}^{\bar n})+\psi_{\parallel}^{\bar nn}+\psi_{\parallel}^{1})+\mathcal{V}(r^2(-\psi_{\parallel}^{\bar nn})$\\$+r^2(\psi_{\parallel}^{1}-\psi_{\parallel}^{\bar n})+\psi_{\parallel}^{\bar nn}+\psi_{\parallel}^{1})))$\\$+(r^2-1)\psi_{\parallel}^{\bar n}(x_3'-1)(x_3+y-1)(\mathcal{A}-\mathcal{V}))$}\\
		
		$H_{T_{b7}}^{SP}$
		&\tabincell{c}{$0$}
		&\tabincell{c}{$\frac{1}{2}x_2(r_c\psi^t(\mathcal{A}(r^2\psi_{\parallel}^{\bar n}+\psi_{\parallel}^{\bar nn}+\psi_{\parallel}^{1})$\\$+\mathcal{V}(r^2(-\psi_{\parallel}^{\bar n})+\psi_{\parallel}^{\bar nn}+\psi_{\parallel}^{1}))-r\psi_{\parallel}^{\bar n}$\\$\psi^L(\mathcal{A}-\mathcal{V})((r^2-1)x_3'+r^2(y-1)))$}
		&\tabincell{c}{$-\frac{1}{r^2-1}x_2(2r\psi_{\parallel}^{\bar n}\psi^L(\mathcal{A}-\mathcal{V})((r^2-1)x_3'+r^2(y-1))$\\$-r_c\psi^t(\mathcal{A}((r^2+1)(\psi_{\parallel}^{\bar nn}+\psi_{\parallel}^{1})+2r^2\psi_{\parallel}^{\bar n})$\\$+\mathcal{V}((r^2+1)(\psi_{\parallel}^{\bar nn}+\psi_{\parallel}^{1})-2r^2\psi_{\parallel}^{\bar n})))$}\\
		
		$H_{T_{c1}}^{LL}$
		&\tabincell{c}{$0$}
		&\tabincell{c}{$(1-x_1)\psi_{\parallel}^{1}(\mathcal{A}+\mathcal{V})(r_c\psi^t+r(1-x_1)$\\$\psi^L-ry\psi^L)+\frac{1}{2}r(r^2-1)\psi_{\parallel}^{\bar n}\psi^L$\\$x_3'(1-x_1-y)(\mathcal{A}-\mathcal{T}-\mathcal{V})$}
		&\tabincell{c}{$\frac{1}{r^2-1}2(r^2+1)(1-x_1)\psi_{\parallel}^{1}(\mathcal{A}+\mathcal{V})$\\$(r_c\psi^t+r(1-x_1)\psi^L-ry\psi^L)$\\$+2r\psi_{\parallel}^{\bar n}\psi^Lx_3'(1-x_1-y)(\mathcal{A}-\mathcal{T}-\mathcal{V})$}\\
		
        \hline\hline
     \end{tabular}
 \end{table}

 \begin{table}[H]
    \scriptsize
    \centering
    TABLE~\ref{tab:amp} (continued)
    \newcommand{\tabincell}[2]{\begin{tabular}{@{}#1@{}}#2\end{tabular}}
        \begin{tabular}[t]{lccc}
        \hline\hline
        $$&$\frac{C_1^L}{16M^3}$&$\frac{C_2^L}{16M^2}$&$\frac{C_3^L}{16M}$\\ \hline
        $H_{T_{c1}}^{SP}$
        &\tabincell{c}{$\frac{1}{2}(x_2(r\psi^L((r^2-1)x_3'(\mathcal{A}(2\psi_{\parallel}^{1}-\psi_{\parallel}^{\bar n})$\\$+\mathcal{V}(2\psi_{\parallel}^{1}+\psi_{\parallel}^{\bar n}))-2r^2y\psi_{\parallel}^{1}\mathcal{A}$\\$-2M(r^2-1)\psi_{\parallel}^{1}(y-2x_3)\phi-2r^2y\psi_{\parallel}^{1}\mathcal{V}$\\$-2x_3\psi_{\parallel}^{n}\mathcal{T}+y\psi_{\parallel}^{n}\mathcal{T})-r_c\psi^t$\\$(-2r^2\psi_{\parallel}^{1}\mathcal{A}-2M(r^2-1)\psi_{\parallel}^{1}\phi$\\$-2r^2\psi_{\parallel}^{1}\mathcal{V}+\psi_{\parallel}^{n}\mathcal{T}))+x_3(ry\psi^L-r_c\psi^t)$\\$(\psi_{\parallel}^{n}\mathcal{T}-2r^2\psi_{\parallel}^{1}\mathcal{A}-2M(r^2-1)\psi_{\parallel}^{1}\phi-2r^2\psi_{\parallel}^{1}\mathcal{V})$\\$+(r^2-1)\psi_{\parallel}^{\bar n}x_3'(\mathcal{A}-\mathcal{V})(ry\psi^L-r_c\psi^t)$\\$+2r(r^2-1)\psi_{\parallel}^{1}\psi^Lx_2x_2'(\mathcal{A}+\mathcal{V})$\\$+2r(r^2-1)x_3\psi_{\parallel}^{1}\psi^Lx_2'(\mathcal{A}+\mathcal{V})+r(r^2-1)$\\$\psi^Lx_3x_3'(\mathcal{A}(2\psi_{\parallel}^{1}-\psi_{\parallel}^{\bar n})+\mathcal{V}(2\psi_{\parallel}^{1}+\psi_{\parallel}^{\bar n}))$\\$+r\psi^L(x_2^2+x_3^2)(2M(r^2-1)\psi_{\parallel}^{1}\phi-\psi_{\parallel}^{n}\mathcal{T}))$}
        &\tabincell{c}{$\frac{1}{2}(r^2-1)\psi_{\parallel}^{\bar n}r_cx_3'\psi^t(-\mathcal{A}+\mathcal{T}+\mathcal{V})$}
        &\tabincell{c}{$-\frac{1}{r^2-1}\psi_{\parallel}^{\bar n}x_3'((r^3-r)\psi^L\mathcal{T}(1-x_1-y)$\\$-(r^2-1)r_c\psi^t(-2\mathcal{A}+\mathcal{T}+2\mathcal{V}))$\\$+(1-x_1)(r_c\psi^t+r(1-x_1)\psi^L-ry\psi^L)$\\$(\psi_{\parallel}^{n}\mathcal{T}-2M(r^2-1)\psi_{\parallel}^{1}\phi)$}\\\\

        $H_{T_{c2}}^{LL}$
        &\tabincell{c}{$0$}
        &\tabincell{c}{$-\frac{1}{2}r^2x_3\psi_{\parallel}^{\bar n}r_c\psi^t(-\mathcal{A}+\mathcal{T}+\mathcal{V})$}
        &\tabincell{c}{$-\frac{1}{r^2-1}2r^2x_3\psi_{\parallel}^{\bar n}r_c\psi^t(-\mathcal{A}+\mathcal{T}+\mathcal{V})$}\\\\

        $H_{T_{c2}}^{SP}$
        &\tabincell{c}{$\frac{1}{2}(r_c\psi^t+r\psi^L(1-x_1-y))$\\$(x_3\psi_{\parallel}^{n}(\mathcal{A}-\mathcal{V})+(r^2-1)\psi_{\parallel}^{\bar n}(1-x_1')\mathcal{T})$}
        &\tabincell{c}{$-\frac{1}{2}rx_3\psi_{\parallel}^{\bar n}\psi^L(\mathcal{A}-\mathcal{T}-\mathcal{V})$\\$((r^2-1)(1-x_1')-r^2y)$}
        &\tabincell{c}{$\frac{1}{r^2-1}x_3(r_c\psi^t((\mathcal{A}-\mathcal{V})(r^2\psi_{\parallel}^{\bar n}+\psi_{\parallel}^{n})-r^2\psi_{\parallel}^{\bar n}\mathcal{T})$\\$+r\psi^L((\mathcal{A}-\mathcal{V})(r^2y\psi_{\parallel}^{\bar n}+x_3\psi_{\parallel}^{n}-y\psi_{\parallel}^{n})$\\$-r^2y\psi_{\parallel}^{\bar n}\mathcal{T}))+\psi_{\parallel}^{\bar n}x_2'(r\psi^L(x_3(-\mathcal{A}+2\mathcal{T}+\mathcal{V})$\\$-y\mathcal{T})+r_c\psi^t\mathcal{T})+\psi_{\parallel}^{\bar n}x_3'(r\psi^L(x_3$\\$(-\mathcal{A}+2\mathcal{T}+\mathcal{V})-y\mathcal{T})+r_c\psi^t\mathcal{T})+rx_2\psi^L$\\$(\frac{1}{r^2-1}x_3\psi_{\parallel}^{n}(\mathcal{A}-\mathcal{V})+\psi_{\parallel}^{\bar n}x_3'\mathcal{T})+r\psi_{\parallel}^{\bar n}\psi^Lx_2x_2'\mathcal{T}$}\\\\

        $H_{T_{c5}}^{LL}$
        &\tabincell{c}{$0$}
        &\tabincell{c}{$\frac{1}{2}(-r\psi^L(\psi_{\parallel}^{\bar n}r_c^2(\mathcal{A}-\mathcal{V})$\\$+(x_1+y-1)((y-x_3)$\\$(\mathcal{A}(\psi_{\parallel}^{n}+\psi_{\parallel}^{\bar nn}-\psi_{\parallel}^{1})-\psi_{\parallel}^{n}\mathcal{T}$\\$-\mathcal{V}(\psi_{\parallel}^{n}-\psi_{\parallel}^{\bar nn}+\psi_{\parallel}^{1}))$\\$-\psi_{\parallel}^{\bar n}x_3'(\mathcal{A}-\mathcal{V})))+r^2x_2\psi_{\parallel}^{\bar n}r_c$\\$\psi^t(\mathcal{A}-\mathcal{V})-r_c\psi^t(\mathcal{A}+\mathcal{V})$\\$((x_3-y)(\psi_{\parallel}^{\bar nn}-\psi_{\parallel}^{1})-2x_2\psi_{\parallel}^{1})-r^3$\\$\psi_{\parallel}^{\bar n}\psi^L(1-x_1-y)(\mathcal{A}-\mathcal{V})(y-x_3'))$}
        &\tabincell{c}{$-\frac{1}{r^2-1}2r\psi_{\parallel}^{\bar n}r_c^2\psi^L(\mathcal{A}-\mathcal{V})+r_c\psi^t$\\$(2x_2(\mathcal{A}(-r^2\psi_{\parallel}^{1}-r^2\psi_{\parallel}^{\bar n}-\psi_{\parallel}^{1})$\\$-\mathcal{V}(r^2\psi_{\parallel}^{1}-r^2\psi_{\parallel}^{\bar n}+\psi_{\parallel}^{1}))-(r^2+1)$\\$(y-x_3)(\psi_{\parallel}^{\bar nn}-\psi_{\parallel}^{1})(\mathcal{A}+\mathcal{V}))+r\psi^L$\\$(x_1+y-1)(2(r^2-1)\psi_{\parallel}^{\bar n}x_3'(\mathcal{A}-\mathcal{V})$\\$+x_3(\mathcal{A}(-(r^2+1)(\psi_{\parallel}^{\bar nn}-\psi_{\parallel}^{1})-2\psi_{\parallel}^{n})$\\$=8+\mathcal{V}(2\psi_{\parallel}^{n}-(r^2+1)(\psi_{\parallel}^{\bar nn}-\psi_{\parallel}^{1}))$\\$+2\psi_{\parallel}^{n}\mathcal{T})+y(\mathcal{A}((r^2+1)(\psi_{\parallel}^{\bar nn}-\psi_{\parallel}^{1})$\\$-2r^2\psi_{\parallel}^{\bar n}+2\psi_{\parallel}^{n})+\mathcal{V}((r^2+1)(\psi_{\parallel}^{\bar nn}-\psi_{\parallel}^{1})$\\$+2r^2\psi_{\parallel}^{\bar n}-2\psi_{\parallel}^{n})-2\psi_{\parallel}^{n}\mathcal{T}))$}\\\\

        $H_{T_{c5}}^{SP}$
        &\tabincell{c}{$\frac{1}{2}(rr_c^2\psi^L(\mathcal{A}(2\psi_{\parallel}^{1}-\psi_{\parallel}^{\bar n})-\psi_{\parallel}^{\bar n}\mathcal{T}$\\$+\mathcal{V}(2\psi_{\parallel}^{1}+\psi_{\parallel}^{\bar n}))+r_c\psi^t(x_3(r^2\mathcal{A}(\psi_{\parallel}^{1}-\psi_{\parallel}^{\bar nn})$\\$-M(r^2-1)\phi(\psi_{\parallel}^{\bar nn}-\psi_{\parallel}^{1})-r^2\mathcal{V}(\psi_{\parallel}^{\bar nn}-\psi_{\parallel}^{1})$\\$-\psi_{\parallel}^{n}\mathcal{T})+x_2(2M(r^2-1)\psi_{\parallel}^{1}\phi-\psi_{\parallel}^{n}(\mathcal{A}+\mathcal{T}-\mathcal{V}))$\\$+y(r^2\mathcal{A}(\psi_{\parallel}^{\bar nn}-\psi_{\parallel}^{1}+\psi_{\parallel}^{\bar n})+M(r^2-1)\phi$\\$(\psi_{\parallel}^{\bar nn}-\psi_{\parallel}^{1})+r^2\mathcal{V}(\psi_{\parallel}^{\bar nn}-\psi_{\parallel}^{1}-\psi_{\parallel}^{\bar n})+\psi_{\parallel}^{n}\mathcal{T})$\\$-(r^2-1)\psi_{\parallel}^{\bar n}x_3'(\mathcal{A}-\mathcal{V}))+r\psi^L(-(y-x_3)$\\$(y(r^2\mathcal{A}(\psi_{\parallel}^{\bar nn}+\psi_{\parallel}^{1})+M(r^2-1)\phi(\psi_{\parallel}^{\bar nn}-\psi_{\parallel}^{1})$\\$+\mathcal{T}(\psi_{\parallel}^{n}-r^2\psi_{\parallel}^{\bar n})+r^2\mathcal{V}(\psi_{\parallel}^{\bar nn}+\psi_{\parallel}^{1}))$\\$-(r^2-1)x_3'((\psi_{\parallel}^{\bar nn}+\psi_{\parallel}^{1})(\mathcal{A}+\mathcal{V})-\psi_{\parallel}^{\bar n}\mathcal{T})$\\$+x_3(M(1-r^2)\phi(\psi_{\parallel}^{\bar nn}-\psi_{\parallel}^{1})-\psi_{\parallel}^{n}\mathcal{T}))$\\$+(r^2-1)x_2'(y-x_3)(\psi_{\parallel}^{\bar nn}+\psi_{\parallel}^{1})(\mathcal{A}+\mathcal{V})$\\$+x_2(x_3(-M(r^2-1)\phi(\psi_{\parallel}^{\bar nn}-\psi_{\parallel}^{1})-\psi_{\parallel}^{n}\mathcal{T})$\\$+M(r^2-1)y\phi(\psi_{\parallel}^{\bar nn}-\psi_{\parallel}^{1})$\\$+(r^2-1)\psi_{\parallel}^{\bar n}x_3'\mathcal{T}+y\mathcal{T}(\psi_{\parallel}^{n}-r^2\psi_{\parallel}^{\bar n}))))$}
        &\tabincell{c}{$\frac{1}{2}(-r\psi_{\parallel}^{\bar n}r_c^2\psi^L(\mathcal{A}-\mathcal{V})$\\$+r_c\psi^t((r^2-1)\psi_{\parallel}^{\bar n}x_2'(\mathcal{A}-\mathcal{V})$\\$+\psi_{\parallel}^{n}(y-x_3)(-\mathcal{A}+\mathcal{T}+\mathcal{V}))$\\$-r\psi_{\parallel}^{\bar n}\psi^L(y-x_3)(\mathcal{A}-\mathcal{V})$\\$((r^2-1)(1-x_1')-r^2y))$}
        &\tabincell{c}{$-\frac{1}{r^2-1}r\psi^L((\mathcal{A}-\mathcal{V})(2\psi_{\parallel}^{\bar n}r_c^2+(x_3-y)$\\$(\psi_{\parallel}^{\bar n}(1-x_1')+\psi_{\parallel}^{n}(1-x_1-y)))$\\$-M(x_3-y)(1-x_1-y)\phi(\psi_{\parallel}^{\bar nn}-\psi_{\parallel}^{1}))$\\$+r^2r_c\psi^t(\psi_{\parallel}^{\bar n}((\mathcal{A}-\mathcal{V})(x_3'-y)+x_2'(\mathcal{V}-\mathcal{A})$\\$+x_2\mathcal{T})+M\phi((x_3-y)(\psi_{\parallel}^{\bar nn}-\psi_{\parallel}^{1})$\\$-2x_2\psi_{\parallel}^{1}))+r_c\psi^t((y-x_3)(\psi_{\parallel}^{n}(\mathcal{A}-2\mathcal{T}-\mathcal{V})$\\$+M\phi(\psi_{\parallel}^{\bar nn}-\psi_{\parallel}^{1}))+x_2(\psi_{\parallel}^{n}(\mathcal{A}+\mathcal{T}-\mathcal{V})$\\$+2M\psi_{\parallel}^{1}\phi)+\psi_{\parallel}^{\bar n}x_2'(\mathcal{A}-\mathcal{V})-\psi_{\parallel}^{\bar n}x_3'(\mathcal{A}-\mathcal{V}))$\\$+r^3\psi^L(x_3-y)(M(1-x_1-y)\phi(\psi_{\parallel}^{\bar nn}-\psi_{\parallel}^{1})$\\$-\psi_{\parallel}^{\bar n}(\mathcal{A}-\mathcal{V})(1-x_1'-y))$}\\\\

        $H_{T_{c7}}^{LL}$
        &\tabincell{c}{$\frac{1}{2}(-rr_c^2\psi^L(\mathcal{A}(\psi_{\parallel}^{\bar nn}-\psi_{\parallel}^{1}+\psi_{\parallel}^{\bar n})$\\$+\mathcal{V}(\psi_{\parallel}^{\bar nn}-\psi_{\parallel}^{1}-\psi_{\parallel}^{\bar n}))$\\$+r_c\psi^t(y(r^2\mathcal{A}(\psi_{\parallel}^{\bar nn}-\psi_{\parallel}^{1}+\psi_{\parallel}^{\bar n})$\\$-M(r^2-1)\phi(\psi_{\parallel}^{\bar nn}-\psi_{\parallel}^{1})$\\$+r^2\mathcal{V}(\psi_{\parallel}^{\bar nn}-\psi_{\parallel}^{1}-\psi_{\parallel}^{\bar n})-\psi_{\parallel}^{n}\mathcal{T})$\\$+(r^2-1)x_3'(\psi_{\parallel}^{\bar nn}-\psi_{\parallel}^{1})(\mathcal{A}+\mathcal{V})$\\$+r^2x_3\psi_{\parallel}^{\bar n}(\mathcal{A}-\mathcal{V})+r^2(\mathcal{V}(\psi_{\parallel}^{1}-\psi_{\parallel}^{\bar nn}+\psi_{\parallel}^{\bar n})$\\$-\mathcal{A}(\psi_{\parallel}^{\bar nn}-\psi_{\parallel}^{1}+\psi_{\parallel}^{\bar n}))$\\$+x_2(M(r^2-1)\phi(\psi_{\parallel}^{\bar nn}-\psi_{\parallel}^{1})+\psi_{\parallel}^{n}\mathcal{T}))$\\$+r\psi^L(y-x_2)(x_3+y-1)$\\$(M(r^2-1)\phi(\psi_{\parallel}^{\bar nn}-\psi_{\parallel}^{1})+\psi_{\parallel}^{n}\mathcal{T}))$}

        &\tabincell{c}{$\frac{1}{2}(r\psi^L(y-x_2)(x_3+y-1)$\\$(\psi_{\parallel}^{\bar nn}+\psi_{\parallel}^{1})(\mathcal{A}+\mathcal{V})$\\$-r_c\psi^t(-x_2\psi_{\parallel}^{n}(\mathcal{T}-\mathcal{A}+\mathcal{V})$\\$+y(\mathcal{A}(-\psi_{\parallel}^{n}+\psi_{\parallel}^{\bar nn}+\psi_{\parallel}^{1})$\\$+\psi_{\parallel}^{n}\mathcal{T}+\mathcal{V}(\psi_{\parallel}^{n}+\psi_{\parallel}^{\bar nn}+\psi_{\parallel}^{1}))$\\$+(x_3-1)(\psi_{\parallel}^{\bar nn}+\psi_{\parallel}^{1})(\mathcal{A}+\mathcal{V})))$}

        &\tabincell{c}{$-\frac{1}{r^2-1}2r\psi_{\parallel}^{\bar nn}r_c^2\psi^L(\mathcal{A}+\mathcal{V})+r_c\psi^t(x_2(\mathcal{A}(r^2$\\$(2\psi_{\parallel}^{\bar nn}-\psi_{\parallel}^{\bar n})+\psi_{\parallel}^{n})-M(r^2-1)\phi(\psi_{\parallel}^{\bar nn}-\psi_{\parallel}^{1})$\\$+\mathcal{V}(r^2(2\psi_{\parallel}^{\bar nn}+\psi_{\parallel}^{\bar n})-\psi_{\parallel}^{n})-2\psi_{\parallel}^{n}\mathcal{T})+y(\mathcal{A}(-3r^2\psi_{\parallel}^{\bar nn}$\\$+r^2(\psi_{\parallel}^{1}+\psi_{\parallel}^{\bar n})-\psi_{\parallel}^{n}+\psi_{\parallel}^{\bar nn}+\psi_{\parallel}^{1})+M(r^2-1)$\\$\phi(\psi_{\parallel}^{\bar nn}-\psi_{\parallel}^{1})+\mathcal{V}(-3r^2\psi_{\parallel}^{\bar nn}+r^2(\psi_{\parallel}^{1}-\psi_{\parallel}^{\bar n})+\psi_{\parallel}^{n}$\\$+\psi_{\parallel}^{\bar nn}+\psi_{\parallel}^{1})+2\psi_{\parallel}^{n}\mathcal{T})+(\mathcal{A}+\mathcal{V})(x_3(\psi_{\parallel}^{\bar nn}+\psi_{\parallel}^{1})$\\$-(r^2-1)x_3'(\psi_{\parallel}^{\bar nn}-\psi_{\parallel}^{1}))+(\mathcal{A}+\mathcal{V})((r^2-1)\psi_{\parallel}^{\bar nn}$\\$-(r^2+1)\psi_{\parallel}^{1}))-r\psi^L(y-x_2)(x_3(\mathcal{A}$\\$(r^2\psi_{\parallel}^{\bar n}+\psi_{\parallel}^{n}+\psi_{\parallel}^{\bar nn}+\psi_{\parallel}^{1})+M(r^2-1)\phi(\psi_{\parallel}^{\bar nn}-\psi_{\parallel}^{1})$\\$+\mathcal{V}(r^2(-\psi_{\parallel}^{\bar n})-\psi_{\parallel}^{n}+\psi_{\parallel}^{\bar nn}+\psi_{\parallel}^{1}))+(y-1)$\\$(\mathcal{A}(+r^2(\psi_{\parallel}^{1}-r^2\psi_{\parallel}^{\bar nn}+\psi_{\parallel}^{\bar n})+\psi_{\parallel}^{n}+\psi_{\parallel}^{\bar nn}+\psi_{\parallel}^{1})$\\$+M(r^2-1)\phi(\psi_{\parallel}^{\bar nn}-\psi_{\parallel}^{1})+\mathcal{V}(r^2(-\psi_{\parallel}^{\bar nn})$\\$+r^2(\psi_{\parallel}^{1}-\psi_{\parallel}^{\bar n})-\psi_{\parallel}^{n}+\psi_{\parallel}^{\bar nn}+\psi_{\parallel}^{1}))$\\$-(r^2-1)x_3'(\psi_{\parallel}^{\bar nn}-\psi_{\parallel}^{1})(\mathcal{A}+\mathcal{V}))$}\\

        \hline\hline

    \end{tabular}
\end{table}

\begin{table}[H]
	\scriptsize
	\centering
	TABLE~\ref{tab:amp} (continued)
	\newcommand{\tabincell}[2]{\begin{tabular}{@{}#1@{}}#2\end{tabular}}
	\begin{tabular}[t]{lccc}
		\hline\hline
		$$&$\frac{C_1^L}{16M^3}$&$\frac{C_2^L}{16M^2}$&$\frac{C_3^L}{16M}$\\ \hline
		$H_{T_{c7}}^{SP}$
		&\tabincell{c}{$\frac{1}{2}(r\psi^L(r_c^2(\psi_{\parallel}^{\bar n}\mathcal{T}-(\psi_{\parallel}^{\bar nn}+\psi_{\parallel}^{1})(\mathcal{A}+\mathcal{V}))$\\$-(y-x_2)(x_3+y-1)(\psi_{\parallel}^{n}\mathcal{A}$\\$+M\phi(\psi_{\parallel}^{\bar nn}+\psi_{\parallel}^{1})-\psi_{\parallel}^{n}\mathcal{V}))$\\$+r^2r_c\psi^t((\psi_{\parallel}^{\bar nn}+\psi_{\parallel}^{1})((\mathcal{A}+\mathcal{V})(y-x_2')$\\$-M(x_3+y-1)\phi)+\psi_{\parallel}^{\bar n}\mathcal{T}(x_2-y))$\\$+r_c\psi^t((x_3+y-1)(\psi_{\parallel}^{n}\mathcal{A}+M\phi(\psi_{\parallel}^{\bar nn}+\psi_{\parallel}^{1})$\\$-\psi_{\parallel}^{n}\mathcal{V})+x_2'(\psi_{\parallel}^{\bar nn}+\psi_{\parallel}^{1})(\mathcal{A}+\mathcal{V}))$\\$+Mr^3\psi^L(y-x_2)(x_3+y-1)\phi(\psi_{\parallel}^{\bar nn}+\psi_{\parallel}^{1}))$}
		&\tabincell{c}{$\frac{1}{2}(r\psi^L(y-x_2)(x_3+y-1)(\psi_{\parallel}^{\bar nn}-\psi_{\parallel}^{1})$\\$(\mathcal{A}+\mathcal{V})-r_c\psi^t(\mathcal{A}((y-x_2)(\psi_{\parallel}^{\bar nn}-\psi_{\parallel}^{1})$\\$-\psi_{\parallel}^{n}(x_3+y-1))+\psi_{\parallel}^{n}\mathcal{T}(x_3+y-1)$\\$+\mathcal{V}(\psi_{\parallel}^{n}(x_3+y-1)$\\$+(y-x_2)(\psi_{\parallel}^{\bar nn}-\psi_{\parallel}^{1}))))$}
		&\tabincell{c}{$-\frac{1}{r^2-1}r\psi^L(2\psi_{\parallel}^{\bar nn}r_c^2(\mathcal{A}+\mathcal{V})+(x_3+y-1)$\\$((y-x_2)(\mathcal{A}(\psi_{\parallel}^{1}-\psi_{\parallel}^{\bar nn})+M\phi(\psi_{\parallel}^{\bar nn}+\psi_{\parallel}^{1})$\\$+\psi_{\parallel}^{n}\mathcal{T}+\mathcal{V}(\psi_{\parallel}^{1}-\psi_{\parallel}^{\bar nn}))+x_2'(\psi_{\parallel}^{\bar nn}+\psi_{\parallel}^{1})$\\$(\mathcal{A}+\mathcal{V})))+r^2r_c\psi^t(\mathcal{A}(x_2'(\psi_{\parallel}^{\bar nn}+\psi_{\parallel}^{1})-2x_3\psi_{\parallel}^{\bar nn}$\\$-3y\psi_{\parallel}^{\bar nn}-y\psi_{\parallel}^{1}+2\psi_{\parallel}^{\bar nn})+M(x_3+y-1)$\\$\phi(\psi_{\parallel}^{\bar nn}+\psi_{\parallel}^{1})+\mathcal{V}(x_2'(\psi_{\parallel}^{\bar nn}+\psi_{\parallel}^{1})-2x_3\psi_{\parallel}^{\bar nn}$\\$-3y\psi_{\parallel}^{\bar nn}-y\psi_{\parallel}^{1}+2\psi_{\parallel}^{\bar nn})-\psi_{\parallel}^{\bar n}\mathcal{T}(x_3+y-1))$\\$-r_c\psi^t(\mathcal{A}(x_2'(\psi_{\parallel}^{\bar nn}+\psi_{\parallel}^{1})+2x_3\psi_{\parallel}^{n}$\\$+x_2(\psi_{\parallel}^{\bar nn}-\psi_{\parallel}^{1})+2(y-1)\psi_{\parallel}^{n}+y(\psi_{\parallel}^{1}-\psi_{\parallel}^{\bar nn}))$\\$+M(x_3+y-1)\phi(\psi_{\parallel}^{\bar nn}+\psi_{\parallel}^{1})$\\$+\mathcal{V}(\psi_{\parallel}^{\bar nn}(x_2'+x_2-y)+\psi_{\parallel}^{1}(x_2'-x_2+y)$\\$-2\psi_{\parallel}^{n}(x_3+y-1))-\psi_{\parallel}^{n}\mathcal{T}(x_3+y-1))$\\$+r^3\psi^L(x_3+y-1)((\psi_{\parallel}^{\bar nn}+\psi_{\parallel}^{1})(M(x_2-y)\phi$\\$-(\mathcal{A}+\mathcal{V})(x_2'-y))+\psi_{\parallel}^{\bar n}\mathcal{T}(y-x_2))$}\\\\
		
		$H_{T_{d1}}^{LL}$
		&\tabincell{c}{$\frac{1}{2}(r^2r_c\psi^t(\psi_{\parallel}^{\bar n}x_3'(\mathcal{A}-\mathcal{V})-2(1-x_1)$\\$\psi_{\parallel}^{1}(\mathcal{A}+M\phi+\mathcal{V}))+r_c\psi^t((1-x_1)$\\$(2M\psi_{\parallel}^{1}\phi+\psi_{\parallel}^{n}\mathcal{T})-\psi_{\parallel}^{\bar n}x_3'(\mathcal{A}-\mathcal{V}))$\\$+r^3\psi^L(2(1-x_1)\psi_{\parallel}^{1}((\mathcal{A}+\mathcal{V})(x_2'+y-1)$\\$+M(y-x_1)\phi)+x_3'((1-x_1)(\mathcal{A}(2\psi_{\parallel}^{1}-\psi_{\parallel}^{\bar n})$\\$+\mathcal{V}(2\psi_{\parallel}^{1}+\psi_{\parallel}^{\bar n}))-(y-1)\psi_{\parallel}^{\bar n}(\mathcal{A}-\mathcal{V})))$\\$+r\psi^L(x_3'((y-1)\psi_{\parallel}^{\bar n}(\mathcal{A}-\mathcal{V})$\\$-(1-x_1)(\mathcal{A}(2\psi_{\parallel}^{1}-\psi_{\parallel}^{\bar n})$\\$+\mathcal{V}(2\psi_{\parallel}^{1}+\psi_{\parallel}^{\bar n})))-(1-x_1)(2\psi_{\parallel}^{1}x_2'$\\$(\mathcal{A}+\mathcal{V})+(y-x_1)(2M\psi_{\parallel}^{1}\phi+\psi_{\parallel}^{n}\mathcal{T}))))$}
		&\tabincell{c}{$-\frac{1}{2}(r^2-1)\psi_{\parallel}^{\bar n}r_cx_3'\psi^t(-\mathcal{A}+\mathcal{T}+\mathcal{V})$}
		&\tabincell{c}{$-\psi_{\parallel}^{\bar n}x_3'(r_c\psi^t(-2\mathcal{A}+\mathcal{T}+2\mathcal{V})$\\$+r\psi^L\mathcal{T}(y-x_1))$\\$-\frac{1}{r^2-1}(1-x_1)(r\psi^L(y-x_1)-r_c\psi^t)$\\$(\psi_{\parallel}^{n}\mathcal{T}-2M(r^2-1)\psi_{\parallel}^{1}\phi)$}\\\\
		
		$H_{T_{d1}}^{SP}$
		&\tabincell{c}{$0$}
		&\tabincell{c}{$(1-x_1)\psi_{\parallel}^{1}(\mathcal{A}+\mathcal{V})$\\$(r\psi^L(y-x_1)-r_c\psi^t)$\\$+\frac{1}{2}(r-1)r(r+1)\psi_{\parallel}^{\bar n}\psi^L$\\$x_3'(y-x_1)(\mathcal{A}-\mathcal{T}-\mathcal{V})$}
		&\tabincell{c}{$\frac{1}{r^2-1}2(r^2+1)(1-x_1)\psi_{\parallel}^{1}(\mathcal{A}+\mathcal{V})$\\$(r\psi^L(y-x_1)-r_c\psi^t)$\\$+2r\psi_{\parallel}^{\bar n}\psi^Lx_3'(y-x_1)(\mathcal{A}-\mathcal{T}-\mathcal{V})$}\\\\
		
		$H_{T_{d2}}^{LL}$
		&\tabincell{c}{$\frac{1}{2}(r\psi^L(y-x_1)-r_c\psi^t)(x_3\psi_{\parallel}^{n}(\mathcal{A}-\mathcal{V})$\\$+(r^2-1)\psi_{\parallel}^{\bar n}(1-x_1')\mathcal{T})$}
		&\tabincell{c}{$-\frac{1}{2}rx_3\psi_{\parallel}^{\bar n}\psi^L(\mathcal{A}-\mathcal{T}-\mathcal{V})$\\$((r^2-1)x_2'+(r^2-1)x_3'+r^2(y-1))$}
		&\tabincell{c}{$-\frac{1}{r^2-1}r^2\psi_{\parallel}^{\bar n}r_c\psi^t((1-x_1')\mathcal{T}-x_3(-\mathcal{A}+\mathcal{T}+\mathcal{V}))$\\$+r_c\psi^t(x_3\psi_{\parallel}^{n}(\mathcal{A}-\mathcal{V})-\psi_{\parallel}^{\bar n}(1-x_1')\mathcal{T})$\\$+r^3\psi_{\parallel}^{\bar n}\psi^L(x_3((1-x_1')(\mathcal{A}-2\mathcal{T}-\mathcal{V})-(y-1)$\\$(-\mathcal{A}+\mathcal{T}+\mathcal{V}))-(1-x_1')\mathcal{T}(x_2+y-1))$\\$+r\psi^L(\psi_{\parallel}^{\bar n}(1-x_1')\mathcal{T}(x_2+2x_3+y-1)$\\$-x_3(\mathcal{A}-\mathcal{V})(\psi_{\parallel}^{\bar n}(1-x_1')+\psi_{\parallel}^{n}(y-x_1)))$}\\\\
		
	    $H_{T_{d2}}^{SP}$
		&\tabincell{c}{$0$}
		&\tabincell{c}{$\frac{1}{2}r^2x_3\psi_{\parallel}^{\bar n}r_c\psi^t(-\mathcal{A}+\mathcal{T}+\mathcal{V})$}
		&\tabincell{c}{$\frac{1}{r^2-1}2r^2x_3\psi_{\parallel}^{\bar n}r_c\psi^t(-\mathcal{A}+\mathcal{T}+\mathcal{V})$} \\\\
		
		$H_{T_{d6}}^{LL}$
		&\tabincell{c}{$\frac{1}{2}(r\psi^L(\mathcal{A}(r_c^2(\psi_{\parallel}^{\bar n}-2\psi_{\parallel}^{1})$\\$+(1-x_1')(x_2+y-1)(\psi_{\parallel}^{\bar nn}-\psi_{\parallel}^{1})$\\$+\psi_{\parallel}^{\bar n}x_2'(y-x_1)+\psi_{\parallel}^{n}(x_2+y-1)(y-x_1))$\\$-\mathcal{V}(r_c^2(2\psi_{\parallel}^{1}+\psi_{\parallel}^{\bar n})-(1-x_1')(x_2+y-1)$\\$(\psi_{\parallel}^{\bar nn}-\psi_{\parallel}^{1})+\psi_{\parallel}^{\bar n}x_2'(y-x_1)$\\$+\psi_{\parallel}^{n}(x_2+y-1)(y-x_1))+\psi_{\parallel}^{\bar n}r_c^2\mathcal{T}$\\$+M(x_2+y-1)(y-x_1)\phi(\psi_{\parallel}^{\bar nn}+\psi_{\parallel}^{1}))$\\$+r^2r_c\psi^t(\mathcal{A}(x_2+y-1)(\psi_{\parallel}^{\bar nn}+\psi_{\parallel}^{1})$\\$+M\phi((x_2+y-1)(\psi_{\parallel}^{\bar nn}+\psi_{\parallel}^{1})$\\$+2x_3\psi_{\parallel}^{1})-\psi_{\parallel}^{\bar n}\mathcal{T}(x_2'+y-1)$\\$+\mathcal{V}(x_2+y-1)(\psi_{\parallel}^{\bar nn}+\psi_{\parallel}^{1}))$\\$-r_c\psi^t(\psi_{\parallel}^{n}\mathcal{A}(y-x_1)$\\$+M\phi((x_2+y-1)(\psi_{\parallel}^{\bar nn}+\psi_{\parallel}^{1})+2x_3\psi_{\parallel}^{1})$\\$+\mathcal{T}(x_3\psi_{\parallel}^{n}-\psi_{\parallel}^{\bar n}x_2')-\psi_{\parallel}^{n}\mathcal{V}(y-x_1))$\\$-r^3\psi^L(\mathcal{A}((x_2+y-1)(\psi_{\parallel}^{\bar nn}-\psi_{\parallel}^{1})(y-x_1')$\\$+\psi_{\parallel}^{\bar n}(y-x_1)(x_2'+y-1))$\\$+M(x_2+y-1)(y-x_1)$\\$\phi(\psi_{\parallel}^{\bar nn}+\psi_{\parallel}^{1})+\mathcal{V}((x_2+y-1)$\\$(\psi_{\parallel}^{\bar nn}-\psi_{\parallel}^{1})(y-x_1')$\\$-\psi_{\parallel}^{\bar n}(y-x_1)(x_2'+y-1))))$}
		&\tabincell{c}{$\frac{1}{2}(r_c\psi^t(\psi_{\parallel}^{n}(x_2+y-1)$\\$(-\mathcal{A}+\mathcal{T}+\mathcal{V})+(r^2-1)\psi_{\parallel}^{\bar n}x_3'\mathcal{T})$\\$+r\psi_{\parallel}^{\bar n}r_c^2\psi^L\mathcal{T}-r\psi_{\parallel}^{\bar n}\psi^L\mathcal{T}$\\$(x_2+y-1)((r^2-1)$\\$(1-x_1')+r^2(y-1)))$}
		&\tabincell{c}{$-\frac{1}{r^2-1}-r^2r_c\psi^t(M\phi((x_2+y-1)$\\$(\psi_{\parallel}^{\bar nn}+\psi_{\parallel}^{1})+2x_3\psi_{\parallel}^{1})-\psi_{\parallel}^{\bar n}(x_3(\mathcal{A}-\mathcal{V})$\\$+\mathcal{T}(x_2'-x_3'+y-1)))+r_c\psi^t(\psi_{\parallel}^{n}$\\$\mathcal{A}(2(x_2+y-1)+x_3)+M\phi((x_2+y-1)$\\$(\psi_{\parallel}^{\bar nn}+\psi_{\parallel}^{1})+2x_3\psi_{\parallel}^{1})+\mathcal{T}(\psi_{\parallel}^{\bar n}(x_3'-x_2')$\\$-\psi_{\parallel}^{n}(x_2-x_3+y-1))$\\$-\psi_{\parallel}^{n}\mathcal{V}(2(x_2+y-1)+x_3))$\\$-r\psi^L(\mathcal{T}(2\psi_{\parallel}^{\bar n}r_c^2+(x_2+y-1)$\\$(\psi_{\parallel}^{\bar n}x_2'+\psi_{\parallel}^{\bar n}x_3'+\psi_{\parallel}^{n}(x_3+y-1)$\\$+x_2\psi_{\parallel}^{n}))+M(x_2+y-1)$\\$(y-x_1)\phi(\psi_{\parallel}^{\bar nn}+\psi_{\parallel}^{1}))$\\$+r^3\psi^L(x_2+y-1)(M(y-x_1)$\\$\phi(\psi_{\parallel}^{\bar nn}+\psi_{\parallel}^{1})+\psi_{\parallel}^{\bar n}\mathcal{T}(y-x_1'))$}\\
		
		\hline\hline
	\end{tabular}
\end{table}

\begin{table}[H]
	\scriptsize
	\centering
	TABLE~\ref{tab:amp} (continued)
	\newcommand{\tabincell}[2]{\begin{tabular}{@{}#1@{}}#2\end{tabular}}
	\begin{tabular}[t]{lccc}
		\hline\hline
		$$&$\frac{C_1^L}{16M^3}$&$\frac{C_2^L}{16M^2}$&$\frac{C_3^L}{16M}$\\ \hline
		$H_{T_{d6}}^{SP}$
		&\tabincell{c}{$0$}
		&\tabincell{c}{$\frac{1}{2}(-r\psi^L((y-x_1)((x_2+y-1)$\\$(\mathcal{A}(\psi_{\parallel}^{n}+\psi_{\parallel}^{\bar nn}+\psi_{\parallel}^{1})-\psi_{\parallel}^{n}\mathcal{T}+\mathcal{V}(-\psi_{\parallel}^{n}+\psi_{\parallel}^{\bar nn}+\psi_{\parallel}^{1}))$\\$-\psi_{\parallel}^{\bar n}x_2'\mathcal{T})-\psi_{\parallel}^{\bar n}r_c^2\mathcal{T})+r_c\psi^t(\mathcal{A}+\mathcal{V})$\\$((x_2+y-1)(\psi_{\parallel}^{\bar nn}+\psi_{\parallel}^{1})+2x_3\psi_{\parallel}^{1})$\\$+r^2x_3\psi_{\parallel}^{\bar n}r_c\psi^t\mathcal{T}-r^3\psi_{\parallel}^{\bar n}\psi^L\mathcal{T}(y-x_1)(x_2'+y-1))$}
		&\tabincell{c}{$-\frac{1}{r^2-1}-r_c\psi^t((r^2+1)(\mathcal{A}+\mathcal{V})((x_2+y-1)$\\$(\psi_{\parallel}^{\bar nn}+\psi_{\parallel}^{1})+2x_3\psi_{\parallel}^{1})+2r^2x_3\psi_{\parallel}^{\bar n}\mathcal{T})+r\psi^L(y-x_1)$\\$(x_2(\mathcal{A}((r^2+1)(\psi_{\parallel}^{\bar nn}+\psi_{\parallel}^{1})+2\psi_{\parallel}^{n})$\\$+\mathcal{V}((r^2+1)(\psi_{\parallel}^{\bar nn}+\psi_{\parallel}^{1})-2\psi_{\parallel}^{n})$\\$-2\psi_{\parallel}^{n}\mathcal{T})+(y-1)(\mathcal{A}((r^2+1)$\\$(\psi_{\parallel}^{\bar nn}+\psi_{\parallel}^{1})+2\psi_{\parallel}^{n})-2\mathcal{T}(\psi_{\parallel}^{n}-r^2\psi_{\parallel}^{\bar n})+\mathcal{V}((r^2+1)$\\$(\psi_{\parallel}^{\bar nn}+\psi_{\parallel}^{1})-2\psi_{\parallel}^{n}))+2(r^2-1)\psi_{\parallel}^{\bar n}x_2'\mathcal{T})-2r\psi_{\parallel}^{\bar n}r_c^2\psi^L\mathcal{T}$}\\
		
		\hline\hline
	\end{tabular}
\end{table}

\end{appendix}

\end{document}